\documentclass[aps,twocolumn,floatfix,showpacs,superscriptaddress,groupedaddress]{revtex4}  % for review and submission
\usepackage{graphicx}  % needed for figures
\usepackage{subfigure} % for side-by-side figures.
\usepackage{bm}        % for math
\usepackage{amsmath}   % for math
\usepackage{cancel} 
\usepackage{verbatim} 
\begin{document}

\title{Dynamical Stability of Slip-stacking Particles}
\author{Jeffrey Eldred}
\affiliation{Department of Physics, Indiana University, Bloomington, Indiana 47405, USA}
\affiliation{Fermi National Accelerator Laboratory, Batavia, Illinois 60510, USA}
\author{Robert Zwaska}
\affiliation{Fermi National Accelerator Laboratory, Batavia, Illinois 60510, USA}
\date{\today}

\begin{abstract}
We study the stability of particles in slip-stacking configuration, used to nearly double proton beam intensity at Fermilab.  We introduce universal area factors to calculate the available phase space area for any set of beam parameters without individual simulation. We find perturbative solutions for stable particle trajectories. We establish Booster beam quality requirements to achieve 97\% slip-stacking efficiency. We show that slip-stacking dynamics directly correspond to the driven pendulum and to the system of two standing-wave traps moving with respect to each other.
\end{abstract}

\maketitle

\section*{Introduction}
Slip-stacking is integral to high-intensity operation at Fermilab and will likely play a central role in upgrades to the accelerator complex \cite{PIP}\cite{Mariani}\cite{Galambos}. Particle loss in the slip-stacking process is a limiting factor on ultimate performance \cite{PIP} \cite{Brown}. Single-particle dynamics associated with slip-stacking contribute directly to the particle losses. This paper analyzes these dynamics at depth, both analytically and numerically. Our numerical results completely characterize the stable phase-space boundary. We use these results to recommend an upgrade to the Fermilab Booster that would substantially reduce slip-stacking losses.

Our analytical results provide insight into slip-stacking by presenting a perturbative general solution and new parameteric resonances. These results should also be of interest to the greater field of dynamical mathematics because, as we demonstrate, the dynamics of slip-stacking are isomorphic to the well-studied dynamics of the driven pendulum. The analysis in this paper is also intended to facilitate application of slip-stacking to other accelerators and non-accelerator systems with analogous dynamics.

\section*{Background}
Slip-stacking is a particle accelerator configuration that permits two high-energy particle beams of different momenta to use the same transverse space in a cyclic accelerator. The two beams are longitudinally focused by two sets of rf cavities with a small frequency difference between them. Each frequency is tuned to the momentum of one of the beams.

The two azimuthal beam distributions are manipulated as a consequence of their difference in rf frequency. The two beams injected on separated portions of azimuth with a small frequency difference will overlap gradually, allowing injection \cite{Brown}. When the cyclic accelerator is filled, the azimuthal distribution of the beams will coincide at a certain tune and can then be accelerated simultaneously. The accelerating rf cavities operate at the average frequency, capturing both beams as one. The potential beam intensity of a synchrotron is doubled through the application of this technique.

A preliminary study explored the beam dynamics in a 2-rf system \cite{Mills}. The slipping of bunched beams was first demonstrated at the CERN SPS \cite{Boussard} but the emittance growth led to unacceptable particle losses. Fermilab has subsequently implemented slip-stacking operationally since 2004 \cite{MacLachlan}\cite{SeiyaBC}\cite{Brown}. Initially, the higher beam intensity was used to increase antiproton production for proton-antiproton collider experiments \cite{Cd}. Subsequently, slip-stacking was applied to neutrino production for Neutrinos at Main Injector (NuMI) experiments \cite{Minos}\cite{Minerva1}\cite{Minerva2}\cite{Nova}.

Beam-loading effects can impact the effectveness of slip-stacking and were addressed in the Main Injector by the development of a beam-loading compensation system with -14dB feedback and -20dB feedforward \cite{SeiyaB}\cite{DeyKK}\cite{Dey}. The beam-loading effects on slip-stacking in the Recycler will be an order of magnitude weaker than in the Main Injector and can be compensated if necessary.  The typical beam-loading voltage is $\sim$2kV \cite{DeyKK} compared to a typical rf cavity voltage of 90kV\cite{Dey}. In the Main Injector the $R_{sh}/Q$ of the rf cavities is $100 \Omega$ \cite{DeyKK}, while in the Recycler the $R_{sh}/Q$ is $13 \Omega$ \cite{Madrak}. This paper focuses on the constraints on the stable phase-space area from the single-particle dynamics of the two-rf system; direct space-charge effects are an order of magnitude weaker.

For the particular case of slip-stacking at Fermilab, the difference between the two rf frequencies must be equal to the product of the harmonic number of the Booster rf and the cycle rate of the Fermilab Booster. The cycle-rate of the Booster is 15-Hz and therefore $\Delta f = h_{B} f_{B} = 1260$ Hz. A possible upgrade to a 20-Hz Booster is also analyzed, for which $\Delta f = 1680$ Hz. A 20-Hz Booster would enable slip-stacking buckets with substantially greater phase-space area. However, the slip-stacking cyclic accelerator (either the Main Injector or the Recycler) must be able to simultaneously accommodate beams in a range of momentum corresponding to their frequency difference. 

\section*{Single-rf Longitudinal Stability}

The motion of a single particle under the influence of a single stationary rf cavity is described in terms of its phase space coordinates $\phi$ and $\delta$. The phase $\phi$ is the phase of the particle relative to the resonating electromagnetic field in the rf cavity. $\delta$ is the fractional deviation from the reference momentum; $\delta=0$ corresponds to a particle whose revolution frequency $f_{rev}$ is a subharmonic of the frequency of the rf cavity $f_{rf} = h f_{rev}$. The phase-slip factor $\eta$ is used to describe how the revolution period $T$ changes with $\delta$ and is given by $\displaystyle \eta \delta = \Delta T / T$ (see \cite{SYLee}).

The equations of motion associated with the trajectory of a single particle under the influence of a single stationary rf cavity \cite{SYLee} are given by:
\begin{equation} \label{dif1srf}
\dot{\phi} = 2 \pi f_{rev} h \eta \delta,~\dot{\delta} = f_{rev} \frac{eV}{\beta^{2}E} \sin(\phi).
\end{equation}
$V$ is the effective voltage of the rf cavity, $e$ is the charge of the particle, $\beta = v/c$ is the velocity fraction of the speed of light, $E$ is the total energy of the particle. Let $V_{\delta}$ be equal to $\frac{eV}{\beta^{2}E}$, the maximum change in $\delta$ during a single revolution.

For small $\phi$, Eq.~\ref{dif1srf} has a stable solution known as a synchrotron oscillation
\begin{equation} \label{RFsol}
\phi = \rho \sin(\omega_{s}t + \psi)
\end{equation}
where the synchrotron frequency $\omega_{s}$ is given by $\displaystyle \omega_{s} = 2\pi f_{rev} \sqrt{\frac{V_{\delta}h\eta}{2\pi}}$. The amplitude $\rho$ and the initial phase $\psi$ are set by initial conditions. More generally, stable oscillatory motion is bound in phase and momentum by the separatrix
\begin{equation} \label{RFsep}
\delta = \pm \frac{2}{h |\eta|} \frac{\omega_{s}}{\omega_{rev}} \cos\left(\frac{\phi}{2}\right).
\end{equation}

The equations of motion given in Eq.~\ref{dif1srf} are isomorphic to that of a simple pendulum. The stable region of phase space within the separatrix is referred to as the rf bucket and the region outside of the separatrix is referred to as slipping with respect to the bucket.

In contrast, the dynamics of slip-stacking are explicitly time-dependent and there is no simple separatrix delineating the bucket boundary. The lack of a clearly defined bucket confounds beam operation as the incoming particles cannot be conventionally inserted into the bucket. We broaden the term bucket to include cases without a separatrix: a particle trajectory is in a particular rf bucket if the particle phase with respect to the rf cavity is bounded and averages to zero. 

\section*{Slip-stacking and the Driven Pendulum}

The equations of motion for a single particle under the influence of two rf cavities with identical voltage and different frequencies are:
\begin{align} \nonumber
\dot{\phi_{A}} & = 2 \pi f_{rev} h \eta \delta_{A} \\ \label{dif1c}
\dot{\delta_{A}} & = 2 f_{rev} V_{\delta} \sin(\phi_{A}) \cos\left(\frac{\omega_{\phi}t + \phi_{D}}{2}\right).
\end{align}
$\omega_{\phi}$ is what we refer to as the phase-slipping frequency, the angular frequency separation between the two rf cavities $\omega_{\phi} = 2 \pi \Delta f$. $\phi_{A}$ is the average of the phases and $\phi_{D}$ is the difference between the phases for the two rf cavities at $t=0$. Without loss of generality, we eliminate $\phi_{D}$ with $t \rightarrow t - \phi_{D} /\omega_{\phi}$. Applying a substitution to Eq.~\ref{dif1c} yields the corresponding second-order equation of motion:
\begin{equation} \label{dif2c}
\ddot{\phi_{A}} =-2 \omega_{s}^{2} \sin(\phi_{A}) \cos\left(\frac{\omega_{\phi}t}{2}\right).
\end{equation}
The corresponding Hamiltonian is then given by:
\begin{equation} \label{Hc}
H =\pi f_{rev} h \eta \delta_{A}^{2} + f_{rev} V_{\delta} \cos(\phi_{A}) \cos\left(\frac{\omega_{\phi}t}{2}\right).
\end{equation}
This Hamiltonian leads to nonlinear, resonant, and chaotic phase-space trajectories.  We find that this Hamiltonian is isomorphic to that of a pendulum under a sinusoidal driving force (in the absence of gravity). The driven pendulum is a type of nonlinear Mathieu equation that is a subject of ongoing research in computational mathematics \cite{Broer}\cite{Xu}. A canonical form for the driven pendulum may be parameterized as \cite{Broer}: 
\begin{equation} \label{dif2p}
\frac{d^{2}}{d\tau^{2}}x = (a + b \cos(\tau)) \sin(x),~a \geq 0,~b \geq 0.
\end{equation}
Eq.~\ref{dif2p} is obtained from Eq.~\ref{dif2c} under the substitution
\begin{equation} \label{trans}
t \rightarrow (2/\omega_{\phi}) \tau, \quad \phi_{A} \rightarrow x + \pi, \quad b = 8 \left( \frac{\omega_{s}}{\omega_{\phi}} \right)^{2}.
\end{equation}
The parameter $a$ corresponds to the force of gravity and $a=0$ in this analogue.

The accelerator literature \cite{Mills}\cite{Boussard}\cite{MacLachlan} has identified the importance of the slip-stacking parameter
\begin{equation} \label{as} 
\alpha_{s} = \omega_{\phi}/\omega_{s}
\end{equation}
as the criterion for effective slip-stacking. The parameter $b$ in Eq.~\ref{trans} is a function of $\alpha_{s}$, indicating that all nontrivial dynamics of slip-stacking depend only on $\alpha_{s}$. For example, if one slip-stacking configuration has phase-slipping frequency $\omega_{\phi}$ and another configuration with the same $\alpha_{s}$ has phase-slipping frequency $\omega_{\phi}^{\prime}$ then the second phase space diagram is isomorphic to the first where the $\delta$ axis must be scaled by $\omega_{\phi}^{\prime}/\omega_{\phi}$. 

\section*{General Perturbative Solution}

We analyze the dynamics within a slip-stacking bucket by shifting the origin from one synchronized to the average frequency and phase of the two rf cavities to one synchronized to the rf cavity with higher frequency:
\begin{equation} \label{dif1b}
\dot{\phi} = 2 \pi f_{rev} h \eta \delta,~\dot{\delta} = f_{rev} V_{\delta} [\sin(\phi) + \sin(\phi - \omega_{\phi}t )].
\end{equation}
Substituting one equation into the other we have:
\begin{equation} \label{dif2b}
\ddot{\phi} = - \omega_{s}^{2} [\sin(\phi) + \sin(\phi)\cos(\omega_{\phi}t) - \cos(\phi) \sin(\omega_{\phi}t)].
\end{equation}

Eq.~\ref{dif2b} can be expanded into powers of $\phi$ to consider the small $\phi$ perturbation:
\begin{align} \nonumber
\ddot{\phi} = - \omega_{s}^{2} \Bigg\{ & \sum_{k=0}^{\infty} \frac{(-1)^{k}}{(2k+1)!} \phi^{2k+1} [1+\cos(\omega_{\phi}t)]  \\ \label{dif2bex}
 & - \sum_{k=0}^{\infty} \frac{(-1)^{k}}{(2k)!} \phi^{2k} \sin(\omega_{\phi}t)\Bigg\}.
\end{align}

We use the Poincare-Lindstedt method (see Ch. 2 of \cite{Lichtenberg}) to find the perturbative solution to Eq.~\ref{dif2bex} as a linear combination of oscillatory terms. We substitute the case $\phi =0$ into Eq.~\ref{dif2bex} and we solve to generate $\phi = - \alpha_{s}^{2} \sin(\omega_{\phi}t)$. Next we use $\phi = - \alpha_{s}^{2} \sin(\omega_{\phi}t)$ to generate $A_{n} \sin(n \omega_{\phi} t)$ terms. The coefficients $A_{n}$ are of the order $\alpha_{s}^{-2n}$ and do not depend on the initial coordinates of the particle. These terms form the particular solution:
\begin{equation} \label{part}
\phi_{p} = \sum_{n=1}^{\infty} A_{n} \sin (n \omega_{\phi}t).
\end{equation}
There is no stable equilibrium point inside of the bucket. The particular solution is analogous to a moving bucket center; we term the trajectory of a particle at the moving bucket center to be quasi-synchronous because the frequency spectrum will depend only on harmonics of $\omega_{\phi}$.

We continue the perturbation to the general case by using $\phi = \phi_{p} {+ \rho \sin(\omega_{s}t+\psi)}$, the sum of the particular solution (Eq.~\ref{part}) and the small-oscillation single-rf solution (Eq.~\ref{RFsol}). Using $\phi =\phi_{p} + \rho \sin(\omega_{s}t+\psi)$ generates terms of the form $B_{m,n} {\sin[m(1+\sigma)\omega_{s}t+n\omega_{\phi}t+m\psi]}$. The shift in the synchrotron oscillation frequency $\sigma$ is a necessary contribution to the coefficient of ${\sin[(1+\sigma)\omega_{s}t+\psi]}$ in Eq.~\ref{dif2bex} to counterbalance the contribution made by the cross-multiplication of higher order $B_{m,n} {\sin[m(1+\sigma)\omega_{s}t+n\omega_{\phi}t+m\psi]}$ terms. For any integer $m>0$ and any integer $n$, the coefficients $B_{m,n}$ are of the order $\rho^{m} \alpha_{s}^{-2|n|}$; except when $m$ is even and $n=0$, in which case the coefficients $B_{m,0}$ are of the order $\rho^{m} \alpha_{s}^{-2}$. Writing out the full perturbative solution, we have:
\begin{align} \nonumber
\phi = & \sum_{n=1}^{\infty} A_{n} \sin (n \omega_{\phi}t) \\ \nonumber
& + \sum_{m=1}^{\infty} \sum_{n=1}^{\infty} B_{m,n} \sin[m(1+\sigma)\omega_{s}t+n\omega_{\phi}t+m\psi] \\ \label{pert} 
& + \sum_{m=1}^{\infty} \sum_{n=1}^{\infty} B_{m,-n} \sin[m(1+\sigma)\omega_{s}t-n\omega_{\phi}t+m\psi].
\end{align}
The trajectory of a particle in a slip-stacking rf bucket is referred to as a rotating solution in the driven-pendulum literature. The particular solution was previously obtained by Zhang and Ma \cite{Zhang}. An alternate perturbative approach for the general solution in implicit form is given in \cite{Lenci}. We are first to find a general and explicit solution.

The perturbative solution for the small oscillations around the moving bucket center can be expressed in coefficients up to order $\alpha_{s}^{-4}$ and $\rho \alpha_{s}^{-2}$. The derivation shown in the appendix leads to the equations of motion:
\begin{align} \nonumber
\phi = & A_{1} \sin(\omega_{\phi}t) + A_{2} \sin(2\omega_{\phi}t) \\ \nonumber
& + \rho \sin[(1+\sigma)\omega_{s}t+\psi] \\ \nonumber
& + B_{1,1} \sin[(1+\sigma)\omega_{s}t+\omega_{\phi}t+\psi] \\ \label{pert1p}
& + B_{1,-1} \sin[(1+\sigma)\omega_{s}t-\omega_{\phi}t+\psi]. \\ \label{pert1d}
\delta = & \frac{1}{2\pi f_{rev} h \eta} \dot{\phi}. \\ \label{pert1A1}
A_{1} =& -\frac{1}{\alpha_{s}^{2}-1}. \\ \label{pert1A2}
A_{2} =& \frac{1}{(2\alpha_{s})^{2} - 1} \left( \frac{A_{1}}{2} \right). \\ \label{pert1B11}
B_{1,\pm 1} =& \frac{\alpha_{s}^{-1}}{\alpha_{s} \pm 2} \left( \frac{\rho}{2} \right). \\ \label{pert1e}
\sigma =& \frac{3}{4} \alpha_{s}^{-4}.
\end{align}
The parameters $\rho$ and $\psi$ are determined by initial conditions, shown explicitly in the appendix. We are the first to discover and calculate $\sigma$, the synchrotron frequency shift from a slip-stacking perturbation. Generally, particles within a slip-stacking bucket will undergo synchrotron oscillations at a higher frequency than the corresponding single-rf bucket.

Substituting the perturbative terms from Eq.~\ref{pert} into Eq.~\ref{dif2b} indicates that a new parametric resonance will occur wherever $m\omega_{s} (1+\sigma) = n \omega_{\phi}$. For example, the $\rho^{m} \alpha_{s}^{-2(n-1)}\sin[m(1+\sigma)\omega t - (n-1) \omega_{\phi}t + m\psi]$ term will be multiplied by $\cos(\omega_{\phi} t)$ in Eq.~\ref{dif2b} and lead to a growth term proportional to $\rho^{m} \alpha_{s}^{-2(n-1)} \sin(m\psi)$. The case where $m \omega_{s} = \omega_{\phi}$ was previously investigated by Mills \cite{Mills}. An analytical description of the stable phase-space boundary would require a complete determination of the cases in which parametric resonances lead to particle loss; this may be the subject of future work.

\section*{Stability Maps \& Area Factors}

\begin{figure}[htp]
	\centering
    \includegraphics[scale=0.42]{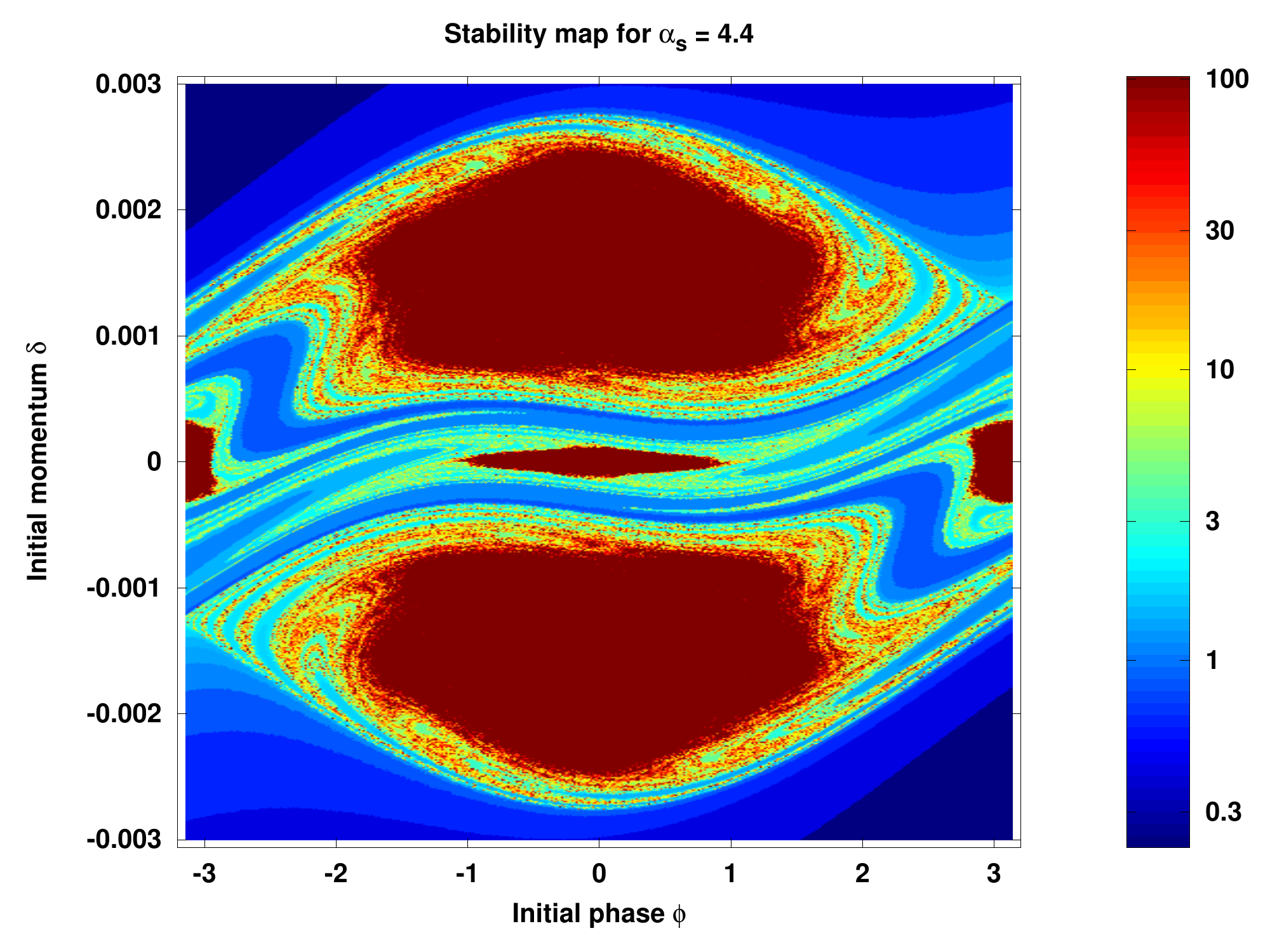}
  \caption{Stability of initial coordinates for $\alpha_{s} = 4.4$. The color corresponds to the number of synchrotron periods a particle with the corresponding initial coordinates survives before it is lost. The two large stable regions correspond to the higher and lower rf buckets where beam is injected and maintained. The two stable regions along the $\delta_{A} = 0$ axis are created by the interaction between the two rf cavities.}
  \label{M}
\end{figure}

\begin{figure}[htp]
\centering
 \begin{subfigure}
  \centering
  \includegraphics[scale=0.4]{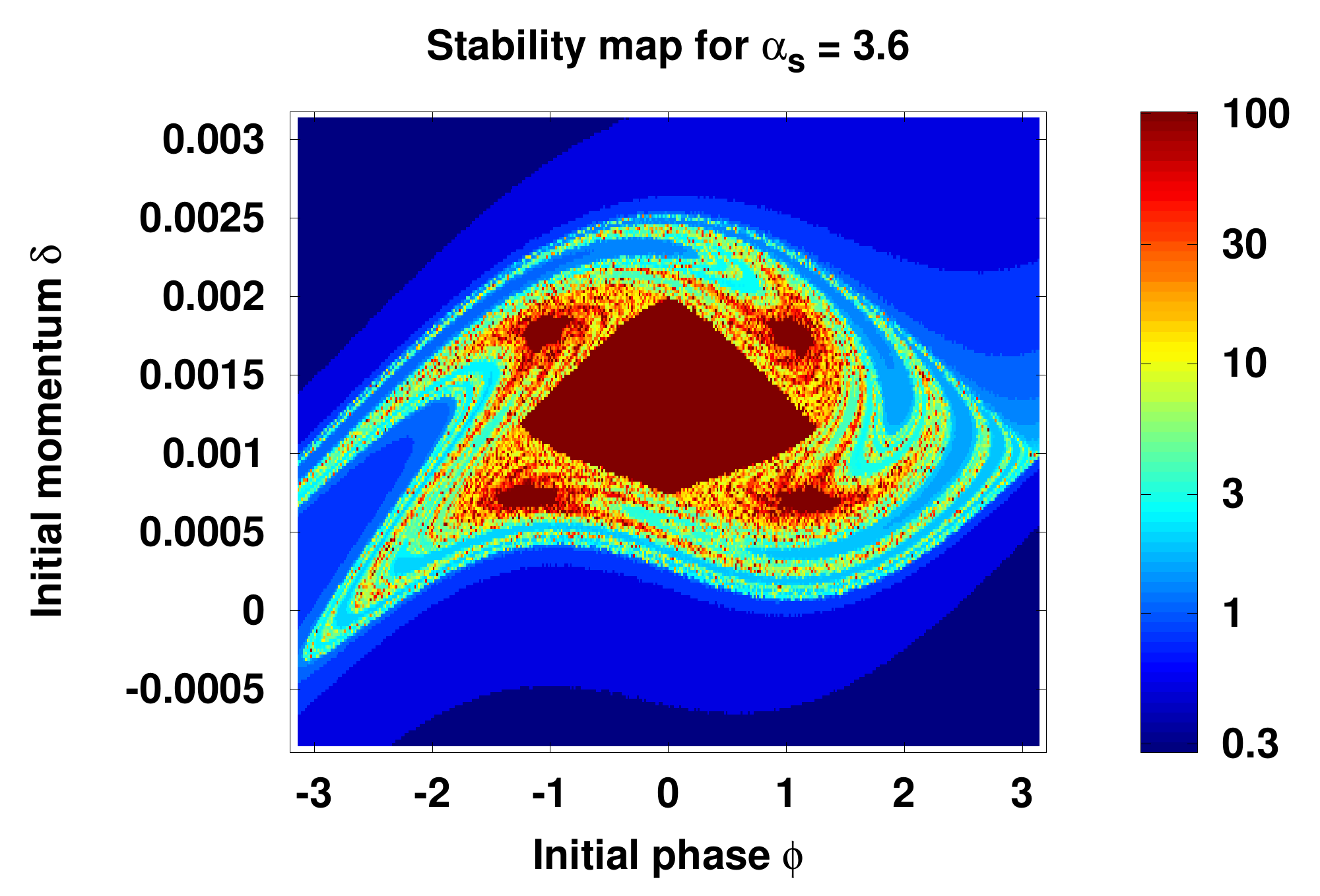}
 \end{subfigure}
 \begin{subfigure}
  \centering
  \includegraphics[scale=0.4]{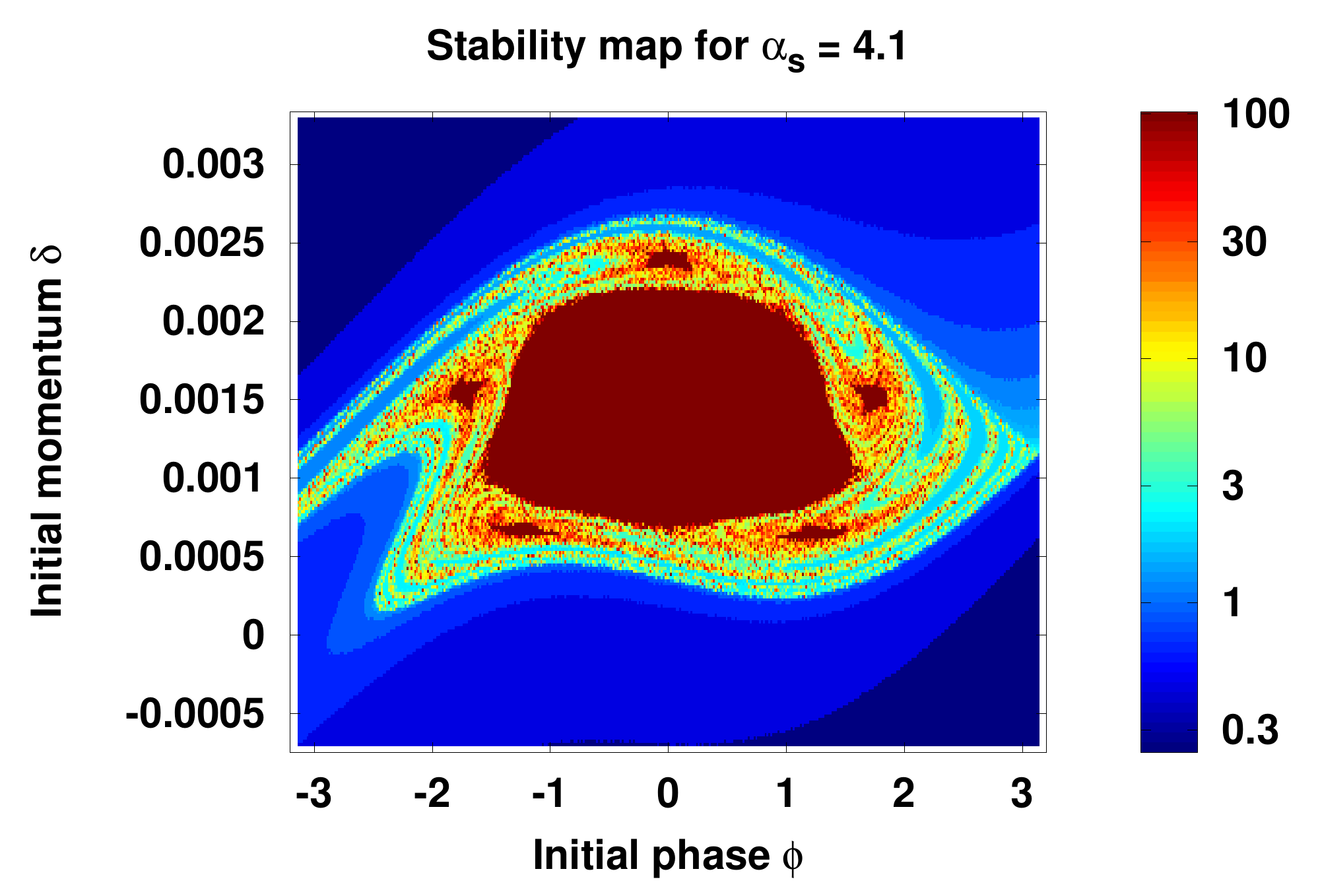}
 \end{subfigure}
 \caption{Stability of initial coordinates for selected values of $\alpha_{s}$.  The color corresponds to the number of synchrotron periods a particle with the corresponding initial coordinates survives before it is lost. On the left, $\alpha_{s} = 3.6$ and four resonance islands can be seen due to the $\omega_{s} (1+\sigma) = 4 \omega_{\phi}$ resonance. On the right, $\alpha_{s} = 4.1$ and five resonance islands can be seen due to the $\omega_{s} (1+\sigma) = 5 \omega_{\phi}$ resonance.}
 \label{res}
\end{figure}

The size and shape of slip-stacking buckets determine which portion of an injected beam distribution is lost. In application, lost particles migrate to an incorrect azimuthal location and consequently collide into the beampipe during injection, extraction, or acceleration \cite{Brown}. We map the stability of initial particle positions by integrating the equations of motion for each position. The integration is iterated for a sufficiently large number of revolutions (at least 30 synchrotron periods). The stability of the particle is tested after every phase-slipping period. A particle is considered lost if its phase with respect to each of the first rf cavity, the second rf cavity, and the average of the two rf cavities, is larger than a certain cut-off (we used $3\pi/2$). The remaining particles therefore belong to one of four stable regions shown in Fig.~\ref{M}: one for the higher frequency, one for the lower frequency, one for the average frequency and average phase, and one with average frequency but $\pi$ offset from the average phase. These two stable regions at the average frequency are the original examples of dynamic stabilization \cite{Butikov}. Fig.~\ref{res} shows the stability of initial coordinates in the higher bucket for $\alpha_{s} = 3.6$ and $\alpha_{s} = 4.1$, in which the effects from slip-stacking resonances are evident. The supplemental material \cite{SM} shows the stability maps of the higher slip-stacking bucket for values of $\alpha_{s}$ from 2 to 8 in descending 0.1 increments.

\begin{figure}[htp]
	\centering
    \includegraphics[scale=0.42]{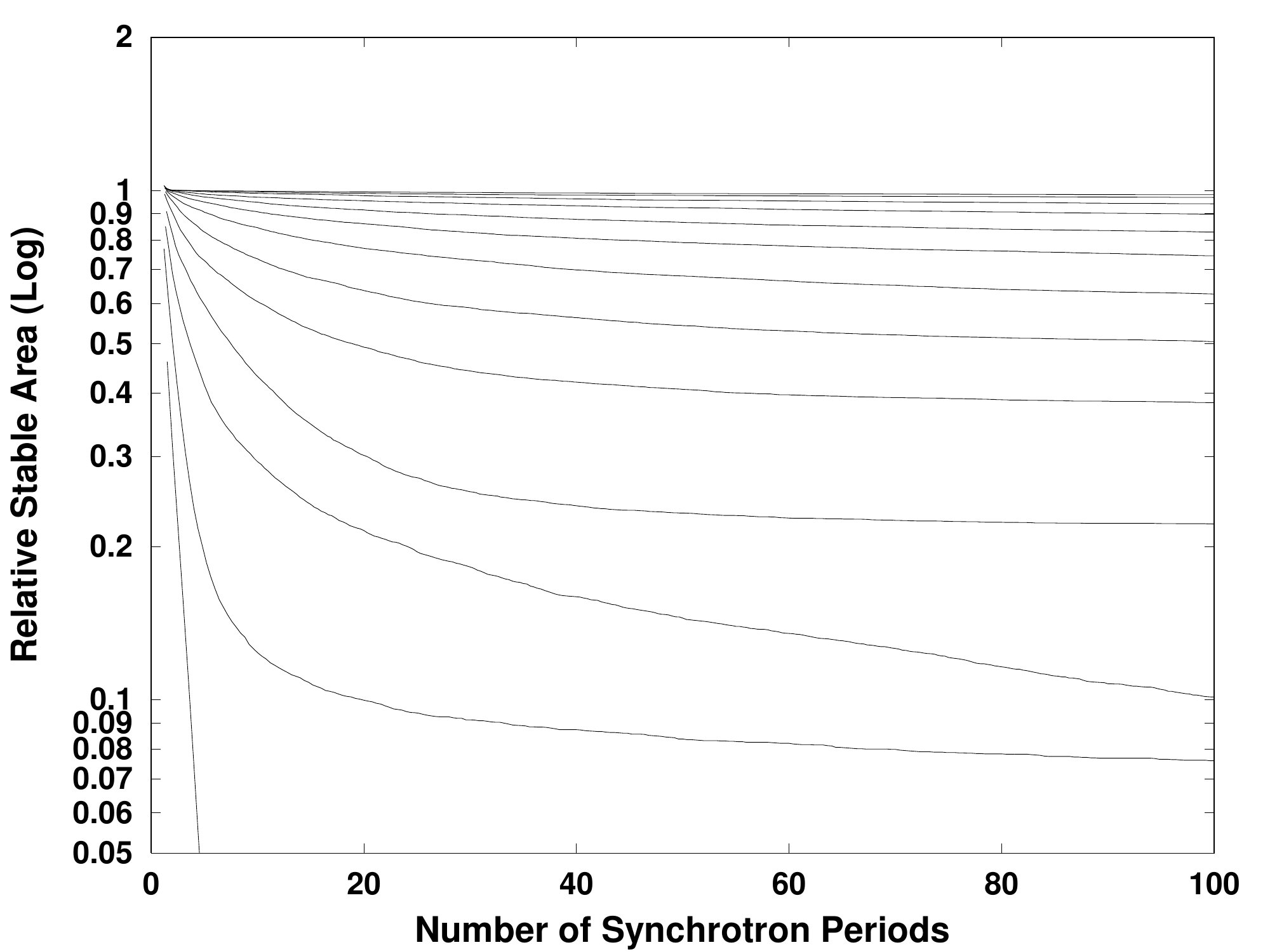}
  \caption{The stable area of the slip-stacking bucket relative to a single rf bucket, is plotted on a log scale and plotted over time. Each curve corresponds to a simulation with a different value of $\alpha_{s}$ with $\alpha_{s} =$ 2.0, 2.5, 3.0, 3.5, 4.0, 4.5, 5.0, 5.5, 6.0, 6.5, 7.0, 7.5, 8.0 (going from the bottom line to the top line). The rapid losses at the beginning corresponds to regions of phase space in which particles rapidly slip by both slip-stacking buckets. In the next phase, the metastable particle loss occurs asymptotically.}
  \label{L}
\end{figure}

We find some trajectories are ``metastable'' because they lead to particle loss only after thousands of revolutions. The stable phase-space area as a function of time is shown in Fig.~\ref{L} for several values of $\alpha_{s}$.

The bucket area is computed as the product of the total number of ultimately surviving points and the cell area. We define the slip-stacking area factor $F(\alpha_{s}) = \mathcal{A}_{s}/\mathcal{A}_{0}$ as the ratio of the slip-stacking bucket area to that of a single-rf bucket with the same rf voltage and frequency. The area factor follows the notation of Lee (Ch. 3.II of \cite{SYLee})  for accelerating beams, in which the ratio of running bucket area to stationary bucket area is used. Particles in the bucket are described by Eq.~\ref{pert} with finite coefficients, therefore the bucket area is conserved. Consequently $F(\alpha_{s})$ does not depend on the initial rf phase difference used to generate the stability map. We write the phase space area ($\phi \cdot \delta$ units) using $F(\alpha_{s})$:
\begin{equation} \label{area}
\mathcal{A}_{s} = \mathcal{A}_{0} F(\alpha_{s}) = \frac{16}{h |\eta|} \frac{\omega_{s}}{\omega_{rev}} F(\alpha_{s}).
\end{equation}

\begin{figure}[htp]
	\centering
    \includegraphics[scale=0.42]{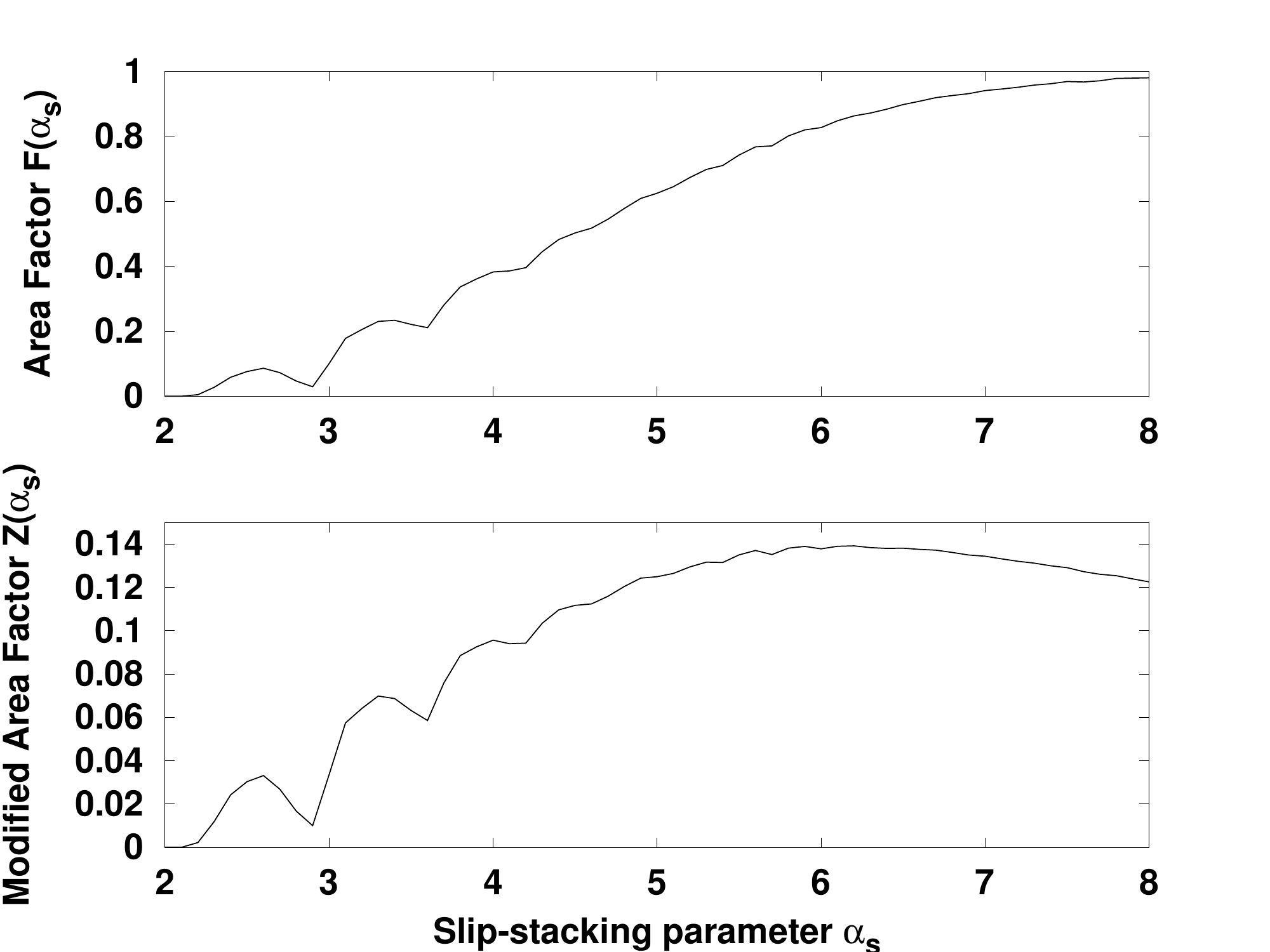}
  \caption{(a) The slip-stacking area factor as a function of $\alpha_{s}$. As $\alpha_{s}$ increases the distance between the rf buckets becomes greater, the buckets become more independent, and the slip-stacking bucket area approaches the single-rf bucket area. \\ (b) The modified slip-stacking area factor as a function of $\alpha_{s}$. The modified slip-stacking area factor is maximized near $\alpha_{s} = 6.2$.}
  \label{Fs}
\end{figure}

Fig.~\ref{Fs}(a) plots the numerically derived slip-stacking area factor $F(\alpha_{s})$. Using Fig.~\ref{Fs}(a) with Eq.~\ref{area} provides the first method for calculating the slip-stacking stable phase-space area without requiring each case to be simulated individually. $F(\alpha_{s})$ increases rapidly above $\alpha_{s} \approx 3$ and asymptotically approaches 1. $F(\alpha_{s})$ has several local minimum where resonances are crossed; this loss of area occurs when large amplitude trajectories have a parametric resonance and therefore does not occur at precise integer values of $\alpha_{s}$.

In application, slip-stacking is tuned to maximize stable phase-space area while holding $\omega_{\phi}$ constant. The value of $\omega_{\phi}$ is generally constrained by gross features of the accelerators, for example the harmonic number and cycle time. The slip-stacking parameter $\alpha_{s}$ is tuned through changing $\omega_{s}$ which is proportional to the square root of the applied rf voltage. Furthermore $\omega_{s}$ changes the bucket area by both the slip-stacking area factor $F(\alpha_{s})$ and the single-rf bucket area, so there is an optimal voltage in which phase space area is maximized. We rewrite Eq.~\ref{area} to separate the parameters that are held constant from those dependent on $\alpha_{s}$:

\begin{equation} \label{area2}
\mathcal{A}_{s} = \frac{16}{h |\eta|} \frac{\omega_{\phi}}{\omega_{rev}} \left( \frac{F(\alpha_{s})}{\alpha_{s}} \right) = \frac{16}{h |\eta|} \frac{\omega_{\phi}}{\omega_{rev}} Z(\alpha_{s}).
\end{equation}

This modified area factor $Z(\alpha_{s})$ is graphed in Fig.~\ref{Fs}(b). $Z(\alpha_{s})$ is maximal near $\alpha_{s} = 6.2$ and when considering other optimization criteria 5.5 to 7 is is a practical tuning range for $\alpha_{s}$.

\section*{Injection Efficiency, Emittance, and Aspect Ratio}

The stability maps can also be used to analyze injection scenarios, by weighting the (scaled) stability maps according to a distribution that represents the number of incoming particles injected into that region of phase-space. We used this technique to identify the greatest longitudinal emittance an incoming Gaussian-distributed beam could have and still achieve 97\% injection efficiency at its optimal value of $\alpha_{s}$. The longitudinal beam emittance is given in Eq.~\ref{lemit} below:
\begin{equation} \label{lemit}
\epsilon = \pi \sigma_{p}\sigma_{T},~\epsilon_{97\%} = 2.17^{2} \pi \sigma_{p}\sigma_{T}
\end{equation}

\begin{figure}[htp]
	\centering
    \includegraphics[scale=0.42]{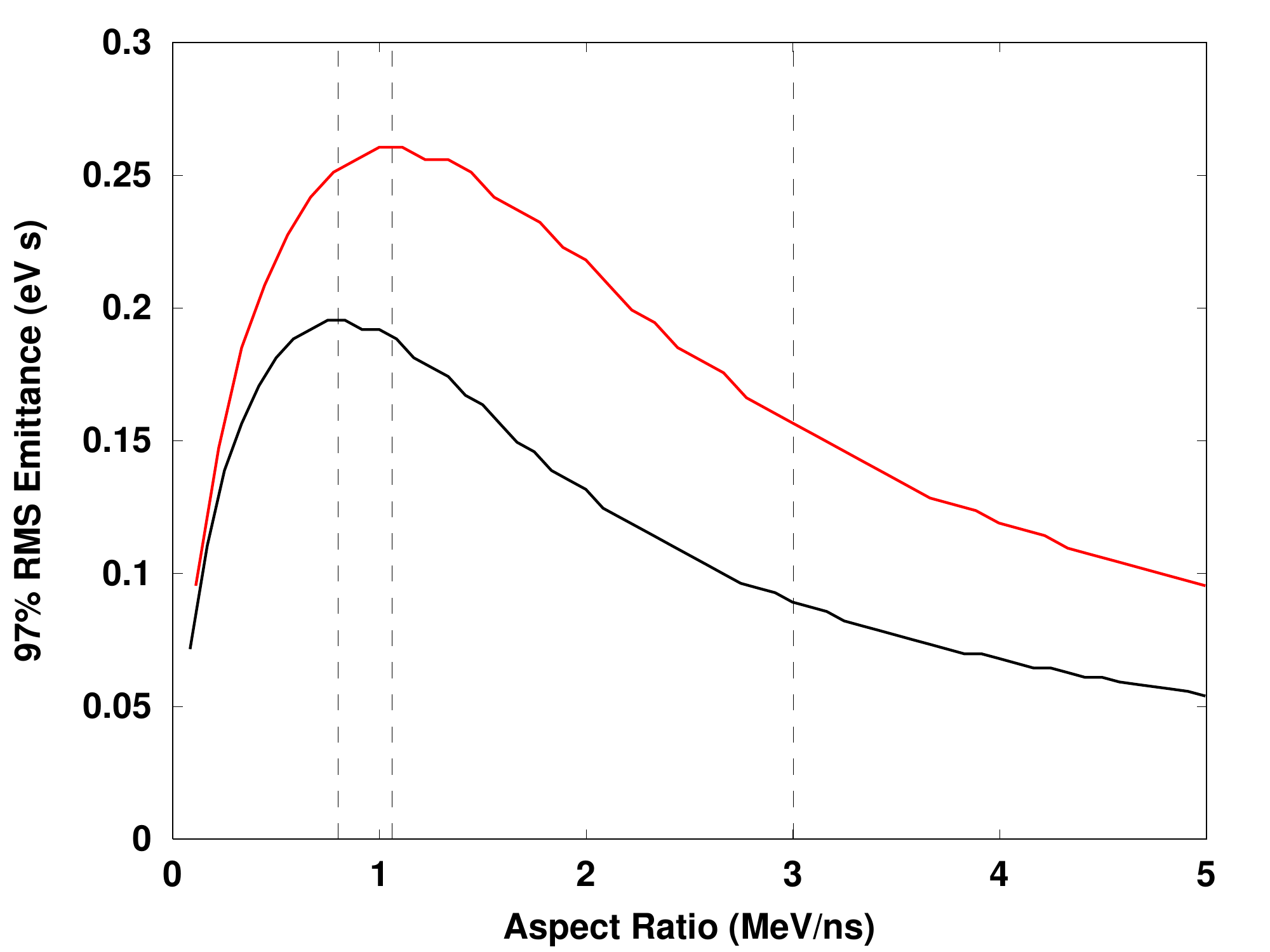}
  \caption{The maximum 97\% emittance at 97\% efficiency (at an optimal value of $\alpha_{s}$) is shown as a function of aspect ratio. The bottom line (black) is for the 15-Hz Booster cycle-rate (status quo) and top line (red) is for 20-Hz Booster cycle-rate (proposed upgrade). The vertical dashed lines represent the nominal aspect ratios given in Table~\ref{Param}.}
  \label{ARe}
\end{figure}

\begin{figure}[htp]
	\centering
    \includegraphics[scale=0.42]{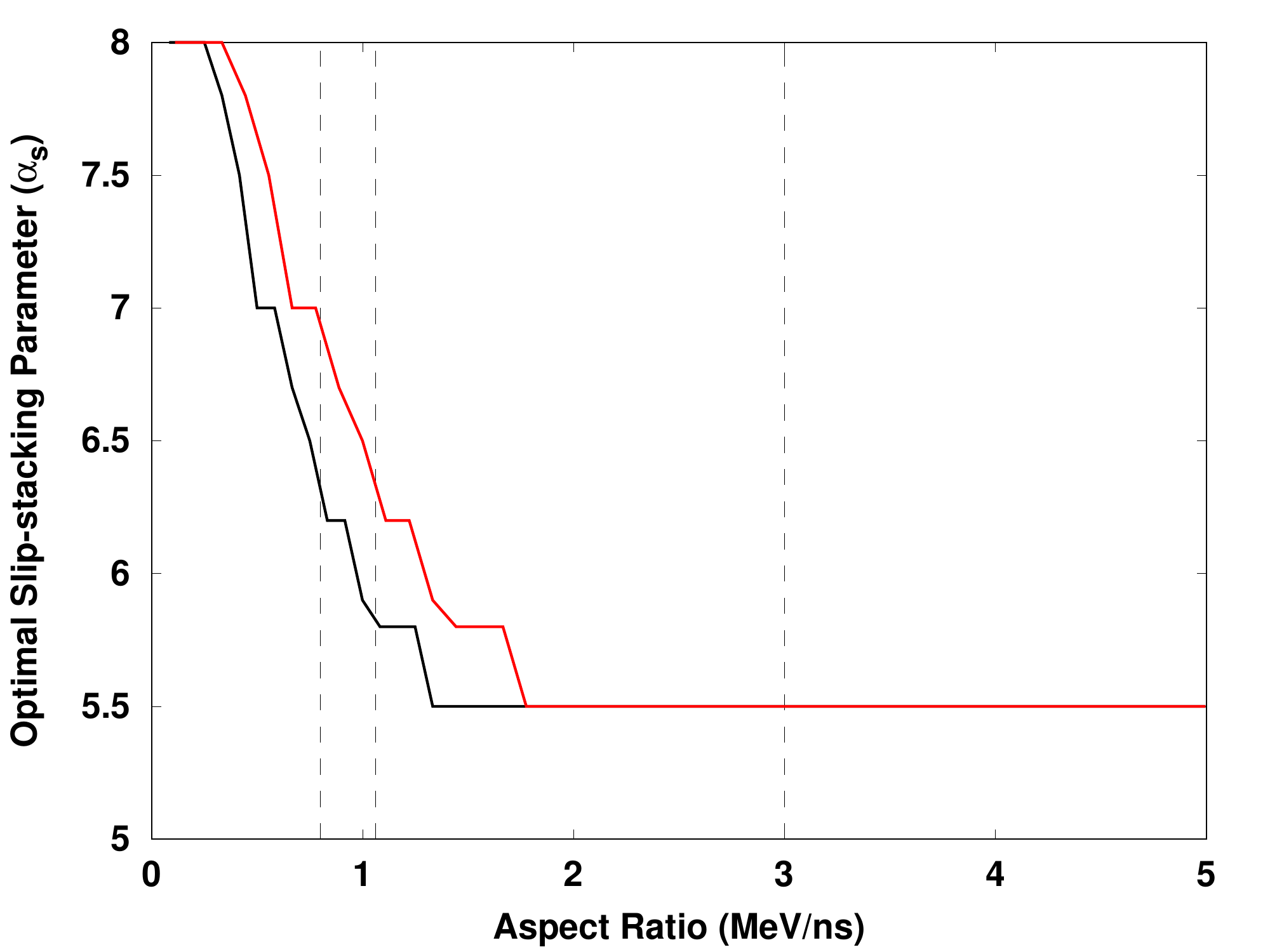}
  \caption{The optimal slip-stacking parameter $\alpha_{s}$ for maximum emittance (at 97\% efficiency) is shown as a function of aspect ratio. Values of $\alpha_{s}$ greater than 8 are not evaluated. The bottom line (black) is for the 15-Hz Booster cycle-rate (status quo) and top line (red) is for 20-Hz Booster cycle-rate (proposed upgrade). The vertical dashed lines represent the nominal aspect ratios given in Table~\ref{Param}.}
  \label{ARa}
\end{figure}

\begin{figure}[htp]
	\centering
    \includegraphics[scale=0.42]{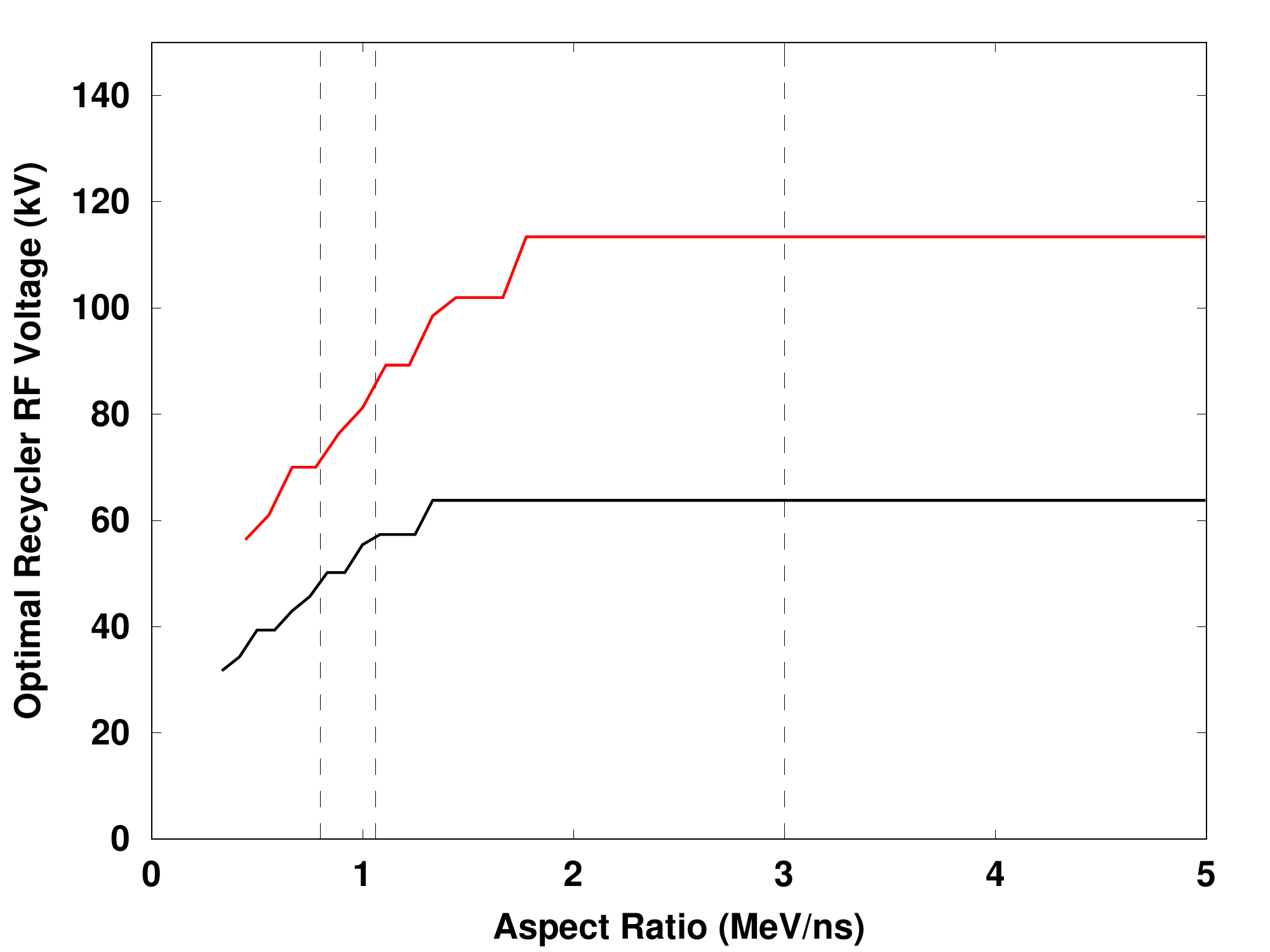}
  \caption{The optimal Recycler rf voltage for maximum emittance (at 97\% efficiency) is shown as a function of aspect ratio. The bottom line (black) is for the 15-Hz Booster cycle-rate (status quo) and top line (red) is for 20-Hz Booster cycle-rate (proposed upgrade). The vertical dashed lines represent the nominal aspect ratios given in Table~\ref{Param}.}
  \label{ARv}
\end{figure}

The current accelerator upgrade proposal, Proton Improvement Plan II (PIP-II)~\cite{PIP}, defines a minimum 97\% slip-stacking efficiency required to maintain current loss levels while increasing intensity. Fig.~\ref{ARe} shows the 97\% longitudinal emittance as a function of aspect ratio and demonstrates the consequences of a mismatched injection into a slip-stacking bucket. Fig.~\ref{ARa} shows the optimal value of $\alpha_{s}$ as a function of aspect ratio. The optimal value of $\alpha_{s}$ determines the optimal rf cavity voltage, shown in Fig.~\ref{ARv}. These results were obtaining using parameter values specific to slip-stacking in the Fermilab Recycler (see Table.~\ref{Param}).

\begin{table}
\begin{tabular}{| l | l |}
\hline
Recycler Kinetic Energy ($E$) & 8 GeV \\
Recycler Reference RF freq. ($f$) & 52.8 MHz \\
Recycler Harmonic number ($h$) & 588 \\
Recycler Phase-slip factor ($\eta$) & -8.6*$10^{-3}$ \\
Maximum Recycler RF Voltage ($V$) & 2 $\times$ 150 kV \\
Booster harmonic number ($h_{B}$) & 84 \\ 
Booster cycle rate ($f_{B}$) & 15/20 Hz \\
Difference in Recycler RF freq. ($\Delta f$) & 1260/1680 Hz \\
\hline
Nominal Booster emittance ($\epsilon_{97\%}$) & 0.12 eV$\cdot$s \\
Nominal Booster Aspect Ratio & 3.00 MeV/ns \\
Nominal Recycler Aspect Ratio (100 kV) & 1.06 MeV/ns \\
Nominal Recycler Aspect Ratio (57 kV) & 0.80 MeV/ns \\
\hline
\end{tabular}
\caption{Recycler and Booster parameters used in analysis.}
\label{Param}
\end{table}

A nominal value for the Booster emittance is 0.12 eV$\cdot$s \cite{SeiyaCD}. The Fermilab Booster uses bunch rotation via quadrupole excitation \cite{Yang}\cite{Ahrens}, with parameters that are actively tuned to minimize losses. With bunch rotation, the aspect ratio of at least 2.6 MeV/ns is achievable at extraction from the Booster \cite{SeiyaCD}. At Recycler rf cavity voltage $V_{0} =$ 100kV, the slip-stacking parameter for the Recycler is $\alpha_{s}(V_{0}) \approx 4.39$ for a 15-Hz Booster cycle-rate and $\alpha_{s}(V_{0}) \approx 5.86$ for a 20-Hz Booster cycle-rate. For other voltages, the Recycler slip-stacking parameter is given by $\alpha_{s}(V) = \alpha_{s}(V_{0}) \sqrt{V/V_{0}}$.

We examine the 97\% efficiency benchmark not only for a 15-Hz Booster cycle-rate but also for a proposed 20-Hz Booster cycle-rate. A 20-Hz Booster cycle-rate would necessarily increase the phase-slipping frequency (rf frequency separation) by a factor of 4/3 and therefore the bucket height would increase by a factor of 4/3 (and have the same $\alpha_{s}$). Slip-stacking losses could be reduced by a factor 4-10 for the same emittance \cite{Eldred}. Alternatively, the 97\% efficiency benchmark is achievable at an emittance up to a factor of $\sim$1.7 greater. Other implications of a 20-Hz Booster cycle-rate are discussed in a Fermilab technical memo \cite{Eldred}. A 20-Hz Booster cycle-rate is clearly superior for high-intensity operation.

The scaling symmetry used to analyze the 20-Hz Booster cycle-rate can generalized. An optimization at phase-slipping frequency $\omega_{\phi}$ and aspect ratio $r$ is equivalent to an optimization at phase-slipping frequency $\omega_{\phi}^{\prime}$ and aspect ratio $(\omega_{\phi}^{\prime}/\omega_{\phi})r$. The same optimal slip-stacking parameter would be obtained at a higher synchrotron frequency $(\omega_{\phi}^{\prime}/\omega_{\phi}) \omega_{s}$, increasing the rf voltage at $(\omega_{\phi}^{\prime}/\omega_{\phi})r$ to $(\omega_{\phi}^{\prime}/\omega_{\phi})^{2}V$.

\section*{Application to Other Physical Systems}
In general, the dynamics discussed in this paper apply to any system governed by (nearly) identical sinusoidal potentials moving with respect to each other (or equivalently, a sinusoidal potential oscillating in amplitude). This is relevant to standing wave traps which are used in optical and acoustic physics and are instances of a controllable sinusoidal potential. Optical lattices are a type of standing wave trap used in ultracold atomic physics. Two optical lattices moving with respect to each other could occupy the same transverse space yet focus two groups of atoms with independent momenta. We know of no such experiment that utilizes this optical slip-stacking laser configuration, but we believe it may be relevant to at least two optical applications: trap accumulation \cite{Davies} and atomic collisions \cite{Jaksch}. If $M$ is the mass of the atom, $V_{H}$ is the potential barrier height, and $v$ is the relative velocity between the two standing wave traps, then the optical slip-stacking parameter is given by:
\begin{equation} \label{opt}
\alpha_{s} = v \sqrt{\frac{M}{2 V_{H}}}.
\end{equation}
The relative velocity $v$ can be calculated from the frequency difference in the standing waves $v = \Delta f$ \cite{Clairon}. Our approach suggests a value of $\alpha_{s}$ equal to at least 5 for efficient stacking.

Acoustic standing waves can be used to trap small spheres or droplets in a sinusoidal potential originally described by Gor'kov \cite{Gorkov}. This technique has grown in sophistication and application \cite{Foresti}\cite{Shields}. Eq.~\ref{opt} would also determine the stability of objects (with mass $M$) in a possible acoustic slip-stacking configuration.

\section*{Conclusion}

In summary, we have provided a framework for addressing both the trajectory and stability of particles in a slip-stacking potential. We introduce the slip-stacking area factor $F(\alpha_{s})$ and the modified area factor $Z(\alpha_{s})$ as tools to calculate the stable slip-stacking bucket area for any combination of accelerator parameters. We introduce the quasi-synchronous particle trajectory and provide a perturbative solution near it. We describe a series of new parametric resonances in slip-stacking. We provide a general method for analyzing slip-stacking injection scenarios and describe the implications for the operation and upgrades of the Fermilab Booster. We identify for the first time how the dynamics of slip-stacking correspond to the driven pendulum and moving standing wave traps.

\section*{Acknowledgments}

This work is supported in part by grants from the US Department of Energy under contract DE-FG02-12ER41800 and the National Science Foundation NSF PHY-1205431. Special thanks to SY Lee for providing a crucial mentoring role immediately prior to the beginning of this research.

\appendix
\section{Derivation of Perturbative Solution}

In this paper we show that the second-order equation of motion for a single particle in a slip-stacking bucket is given by Eq.~\ref{dif2b} and the perturbative solution is given by Eq.~\ref{pert}. Recall also, that the coefficients $A_{n}$ are of order $\alpha_{s}^{-2n}$ and $B_{m,n}$ are of order $\rho^{m} \alpha_{s}^{-2|n|}$ except the $B_{2k,0}$ coefficients which are of the order $\rho^{2k} \alpha_{s}^{-2}$. The coefficient $B_{1,0}$ is defined to be equal to $\rho$. The parameters $\rho$ and $\psi$ are set by initial conditions.

For clarity, we adopt a short-hand notation for the oscillatory terms as follows:
\begin{align} \nonumber
& \sin[m(1+\sigma)\omega_{s}t+n\omega_{\phi}t+m\psi] \equiv s_{m,n}, \\ \label{sh}
& \cos[m(1+\sigma)\omega_{s}t+n\omega_{\phi}t+m\psi] \equiv c_{m,n}.
\end{align}

In this appendix, we explicitly obtain the perturbative solution up to order $\alpha_{s}^{-4}$ and $\rho \alpha_{s}^{-2}$ (or equivalently, all coefficients up to order $\alpha_{s}^{-5}$ with $\rho \sim \alpha_{s}^{-3}$). Therefore, we use the coefficients $A_{1}$, $A_{2}$, $\rho$, $B_{1,1}$ and $B_{1,-1}$; all other coefficients are neglected at this precision. We start by assuming a solution form and will demonstrate it to be self-consistent:
\begin{equation} \label{ansatz}
\phi = A_{1} s_{0,1} + A_{2} s_{0,2} + \rho s_{1,0} + B_{1,1} s_{1,1} + B_{1,-1} s_{1,-1}.
\end{equation}

It will be sufficient to substitute this expression into the form of Eq.~\ref{dif2b} expanded up to second order in $\phi$:
\begin{equation} \label{dif2bex2}
\omega_{\phi}^{-2} \ddot{\phi} = - \alpha_{s}^{-2} \left[ \phi (1+ c_{0,1} ) - s_{0,1} + \frac{1}{2} \phi^{2} s_{0,1} \right].
\end{equation}
Here we have divided both sides by $\omega_{\phi}^{2}$ to make the order of the perturbation terms more explicit.

Next we can substitute the solution given in Eq.~\ref{ansatz} into Eq.~\ref{dif2bex2} to calculate the coefficients. 
We write out the left-hand side (LHS):
\begin{align} \nonumber
&-A_{1} s_{0,1} -4 A_{1} s_{0,2} \\ \nonumber
&- \alpha_{s}^{-2} (1+\sigma)^{2} \rho s_{1,0} \\ \nonumber
&- [\alpha_{s}^{-1} (1+\sigma) + 1]^{2} B_{1,1} s_{1,1} \\ \label{lhs}
&- [\alpha_{s}^{-1} (1+\sigma) - 1]^{2} B_{1,-1} s_{1,-1} =
\end{align}

We write out the right-hand side (RHS):
\begin{align} \nonumber
= -\alpha_{s}^{-2} \bigg(& A_{1}s_{0,1} + A_{1}s_{0,1}c_{0,1} +\frac{1}{2}A_{1}^{2} s_{0,1} s_{0,1} s_{0,1} \\ \nonumber
& + A_{2}s_{0,2} + A_{2}s_{0,2}c_{0,1} - s_{0,1} \\ \nonumber
& + \frac{1}{2}\cancel{A_{2}^{2}} s_{0,2} s_{0,2} s_{0,1} + \cancel{A_{1}A_{2}} s_{0,1} s_{0,2} s_{0,1} \\ \nonumber
& + \rho s_{1,0} + \rho s_{1,0} c_{0,1} \\ \nonumber
& + B_{1,1} s_{1,1} + B_{1,1} s_{1,1} c_{0,1} \\ \nonumber
& + B_{1,-1} s_{1,-1} + B_{1,-1} s_{1,-1} c_{0,1} \\ \nonumber
& + \frac{1}{2}\cancel{(\rho s_{1,0} + B_{1,1} s_{1,1} + B_{1,1} s_{1,-1})^{2}} s_{0,1} \\ \nonumber
& + A_{1} \rho s_{0,1} s_{1,0} s_{0,1} + \cancel{A_{2} \rho} s_{0,2} s_{1,0} s_{0,1} \\ \nonumber
& + \cancel{A_{1} B_{1,1}} s_{0,1} s_{1,1} s_{0,1} + \cancel{A_{1} B_{1,-1}} s_{0,1} s_{1,-1} s_{0,1} \\ \label{rhs}
& + \cancel{A_{2} B_{1,1}} s_{0,2} s_{1,1} s_{0,1} + \cancel{A_{2} B_{1,-1}} s_{0,2} s_{1,-1} s_{0,1}\bigg).
\end{align}
where the crossed-out terms are higher order than the precision of this analysis. We rewrite the RHS without these immediately negligible terms, using the trigonometric product-to-sum rules, and grouping by the oscillatory term:
\begin{align} \nonumber
= -\alpha_{s}^{-2} \bigg[ & \left(A_{1} - 1 + \frac{1}{2}A_{2} + \frac{3}{8} A_{1}^{2}\right) s_{0,1} \\ \nonumber
& + \left(\frac{1}{2}A_{1} + A_{2}\right)s_{0,2} \\ \nonumber
& + \left(\rho + \frac{1}{2} B_{1,1} + \frac{1}{2} B_{1,-1} + \frac{1}{2}A_{1} \rho \right)s_{1,0} \\ \nonumber
& + \left(\frac{1}{2}\rho +B_{1,1}\right)s_{1,1} + \left(\frac{1}{2}\rho +B_{1,-1}\right)s_{1,-1} \\ \nonumber
& + \left(\frac{1}{2}A_{2} - \frac{1}{8}A_{1}^{2}\right) s_{0,3} \\ \nonumber
& + \left(\frac{1}{2}B_{1,1} - \frac{1}{4} A_{1} \rho \right) s_{1,2} \\ \label{rhs2}
& + \left(\frac{1}{2}B_{1,-1} - \frac{1}{4} A_{1} \rho \right) s_{1,-2} \bigg].
\end{align}
We then equate Eq.~\ref{lhs} with Eq.~\ref{rhs2} for all time. Each oscillatory term corresponds to its own equation:
\begin{align} \label{eq01}
-A_{1} &= -\alpha_{s}^{-2} \left(A_{1} - 1 + \frac{1}{2}A_{2} + \frac{3}{8} A_{1}^{2}\right). \\  \label{eq02}
-4 A_{2} &= -\alpha_{s}^{-2} \left(\frac{1}{2}A_{1} + A_{2}\right). \\ \label{eq10}
- \alpha_{s}^{-2} (1+\sigma)^{2} \rho &=  -\alpha_{s}^{-2} \left(\rho + \frac{1}{2} B_{1,1} + \frac{1}{2} B_{1,-1} + \frac{1}{2}A_{1} \rho \right).
\end{align}
\begin{align}\label{eq11}
- [\alpha_{s}^{-1} (1+\sigma) + 1]^{2} B_{1,1} &= -\alpha_{s}^{-2} \left(\frac{1}{2}\rho +B_{1,1}\right). \\ \label{eq1n1}
- [\alpha_{s}^{-1} (1+\sigma) - 1]^{2} B_{1,-1} &=  -\alpha_{s}^{-2} \left(\frac{1}{2}\rho +B_{1,-1}\right).
\end{align}
\begin{align}\label{eq03}
0 &= -\alpha_{s}^{-2} \left(\frac{1}{2}A_{2} - \frac{1}{8}A_{1}^{2}\right). \\ \label{eq12}
0 &= -\frac{1}{2} \alpha_{s}^{-2} B_{1,1} + \frac{1}{4} \alpha_{s}^{-2} A_{1} \rho. \\  \label{eq1n2}
0 &= -\frac{1}{2} \alpha_{s}^{-2} B_{1,-1} + \frac{1}{4} \alpha_{s}^{-2} A_{1} \rho.
\end{align}
Solving Eq.~\ref{eq02} for $A_{2}$ we obtain:
\begin{equation} \label{a2}
A_{2} = \frac{1}{4 - \alpha_{s}^{-2}} \left( \frac{A_{1}}{2} \right) = \frac{1}{(2\alpha_{s})^{2} - 1} \left( \frac{A_{1}}{2} \right).
\end{equation}
Solving Eq.~\ref{eq01} for the linear $A_{1}$ terms we obtain:
\begin{equation} \label{a1a}
A_{1} = \frac{\alpha_{s}^{-2}}{1 - \alpha_{s}^{-2}} \left(-1 + \frac{1}{2}A_{2} + \frac{3}{8} A_{1}^{2}\right).
\end{equation}
Since $A_{1}$ of the order $\alpha_{s}^{-2}$ (as expected) then the $\alpha_{s}^{-2} A_{2}$ term and the $\alpha_{s}^{-2} A_{1}^{2}$ are of order $\alpha_{s}^{-6}$ and are neglected. Rewriting Eq.~\ref{eq01} to reflect this, we have:
\begin{equation} \label{a1b}
A_{1} = - \frac{\alpha_{s}^{-2}}{1 - \alpha_{s}^{-2}} = - \frac{1}{\alpha_{s}^{2}-1}.
\end{equation}

All expressions on the RHS side of Eq.~\ref{eq03} are of order $\alpha_{s}^{-6}$, therefore we are self-consistent to exclude the $A_{3}$ term.

We solve Eq.~\ref{eq11} and Eq.~\ref{eq1n1} for $B_{1,\pm 1}$ and obtain:
\begin{align} \nonumber 
B_{1,\pm 1} &= \frac{ \alpha_{s}^{-2} }{[\alpha_{s}^{-1} \pm 1]^{2} -\alpha_{s}^{-2}} \left( \frac{\rho}{2} \right) \\ \nonumber
&= \frac{\alpha_{s}^{-2}}{1 \pm 2 \alpha_{s}^{-1}} \left( \frac{\rho}{2} \right) \\ \label{b11}
&= \frac{\alpha_{s}^{-1}}{\alpha_{s} \pm 2} \left( \frac{\rho}{2} \right).
\end{align}
where $\sigma$ makes a negligible contribution to $B_{1,\pm 1}$.

All expressions on the RHS of Eq.~\ref{eq12} and Eq.~\ref{eq1n2} are of order $\alpha_{s}^{-4} \rho$, therefore we are self-consistent to exclude the $B_{1,2}$ and $B_{1,-2}$ terms.

For Eq.~\ref{eq10} we call the $\sigma^{2}$ terms negligible, subtract the bare $\rho$ term from each side and solve for $\sigma$ to obtain the shift in synchrotron frequency:
\begin{equation} \label{s1}
\sigma = \frac{1}{4} \left(\frac{B_{1,1}+B_{1,-1}}{\rho}+A_{1}\right).
\end{equation}
To calculate Eq.~\ref{s1} first we must calculate:
\begin{align} \nonumber
\frac{B_{1,1}+B_{1,-1}}{\rho} & = \frac{\alpha_{s}^{-1}}{2} \left( \frac{1}{\alpha_{s}+2} + \frac{1}{\alpha_{s}-2} \right) \\ \nonumber
& = \frac{\alpha_{s}^{-1}}{2} \left( \frac{\alpha_{s}-2+\alpha_{s}+2}{\alpha_{s}^{2}-4} \right) \\ \label{BpB}
& = \frac{1}{\alpha_{s}^{2}-4}.
\end{align}
We substitute Eq.~\ref{BpB} into Eq.~\ref{s1} to obtain:
\begin{align} \nonumber
\sigma &= \frac{1}{4} \left( \frac{1}{\alpha_{s}^{2}-4} - \frac{1}{\alpha_{s}^{2}-1} \right) \\ \nonumber
&= \frac{1}{4} \left( \frac{\alpha_{s}^{2}-1 - \alpha_{s}^{2}+4}{(\alpha_{s}^{2}-4)(\alpha_{s}^{2}-1)} \right) \\
&= \frac{3}{4} \alpha_{s}^{-4}.
\end{align}
To summarize, we write these coefficients together:
\begin{align}  \label{fa1}
A_{1} &= - \frac{1}{\alpha_{s}^{2}-1} = - \alpha_{s}^{-2} (1 + \alpha_{s}^{-2}). \\  \label{fa2}
A_{2} &= \frac{1}{(2\alpha_{s})^{2} - 1} \left( \frac{A_{1}}{2} \right) = - \frac{1}{8} \alpha_{s}^{-4}. \\  \label{fb11}
B_{1,\pm 1} &= \frac{\alpha_{s}^{-1}}{\alpha_{s} \pm 2} \left( \frac{\rho}{2} \right). \\ \label{fs}
\sigma &= \frac{1}{4} \left(\frac{B_{1,1}+B_{1,-1}}{\rho}+A_{1}\right) = \frac{3}{4} \alpha_{s}^{-4}.
\end{align}
\section*{Derivation of $\rho$ and $\psi$ from Initial Conditions}
At this the synchrotron amplitude $\rho$ and initial synchrotron phase $\psi$ are still undetermined, but we can express them in terms of the initial coordinates. At time $t=0$ (which is fixed to a time in which the relative phase between the rf cavities is zero), let $\delta = \delta_{0}$ and $\phi = \phi_{0}$.

We have shown that $\phi$ takes the form give in Eq.~\ref{ansatz}. We evaluate this expression for $\phi$ at $t=0$, leave the short-hand notation (Eq.~\ref{sh}), and find $\phi_{0}$ in terms of $\rho$ and $\psi$:
\begin{align} \label{phi0}
\phi_{0} =( \rho + B_{1,1} + B_{1,-1}) \sin(\psi).
\end{align}
We can calculate $\delta$ from our solution for $\phi$ by taking the derivative:
\begin{align}
\delta &= \frac{1}{2\pi f_{rev} h \eta} & & \! \! \! \! \dot{\phi}. \\ \nonumber
\delta &= \frac{\omega_{\phi}}{2\pi f_{rev} h \eta} \bigg\{ \! \! \! \! \! \! & & A_{1} c_{0,1} + 2 A_{2} c_{0,2} + \alpha_{s}^{-1} (1+\sigma) \rho c_{1,0} \\ \nonumber
& & & + [\alpha_{s}^{-1} (1+\sigma) + 1] B_{1,1} c_{1,1} \\
& & & + [\alpha_{s}^{-1} (1+\sigma) - 1] B_{1,-1} c_{1,-1} \bigg\}.
\end{align}
We evaluate this expression for $\delta$ at $t=0$, leave the short-hand notation (Eq.~\ref{sh}), and find $\delta_{0}$ in terms of $\rho$ and $\psi$:
\begin{align} \nonumber
\delta_{0} = \frac{\omega_{\phi}}{2\pi f_{rev} h \eta} \bigg\{ A_{1} &+ 2 A_{2} \\ \nonumber
 + \bigg[& \left( \frac{1+\sigma}{\alpha_{s}}\right) ( \rho + B_{1,1} + B_{1,-1} ) \\ \label{delta0} 
& + \alpha_{s} (B_{1,1} - B_{1,-1})\bigg]\cos(\psi) \bigg\}.
\end{align}
Next we solve Eq.~\ref{phi0} and Eq.~\ref{delta0} for $\Phi_{0} = \rho\sin(\psi)$ and $\Delta_{0} = \rho\cos(\psi)$:
\begin{align}
\label{P1}
\Phi_{0} = & \left(1 + \frac{B_{1,1} + B_{1,-1}}{\rho}\right)^{-1} \phi_{0}.
\end{align}
\begin{align} \nonumber
\Delta_{0} = \alpha_{s} \bigg[& 1 + \frac{B_{1,1} + B_{1,-1}}{\rho} + \alpha_{s} \frac{B_{1,1} - B_{1,-1}}{\rho} \\ \nonumber
& + \sigma \left(1 + \frac{B_{1,1} + B_{1,-1}}{\rho}\right) \bigg]^{-1} \\ \label{D1}
\times \bigg(& \frac{2\pi f_{rev} h \eta}{\omega_{\phi}} \delta_{0} - A_{1} - 2 A_{2} \bigg).
\end{align}
$\phi_{0}$ and $\delta_{0}$ have been translated by the initial position of the bucket center and rescaled to obtain the expressions for $\Phi_{0}$ and $\Delta_{0}$. Using $\Phi_{0} = \rho\sin(\psi)$ and $\Delta_{0} = \rho\cos(\psi)$, the solution for $\rho$ and $\psi$ be written as: 
\begin{align} \label{rho}
\rho &= \sqrt{ \Phi_{0}^{2} + \Delta_{0}^{2}}. \\ \label{psi}
\psi &= \operatorname{sgn}(\Phi_{0})\arccos\left( \frac{\Delta_{0}}{\rho} \right).
\end{align}
Eq.~\ref{P1} and Eq.~\ref{D1} can be further simplified by writing the $B_{1,1}$ and $B_{1,-1}$ terms explicitly in terms of $\alpha_{s}$. We calculate:
\begin{align} \nonumber
\alpha_{s} \frac{B_{1,1} - B_{1,-1}}{\rho} &= \frac{1}{2} \left( \frac{1}{\alpha_{s}+2} - \frac{1}{\alpha_{s}-2} \right) \\ \nonumber
&= \frac{1}{2} \left( \frac{\alpha_{s}-2-\alpha_{s}-2}{\alpha_{s}^{2}-4} \right) \\ \label{BmB}
&= \frac{-2}{\alpha_{s}^{2}-4}.
\end{align}
We apply Eq.~\ref{BpB} and Eq.~\ref{BmB} to Eq.~\ref{P1} and Eq.~\ref{D1} to obtain:
\begin{align} \label{P2}
\Phi_{0} =& \left(1 + \frac{1}{\alpha_{s}^{2}-4}\right)^{-1} \! \! \! \! \phi_{0} = \frac{\alpha_{s}^{2}-4}{\alpha_{s}^{2}-3} \phi_{0}. \\ \nonumber
\Delta_{0} =& \alpha_{s} \left[1 + \frac{1}{\alpha_{s}^{2}-4} - \frac{2}{\alpha_{s}^{2}-4} + \sigma \left(1 + \frac{1}{\alpha_{s}^{2}-4} \right) \right]^{-1} \\
&\times \left( \frac{2\pi f_{rev} h \eta}{\omega_{\phi}} \delta_{0} - A_{1} - 2 A_{2} \right). \\ \label{D2}
\Delta_{0} =& \frac{\alpha_{s}(\alpha_{s}^{2}-4)}{(\alpha_{s}^{2}-5)+\sigma \alpha_{s}^{2}} \left( \frac{2\pi f_{rev} h \eta}{\omega_{\phi}} \delta_{0} - A_{1} - 2 A_{2} \right).
\end{align}
$\Phi_{0}$ and $\Delta_{0}$ are fully expanded as follows:
\begin{align} \label{P3}
\Phi_{0}  = & \phi_{0} (1 - \alpha_{s}^{-2} -3 \alpha_{s}^{-4}). \\ \nonumber
\Delta_{0} = & \alpha_{s} \left( \frac{2\pi f_{rev} h \eta}{\omega_{\phi}} \delta_{0} + \alpha_{s}^{-2} + \frac{5}{4}\alpha_{s}^{-4} \right) \\ \label{D3}
& \times \left(1+\alpha_{s}^{-2} + \frac{17}{4}\alpha_{s}^{-4}\right).
\end{align}
Eq.~\ref{P3} and Eq.~\ref{D3} can then be substituted into Eq.~\ref{rho} and Eq.~\ref{psi} to obtain $\rho$ and $\psi$ respectively. Then $\rho$ can be substituted into Eq.~\ref{fb11} to obtain $B_{1,1}$ and $B_{1,-1}$.


\begin{thebibliography}{99}

  \bibitem{PIP} P.~Derwent {\sl et al.} Proton Improvement Plan-II December 2013, 2013.
  
  \bibitem{Mariani} C. Mariani (LBNE/DUSEL Collaborations), in Proceedings of the Neutrino Oscillation Workshop, Otranto, Italy, 2011, edited by P.~Bernardini, G.~Fogli, and E.~Lisi.
  
  \bibitem{Galambos} J.~Galambos, M.~Bai, and S.~Nagaitsev, Snowmass Workshop on Frontier Capability Summary Report, 2013.
  
  \bibitem{Brown} B.~C.~Brown, P.~Adamson, D.~Capista, W.~Chou, I.~Kourbanis, D.~K~Morris, K.~Seiya, G.~H.~Wu, and M.~J.~Yang, Phys. Rev. ST Accel. Beams {\bf 16}, 071001 (2013).

  \bibitem{Mills} F.~E.~Mills, Brookhaven National Laboratory Report No. 15936, 1971.

  \bibitem{Boussard} D.~Boussard and Y.~Mizumachi, IEEE Trans. Nucl. Sci. {\bf 26}, 3623 (1979).

  \bibitem{MacLachlan} J.~A.~MacLachlan, Fermi National Accelerator Laboratory Report No. 0711, 2001.
  
  \bibitem{SeiyaBC} K.~Seiya, T.~Berenc, B.~Chase, W.~Chou, J.~Dey, P.~Joireman, I.~Kourbanis, J.~Reid, and D.~Wildman, in Proceedings of HB2006, Tsukuba, Japan, 2006, edited by Y.~H.~Chin, H.~Yoshikawa, and M.~Ikegami.
  
  \bibitem{Cd} T.~Aaltonen {\sl et al.} (CDF and D0 Collaborations), Phys. Rev. D {\bf 86}, 092003 (2012).
  
  \bibitem{Minos} P.~Adamson {\sl et al.} (MINOS Collaboration), Phys. Rev. Lett. {\bf 110}, 251801 (2013).
  
  \bibitem{Minerva1} L.~Fields {\sl et al.} (MINERvA Collaboration), Phys. Rev. Lett. {\bf 111}, 022501 (2013).
    
  \bibitem{Minerva2} G.~A.~Fiorentini {\sl et al.} (MINERvA Collaboration), Phys. Rev. Lett. {\bf 111}, 022502 (2013).
  
  \bibitem{Nova} M.~Muether, Nucl. in Proceedings of Neutrino Oscillation Workshop, Lecce, Italy, 2002, edited by P.~Bernardini, G.~Fogli, E.~Lisi.

  \bibitem{SeiyaB} K.~Seiya {\sl et al.}, in Proceedings of Particle Accelerator Conference, 2005, edited by C.~Horak.
  
  \bibitem{DeyKK} J.~Dey, K.~Koba, I.~Kourbanis, and J.~Reid, in Proceedings of Particle Accelerator Conference, 2007, edited by C.~Petit-Jean-Genaz.
  
  \bibitem{Dey} J.~Dey and I.~Kourbanis, in Proceedings of Particle Accelerator Conference, 2005, edited by C.~Horak.
  
  \bibitem{Madrak} R.~Madrak and D.~Wildman, in Proceedings of North American Particle Accelerator Conference, 2013, edited by T.~Satogata, C.~Petit-Jean-Genaz, and V.~Schaa.
  
  \bibitem{SYLee} S.~Y.~Lee, {\it Accelerator Physics}, 3rd Ed. (World Scientific, Singapore, 2012).
  
  \bibitem{Broer} H.~W.~Broer, I.~Hoveijn, M.~van~Noort, C.~Sim\'o, and G.~Vegter, J. Dyn. Diff. Eq. {\bf 16}, 897 (2004).
  
  \bibitem{Xu} X.~Xu, M.~Wiercigroch, and M.~P.~Cartmell, Chaos, Solitons Fractals {\bf 23}, 1537 (2005).
 
  \bibitem{Lichtenberg} A.~J.~Lichtenberg and M.~A.~Lieberman, {\it Regular and Chaotic Dynamics}, 2nd Ed. (Springer, California, 1992).

  \bibitem{Zhang} H.~Zhang and T.~W.~Ma, Nonlinear Dyn. {\bf 70}, 2433 (2012).
  
  \bibitem{Lenci} S.~Lenci, E.~Pavlovskaia, G.~Rega, and M.~Wiercigroch, J. Sound Vib. {\bf 310}, 243 (2008).
  
  \bibitem{Butikov} E.~I.~Butikov, J. Phys. A {\bf 44}, 295202 (2013).
  
  \bibitem{SM} See supplemental material for complete series of stability maps.

  \bibitem{SeiyaCD} K.~Seiya, B.~Chase, J.~Dey, P.~Joireman, I.~Kourbanis, and J.~Reid, in Proceedings of CARE-HHH-APD Workshop BEAM'07, 2007, edited by W.~Scandale  and F.~Zimmermann.
  
  \bibitem{Yang} X.~Yang, A.~I.~Drozhdin, and W.~Pellico, in Proceedings of Particle Accelerator Conference, 2007, edited by C.~Petit-Jean-Genaz.
  
  \bibitem{Ahrens} L.~A.~Ahrens {\sl et al.}, in Proceedings of Particle Accelerator Conference, 1999, edited by A.~Luccio and W.~MacKay.
  
  \bibitem{Eldred} J.~Eldred and R.~Zwaska, Fermi National Accelerator Laboratory Report TM-2587-APC, 2014.
  
  \bibitem{Davies} H.~J.~Davies and C.~S.~Adams, J. Phys. B {\bf 33}, 4079 (2000).
  
  \bibitem{Jaksch} D.~Jaksch, H.~J.~Briegel, J.~I.~Cirac, C.~W.~Gardiner, and P.~Zoller, Phys. Rev. Lett. {\bf 82}, 1975 (1999).
  
  \bibitem{Clairon} A.~Clairon, C.~Salomo, S.~Guellati, and W.D.~Phillip, Europhys. Lett. {\bf 16}, 165 (1991).
  
  \bibitem{Gorkov} L.~P.~Gor'kov, Sov. Phys. Dokl. {\bf 6}, 773 (1962).
  
  \bibitem{Foresti} D.~Foresti and D.~Poulikakos, Phys. Rev. Lett. {\bf 112}, 024301 (2014).

  \bibitem{Shields} C.~W.~Shields, L.~M.~Johhson, L.~Gao, and G.~P.~L\'opez, Langmuir {\bf 30}, 3923 (2014).

\end{thebibliography}
\end{document}

\title{Supplemental Material for: \\ Dynamical Stability of Slip-stacking}
\author{Jeffrey Eldred}
\affiliation{Department of Physics, Indiana University, Bloomington, Indiana 47405, USA}
\affiliation{Fermi National Accelerator Laboratory, Batavia, Illinois 60510, USA}
\author{Robert Zwaska}
\affiliation{Fermi National Accelerator Laboratory, Batavia, Illinois 60510, USA}
\date{\today}

\maketitle

\section*{Stability Maps}
This supplemental material shows the stability maps of the higher slip-stacking bucket for values of $\alpha_{s}$ from 2 to 8 in descending 0.1 increments. Recall that stability map is a visualization of a numerical integration of the following equation:
\begin{equation} \label{dif1b}
\dot{\phi} = 2 \pi f_{rev} h \eta \delta,~\dot{\delta} = f_{rev} V_{\delta} [\sin(\phi) + \sin(\phi - \omega_{p}t )].
\end{equation}
Where $\phi$ is the phase of the particle relative to the rf cavity and $\delta$ is the fractional deviation from a reference momentum. Each pixel in a stability map represents a different set of initial coordinates and the color corresponds to the number of synchrotron periods a particle survives before it is lost. The stability maps are scaled to accurately reflect the bucket shapes obtained in the Fermilab Recycler with a 15-Hz Booster cycle-rate. The corresponding Recycler rf voltage is listed for each plot.

Some stability maps are particularly noteworthy. Recall that $\alpha_{s} = 6.2$ maximizes the phase-space area and $\alpha_{s} = 5.5$ maximizes the Booster injection efficiency. In these figures one can clearly see the 7th-order resonance ($\alpha_{s} =$~5.0~to~5.1), the 6th-order resonance ($\alpha_{s} =$~4.5~to~4.8), the 5th-order resonance ($\alpha_{s} =$~4.0~to~4.3), the 4th-order resonance ($\alpha_{s} =$~3.5~to~3.7), and the 3.5-order resonance ($\alpha_{s} =$~3.2), the 3rd-order resonance ($\alpha_{s} =$~2.9~to~3.0), the 2.5-order resonance ($\alpha_{s} =$~2.5), and the 2nd-order resonance ($\alpha_{s} =$~2.2~to~2.3). For $\alpha_{s} = 2.1$ and lower, there is no stable phase-space area.

 \centering
 \includegraphics[scale=0.51]{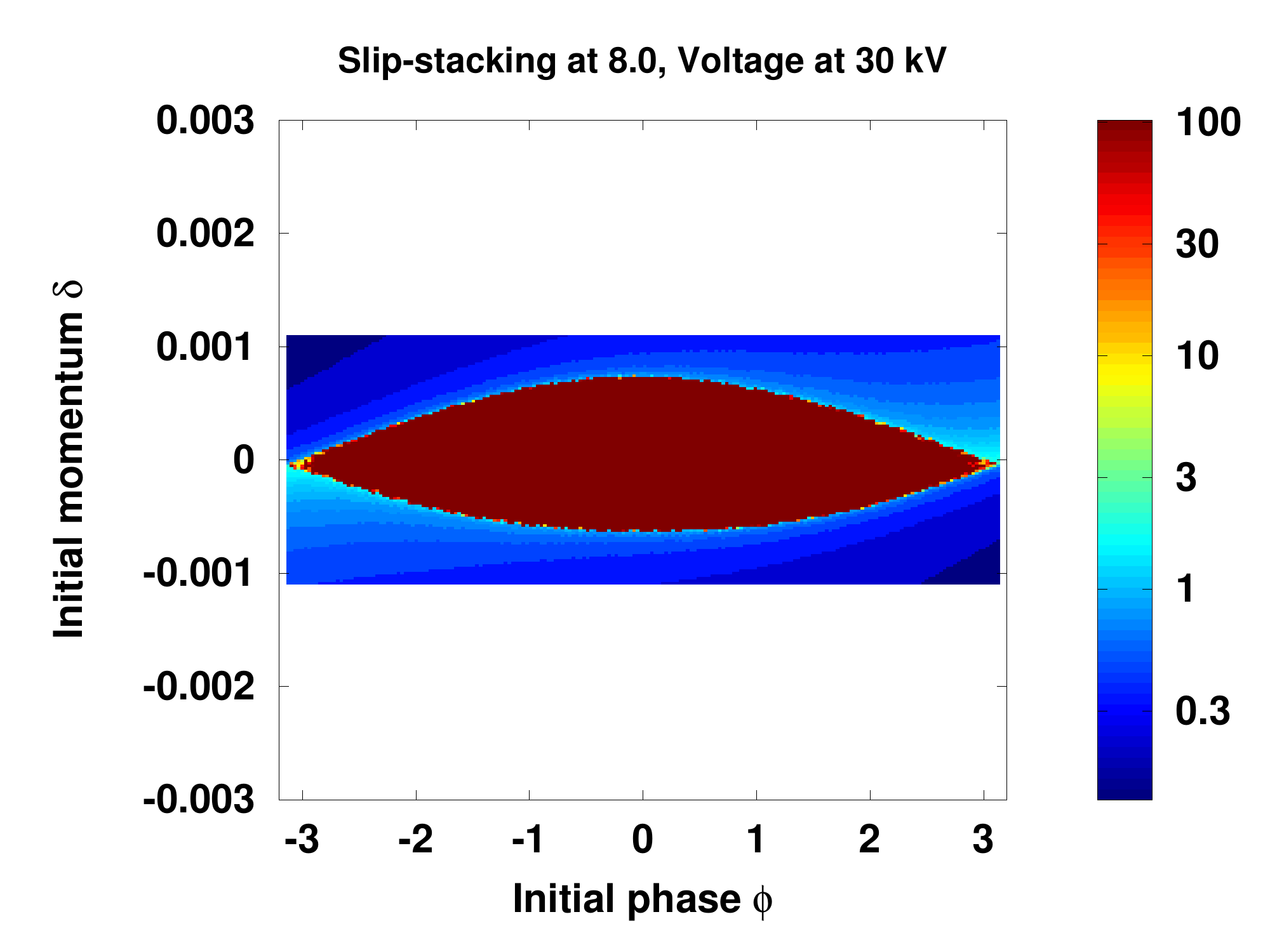}
 \centering
 \includegraphics[scale=0.51]{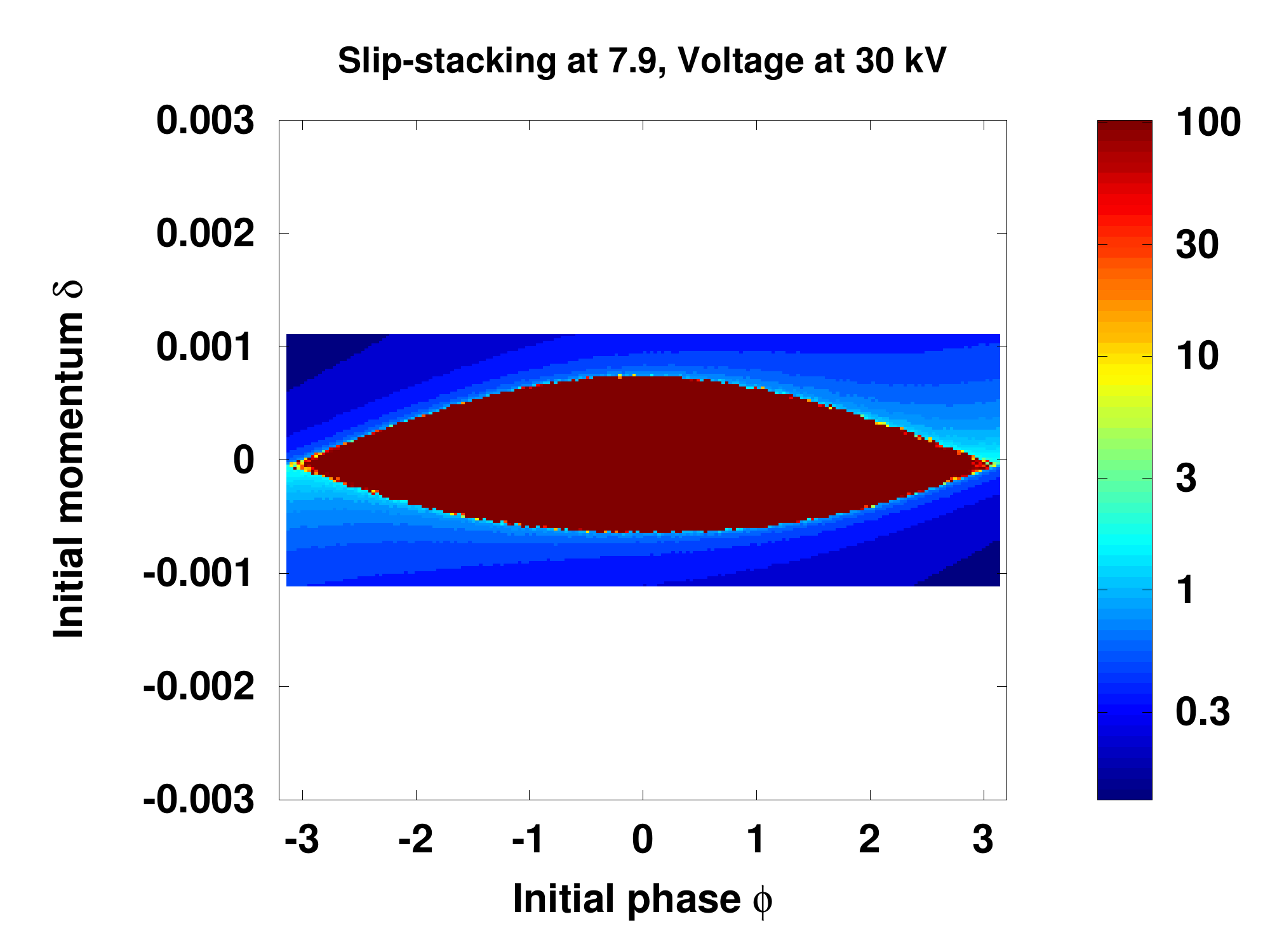}
 \centering
 \includegraphics[scale=0.51]{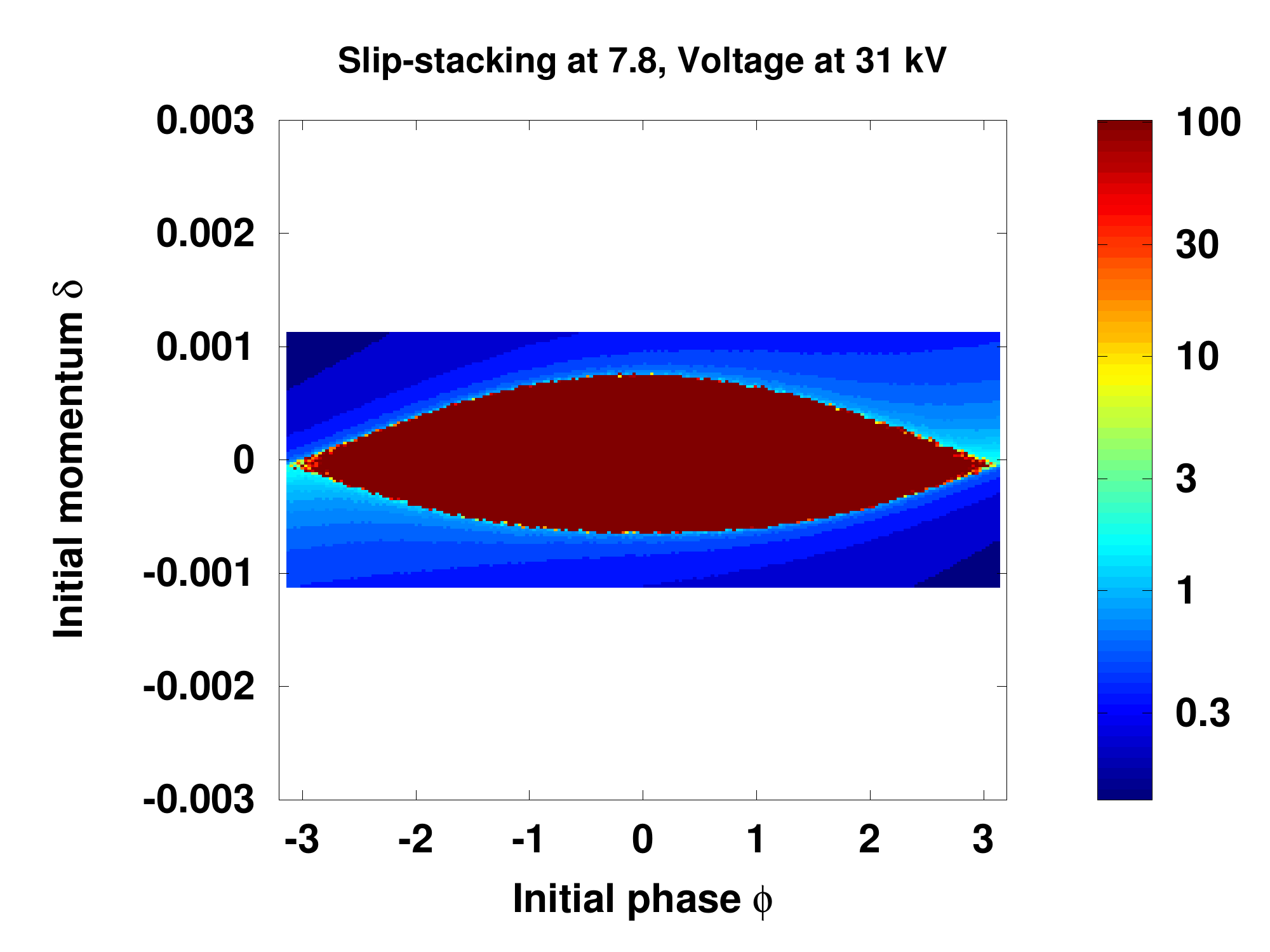}
 \centering
 \includegraphics[scale=0.51]{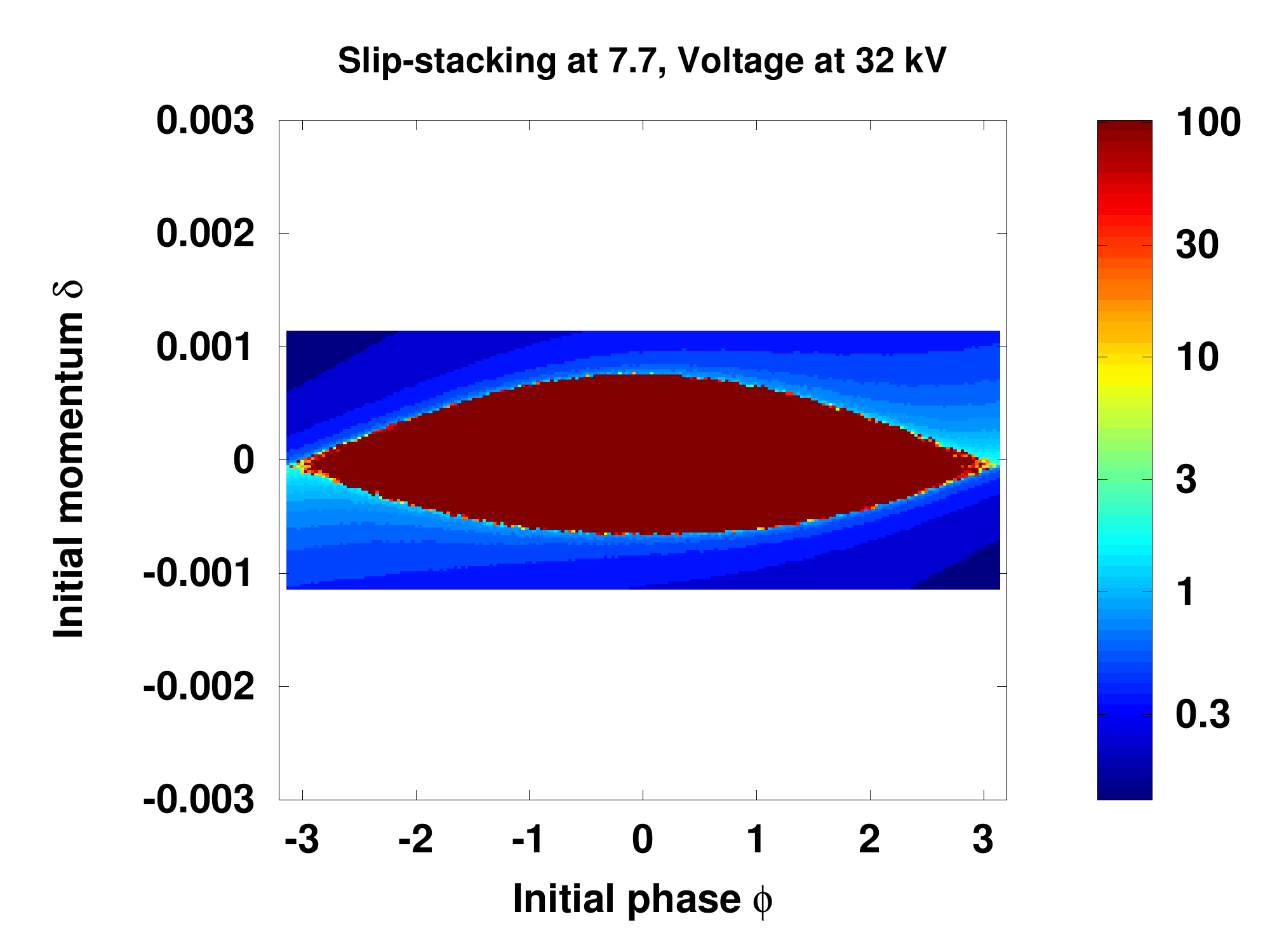}
 \centering
 \includegraphics[scale=0.51]{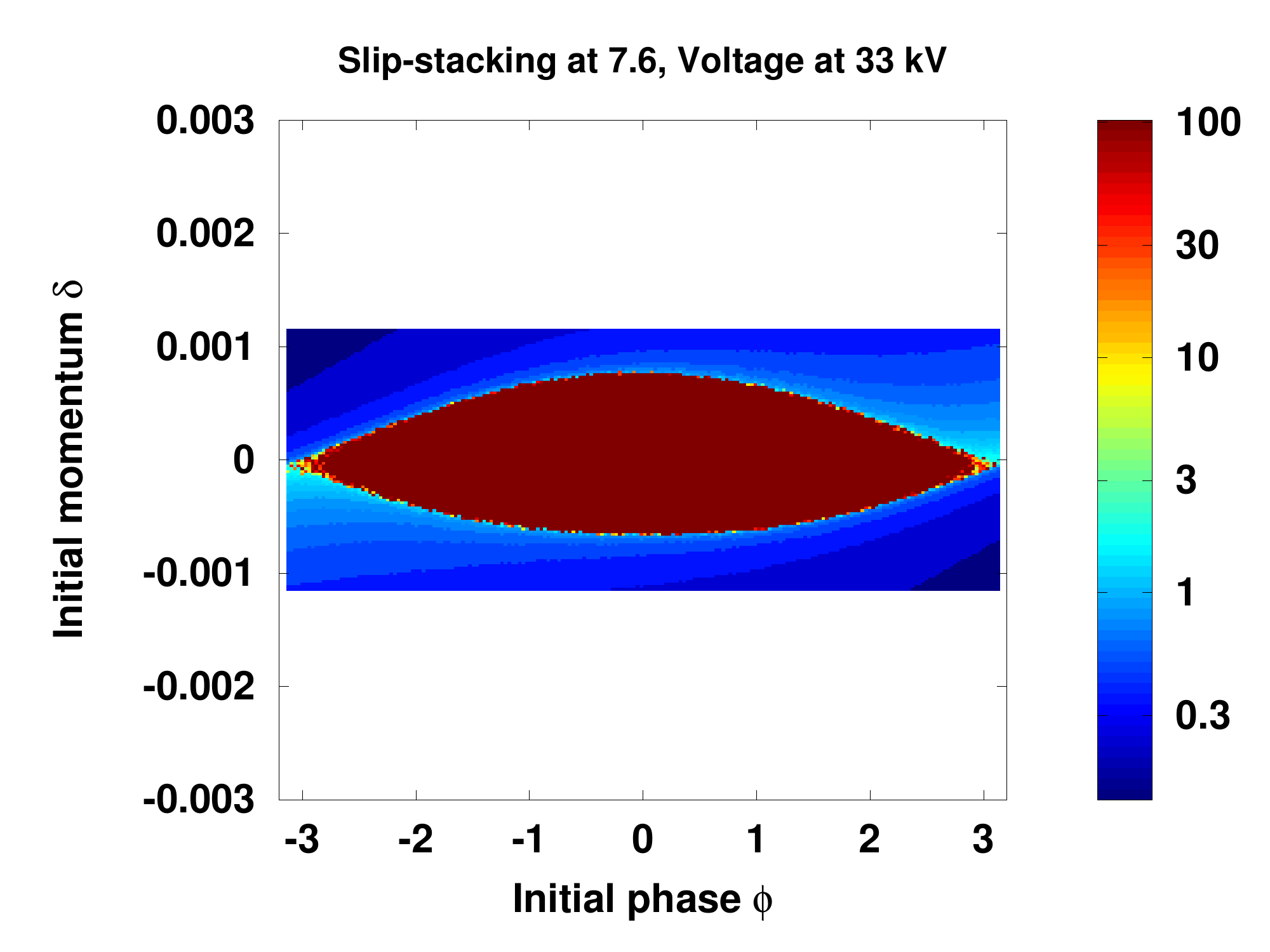}
 \centering
 \includegraphics[scale=0.51]{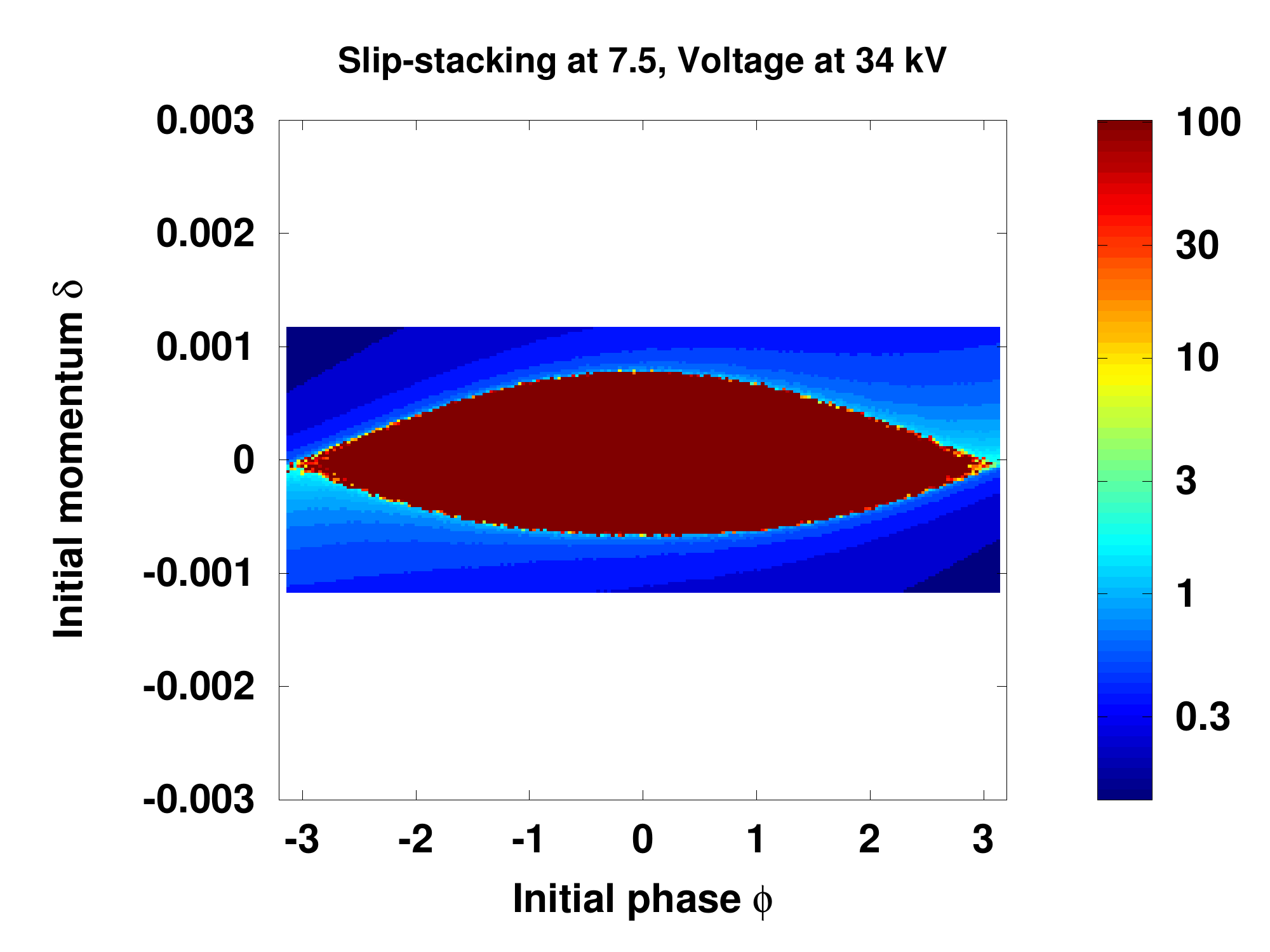}
 \centering
 \includegraphics[scale=0.51]{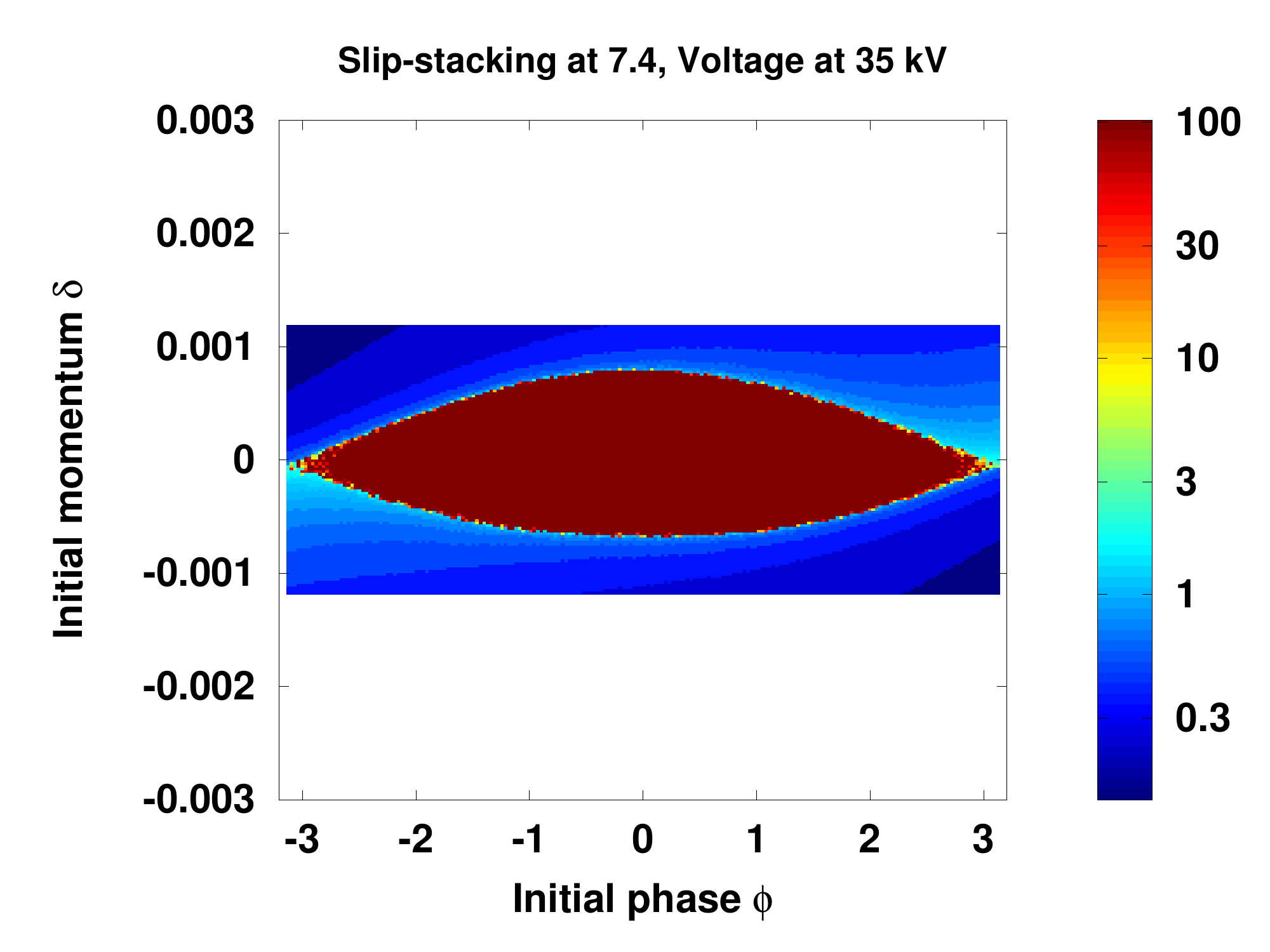}
 \centering
 \includegraphics[scale=0.51]{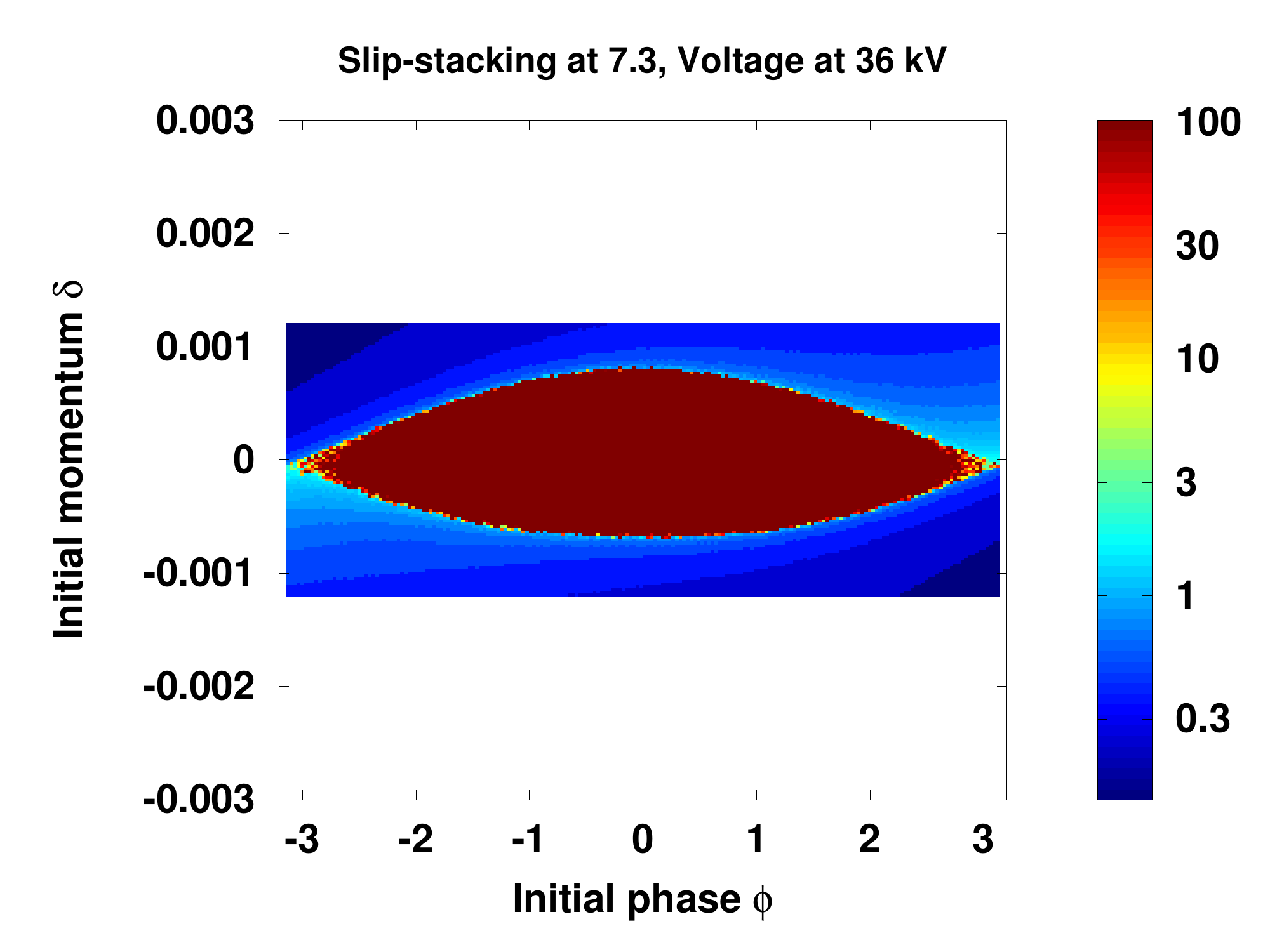}
 \centering
 \includegraphics[scale=0.51]{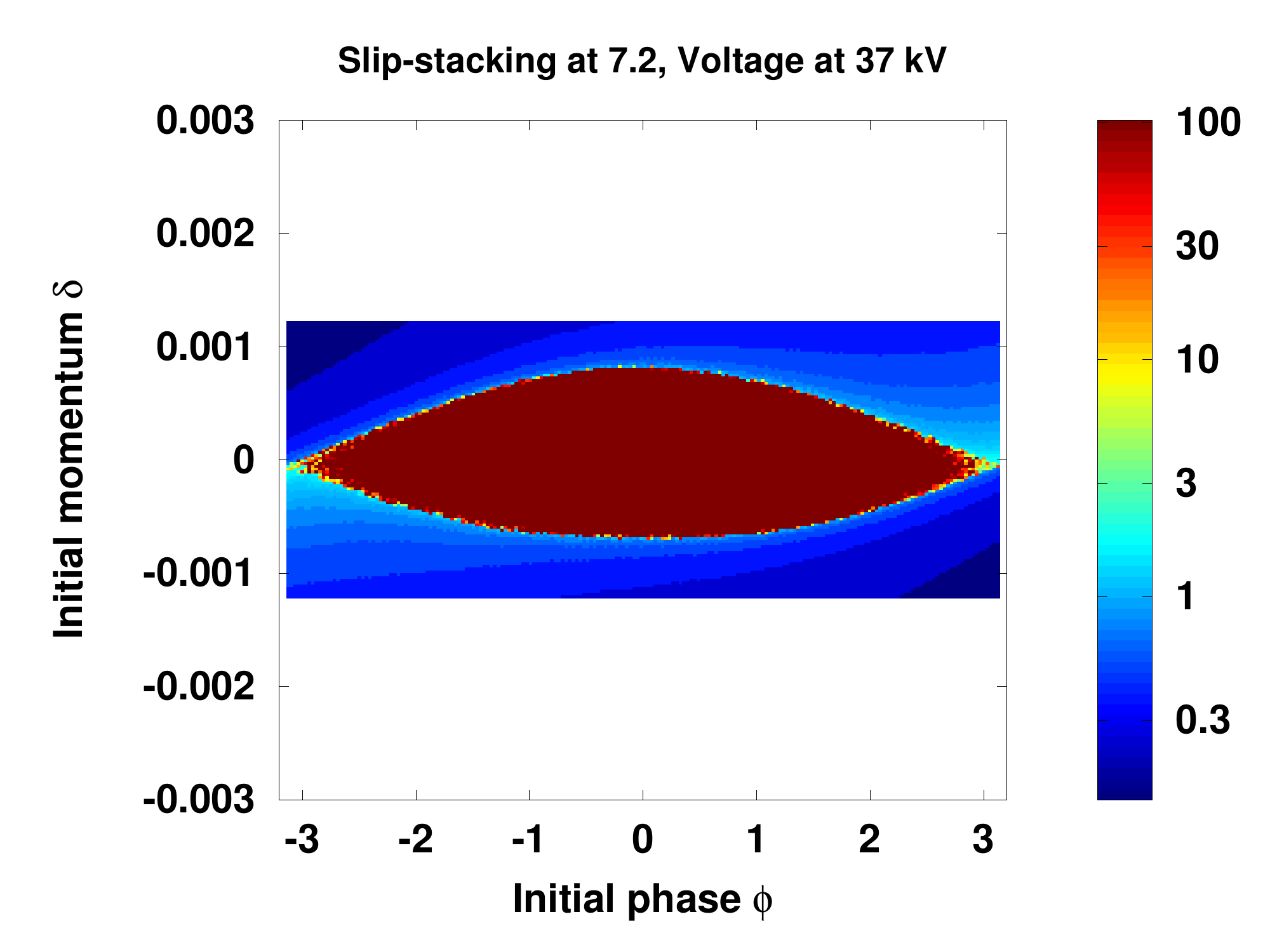}
 \centering
 \includegraphics[scale=0.51]{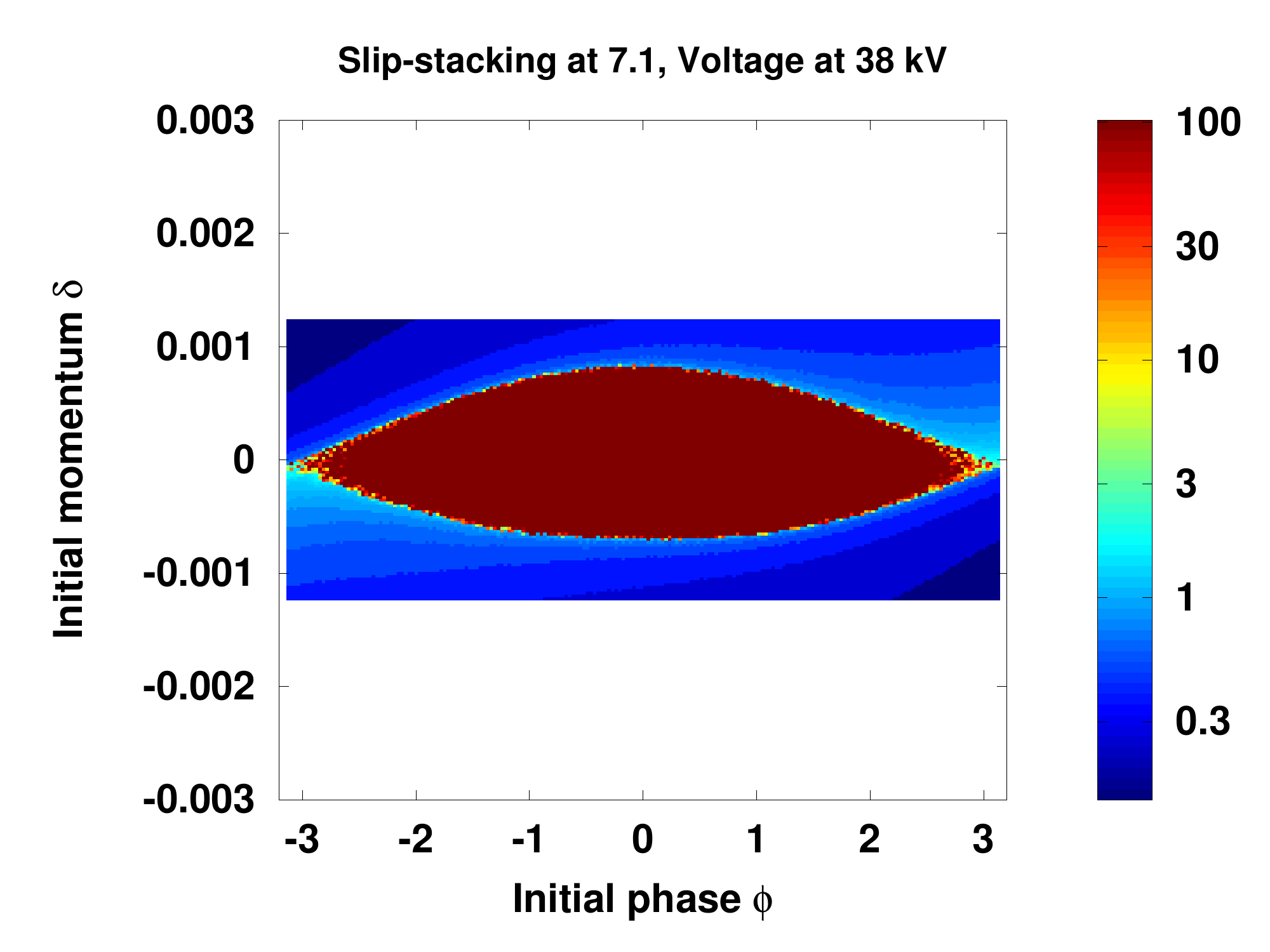}
 \centering
 \includegraphics[scale=0.51]{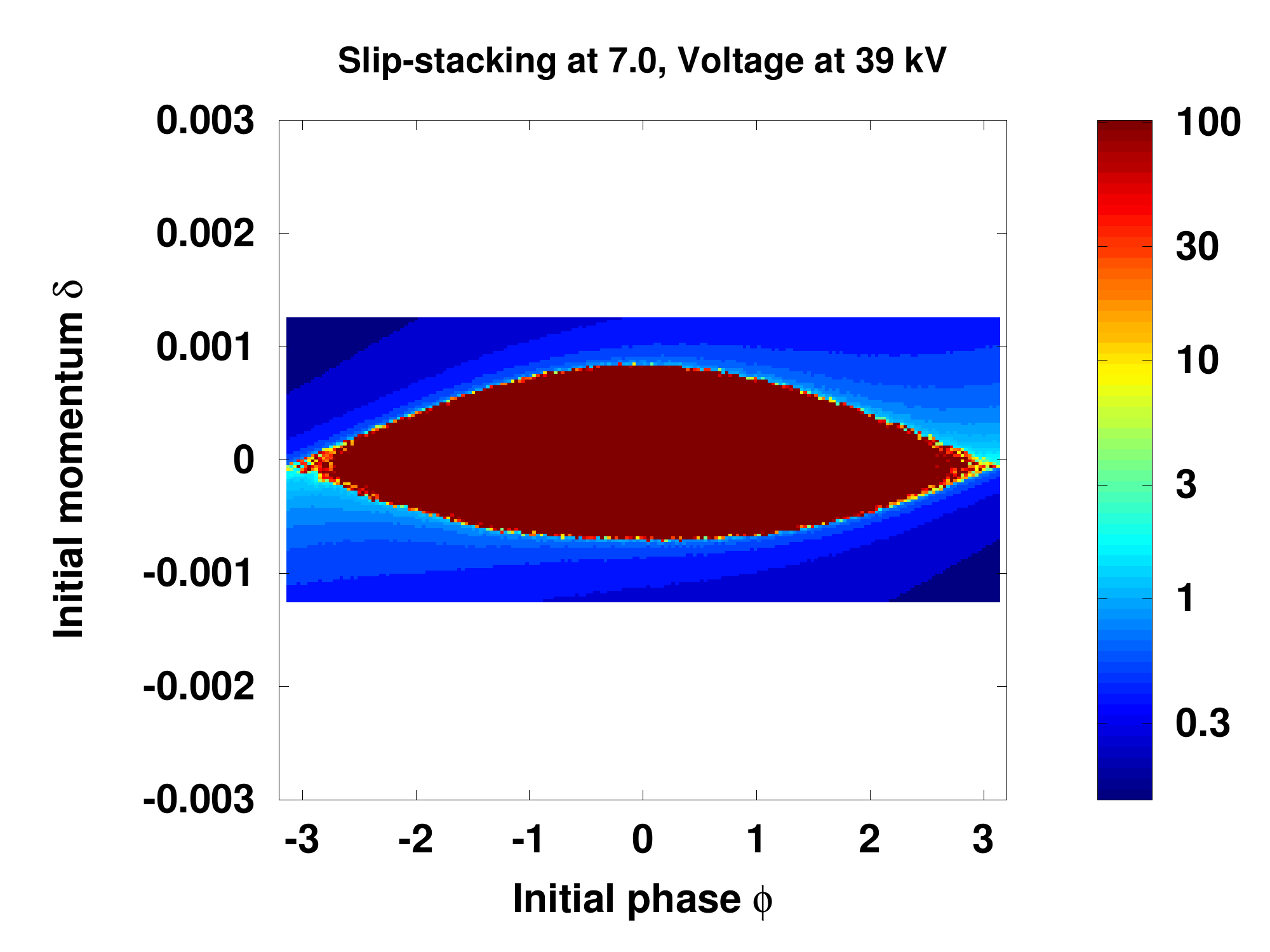}
 \centering
 \includegraphics[scale=0.51]{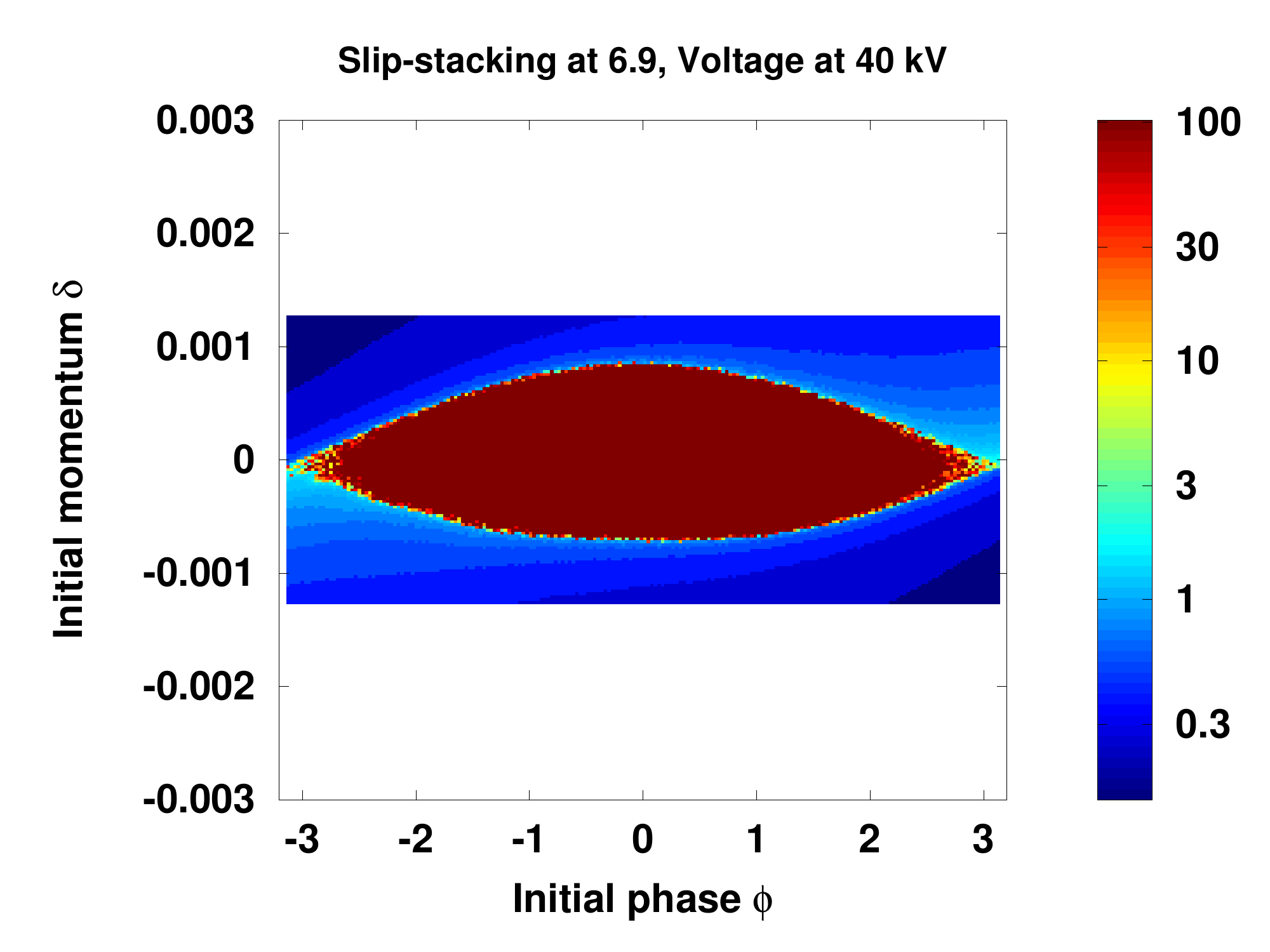}
 \centering
 \includegraphics[scale=0.51]{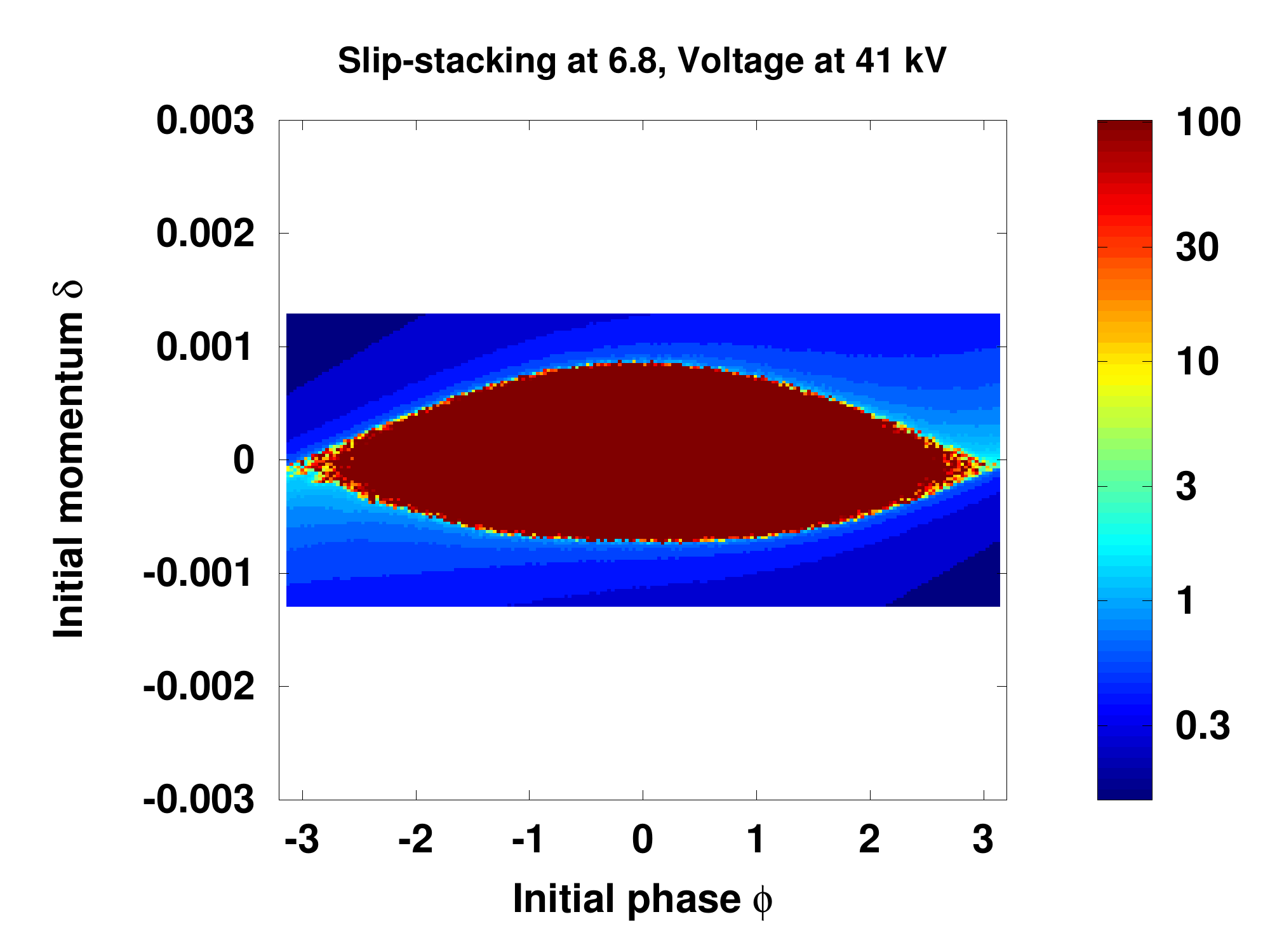}
 \centering
 \includegraphics[scale=0.51]{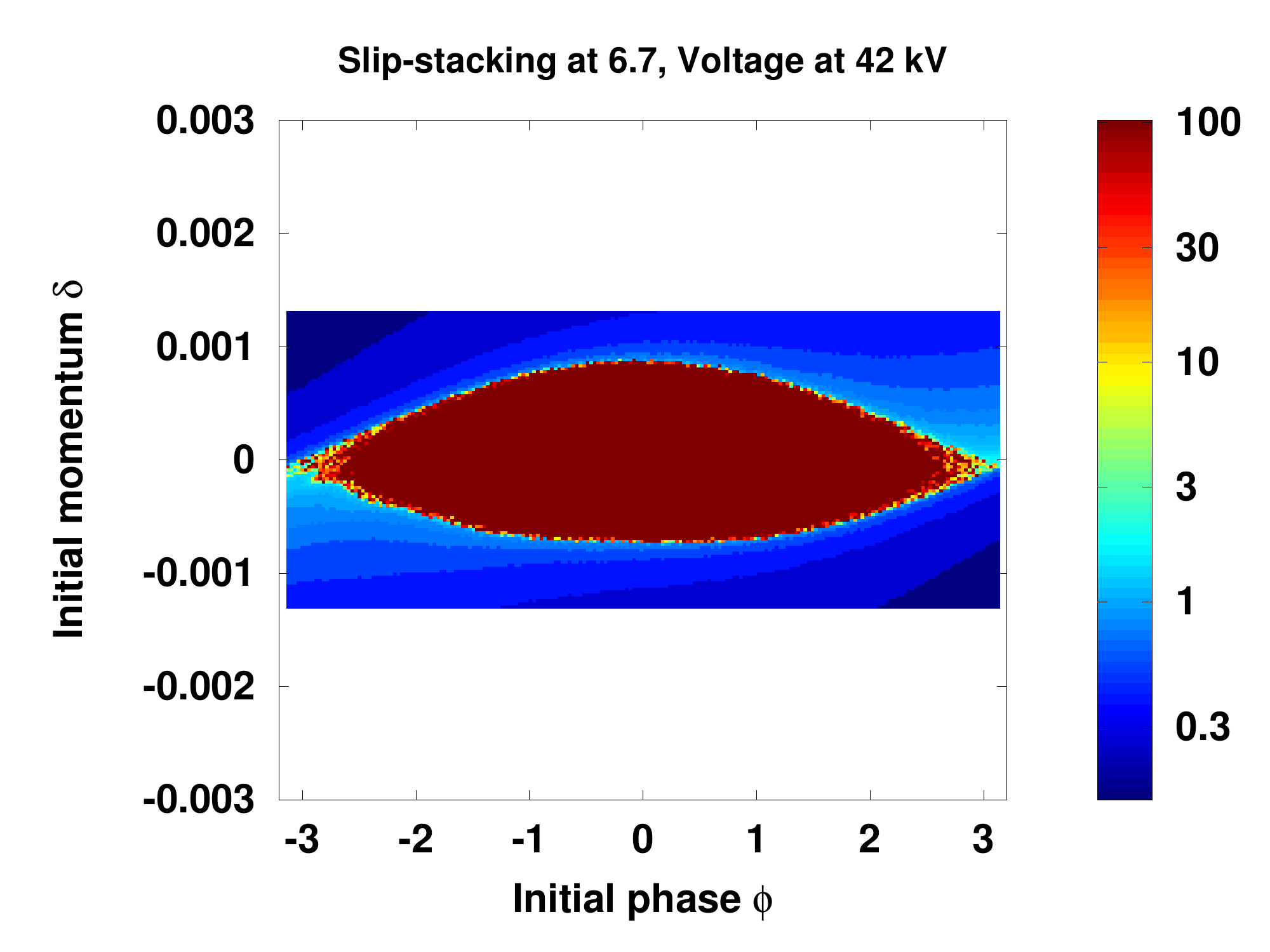}
 \centering
 \includegraphics[scale=0.51]{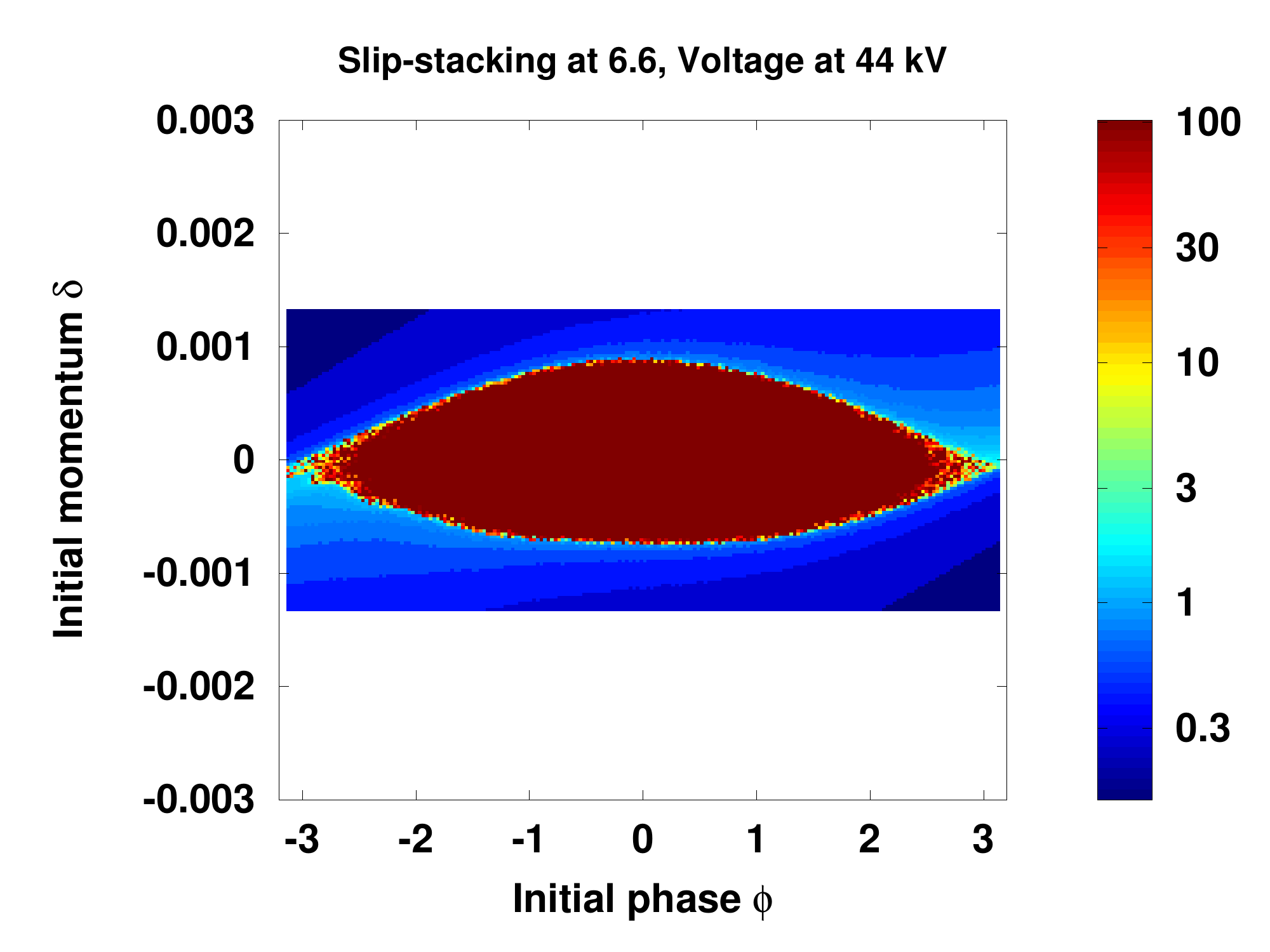}
 \centering
 \includegraphics[scale=0.51]{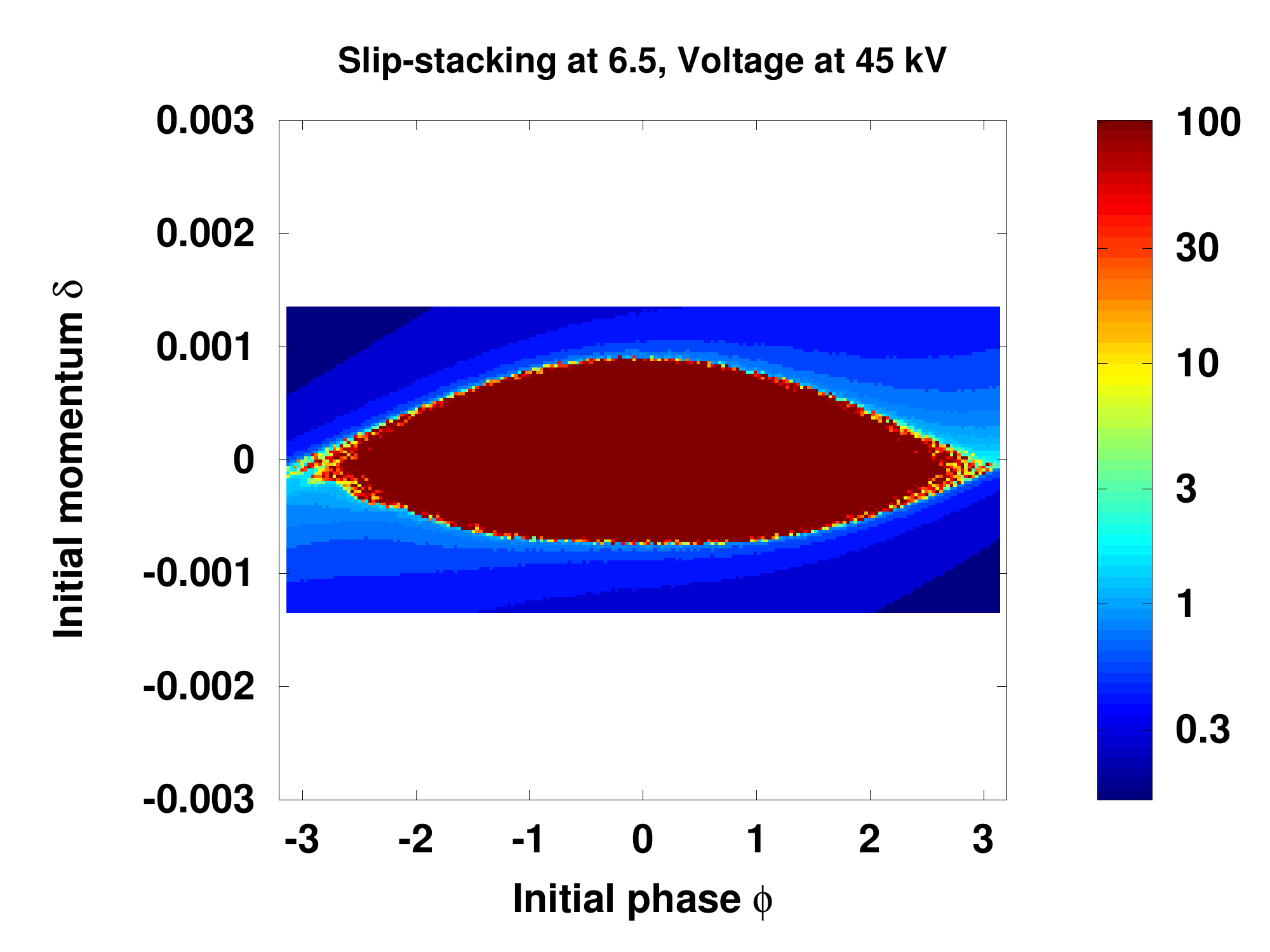}
 \centering
 \includegraphics[scale=0.51]{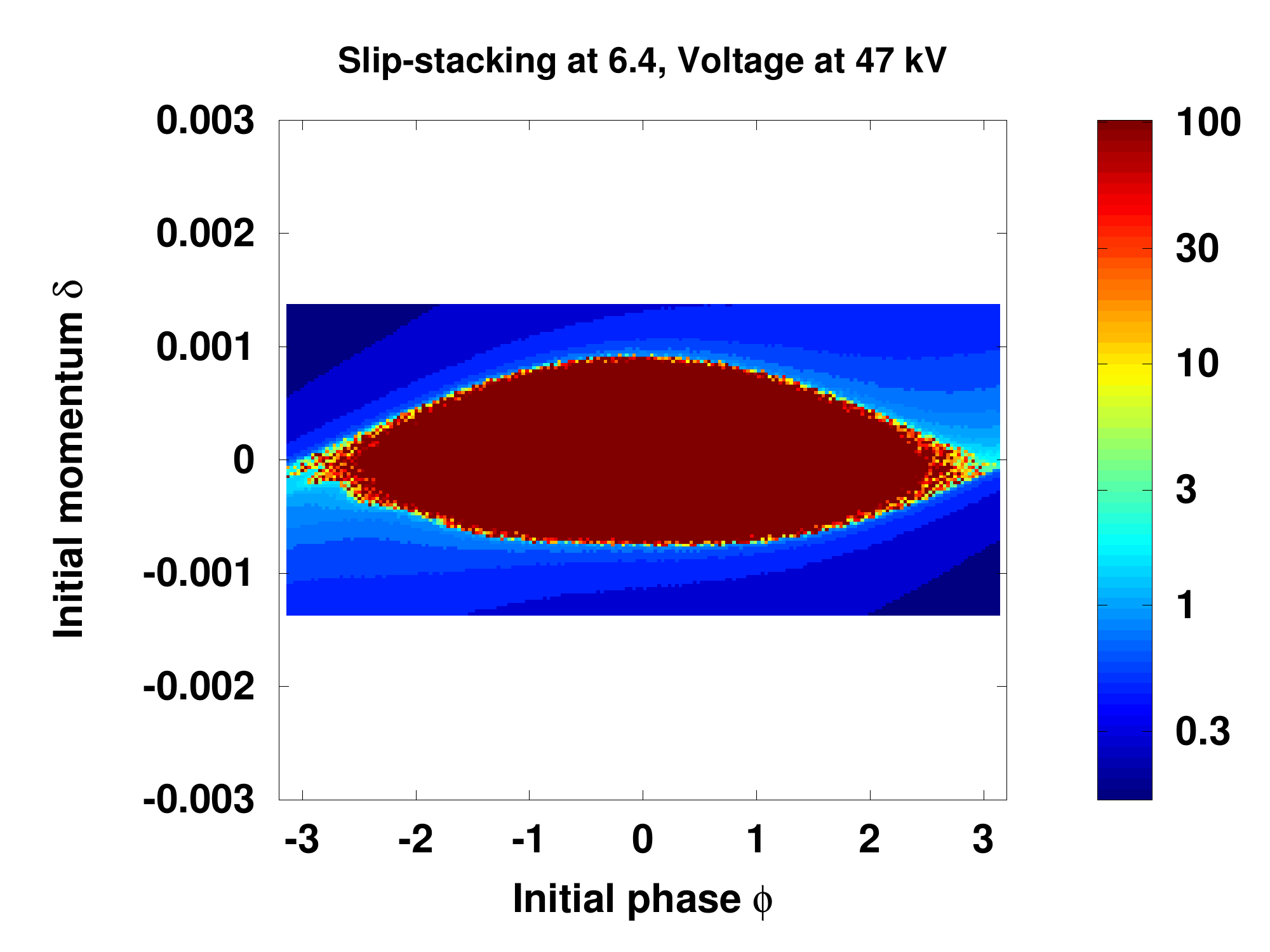}
 \centering
 \includegraphics[scale=0.51]{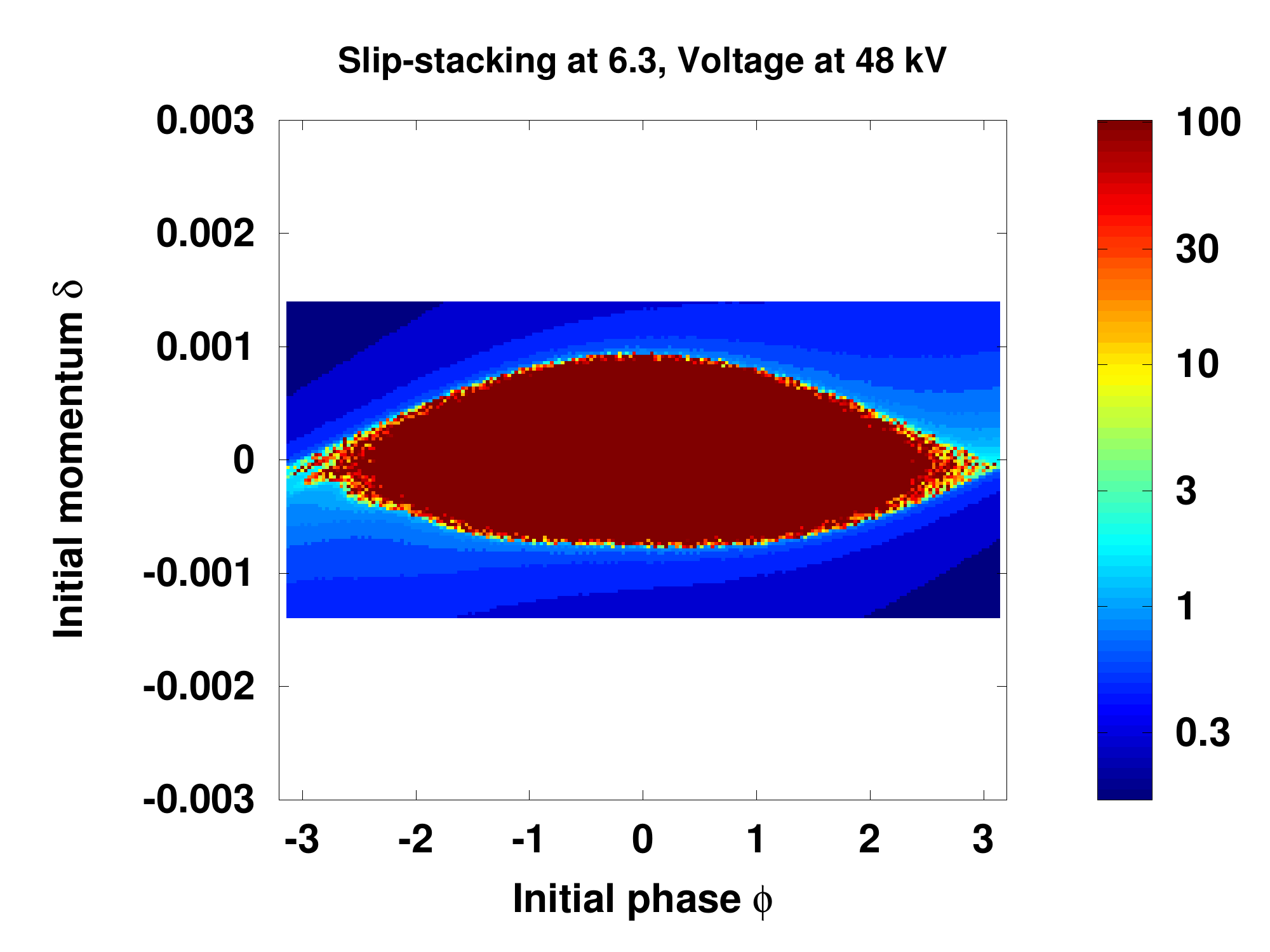}
 \centering
 \includegraphics[scale=0.51]{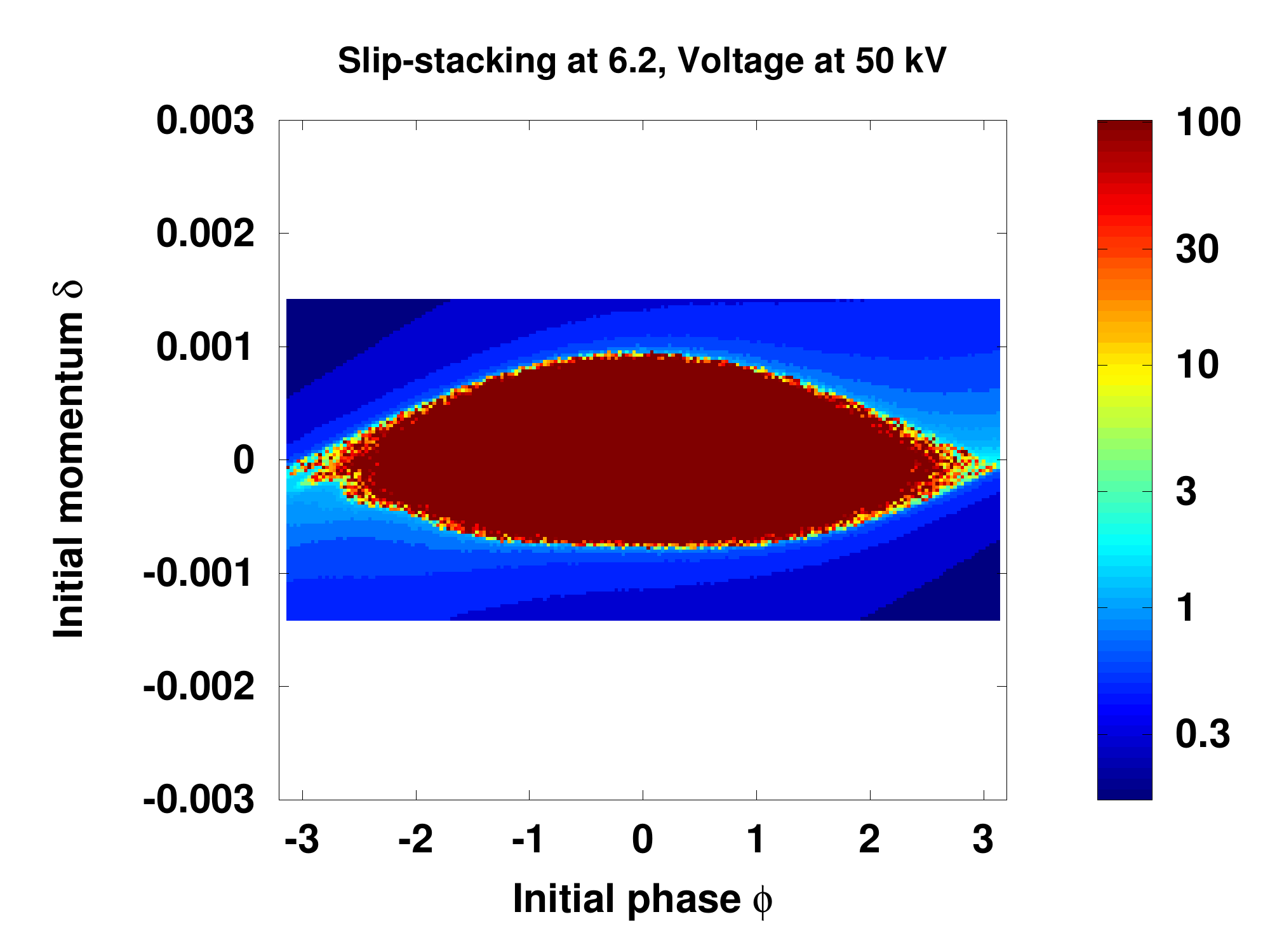}
 \centering
 \includegraphics[scale=0.51]{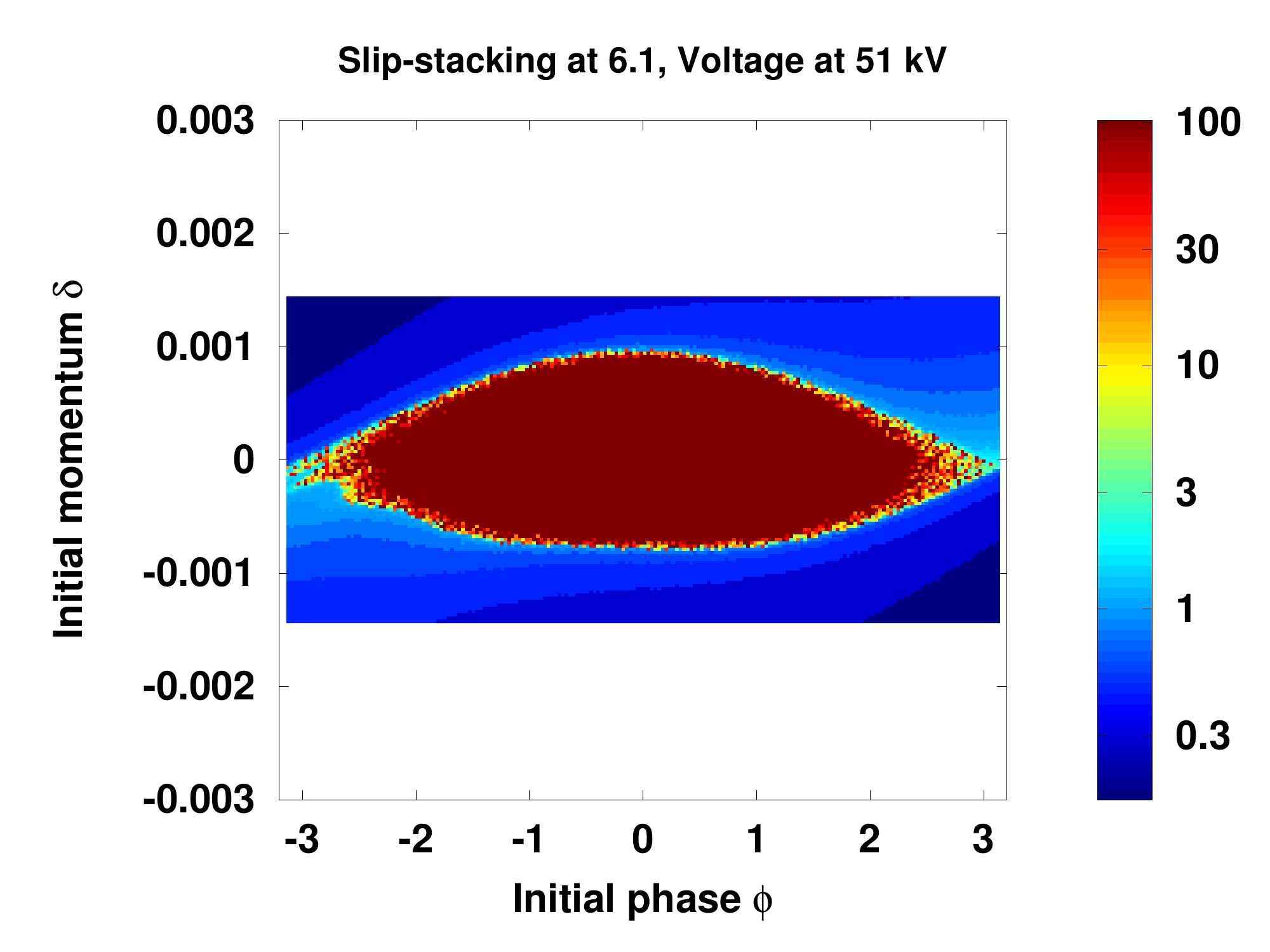}
 \centering
 \includegraphics[scale=0.51]{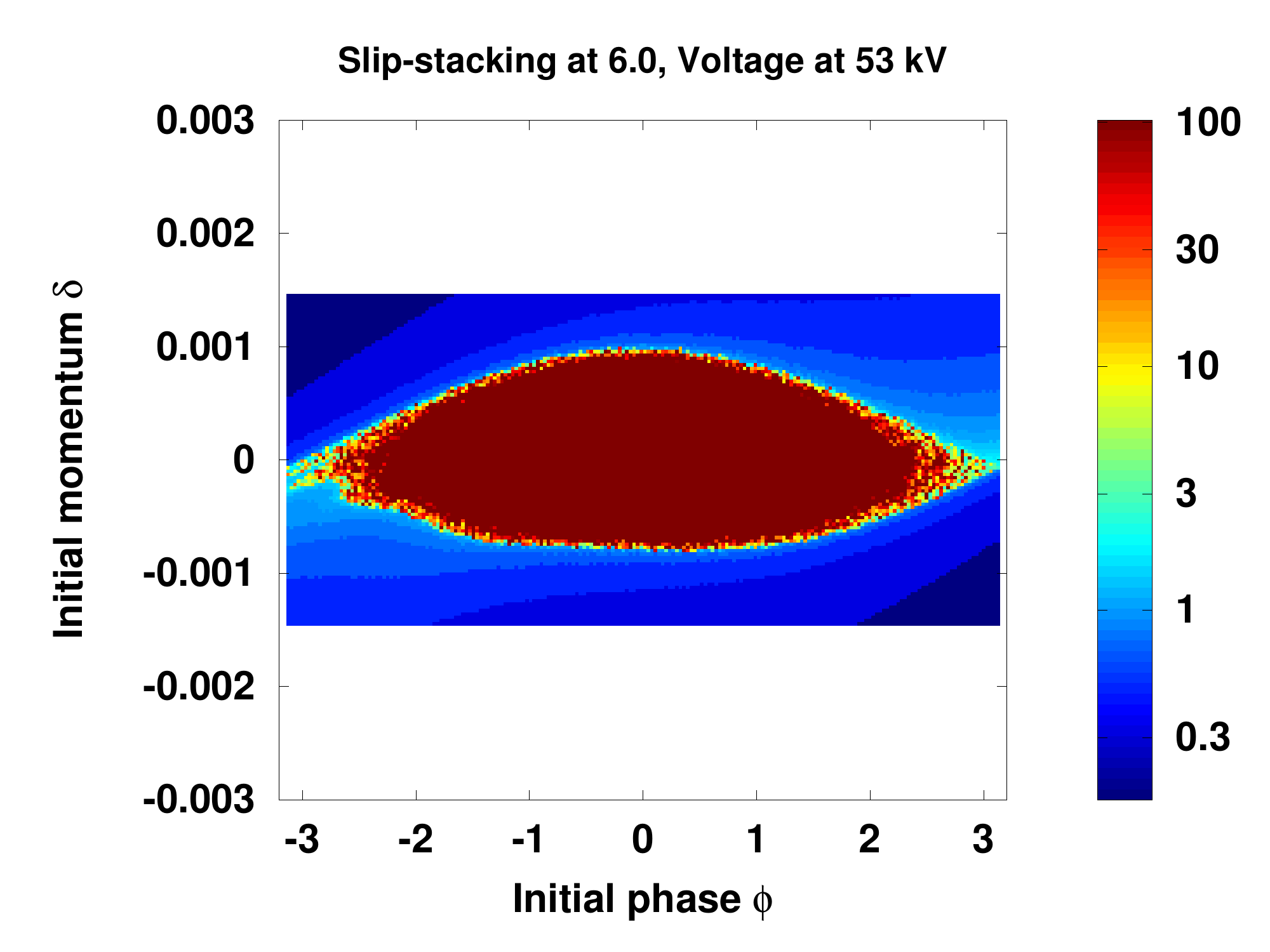}
 \centering
 \includegraphics[scale=0.51]{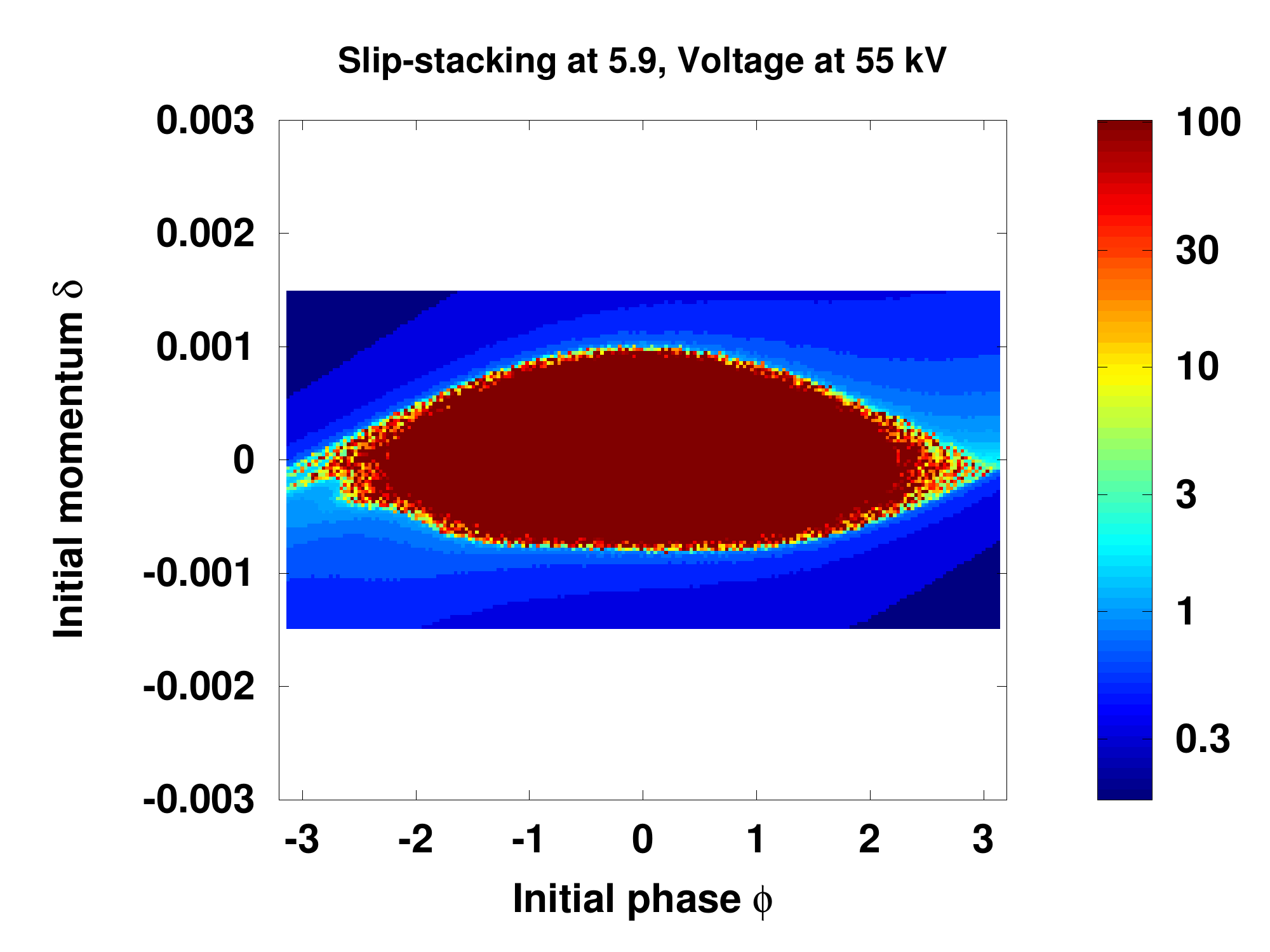}
 \centering
 \includegraphics[scale=0.51]{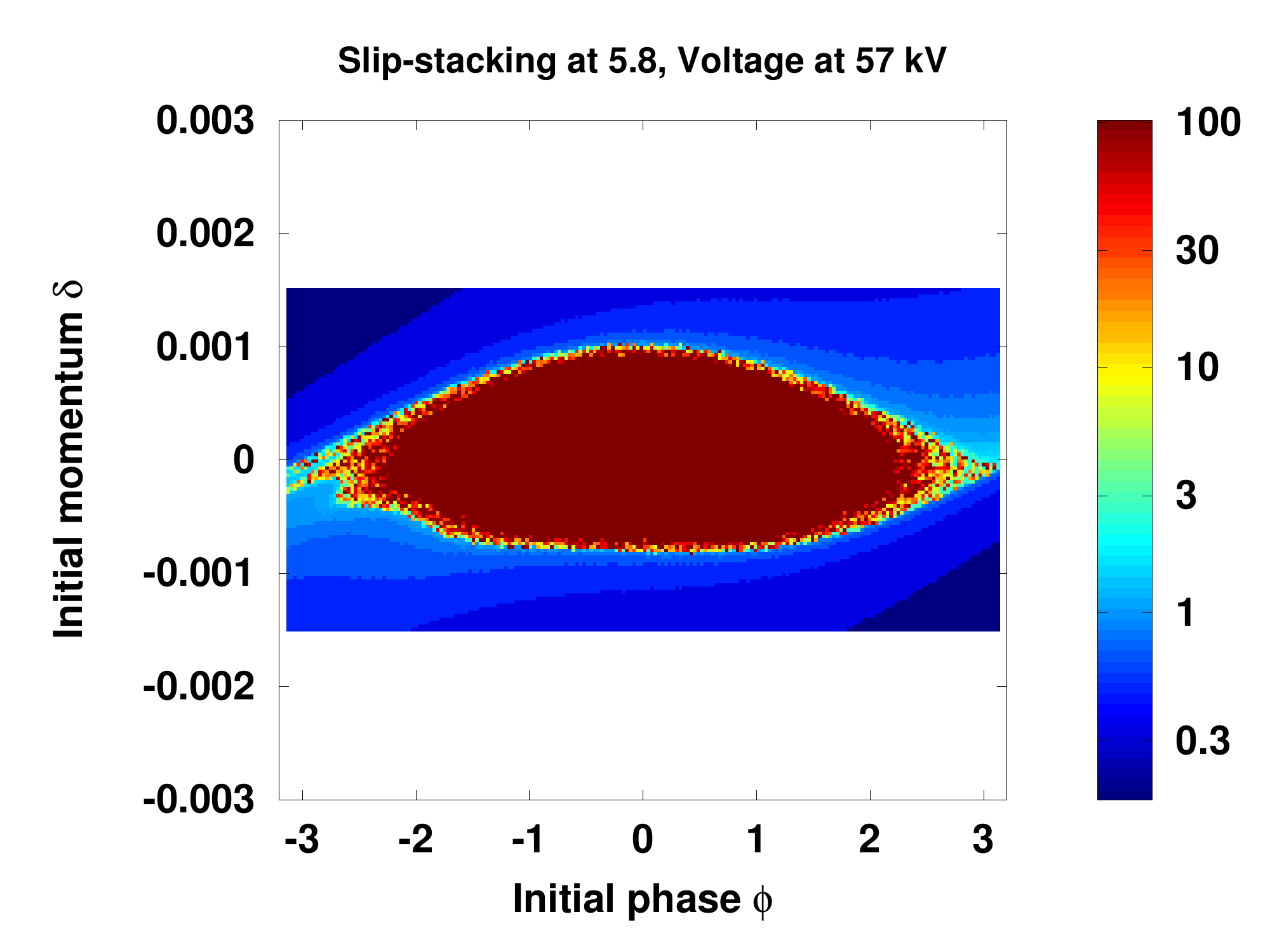}
 \centering
 \includegraphics[scale=0.51]{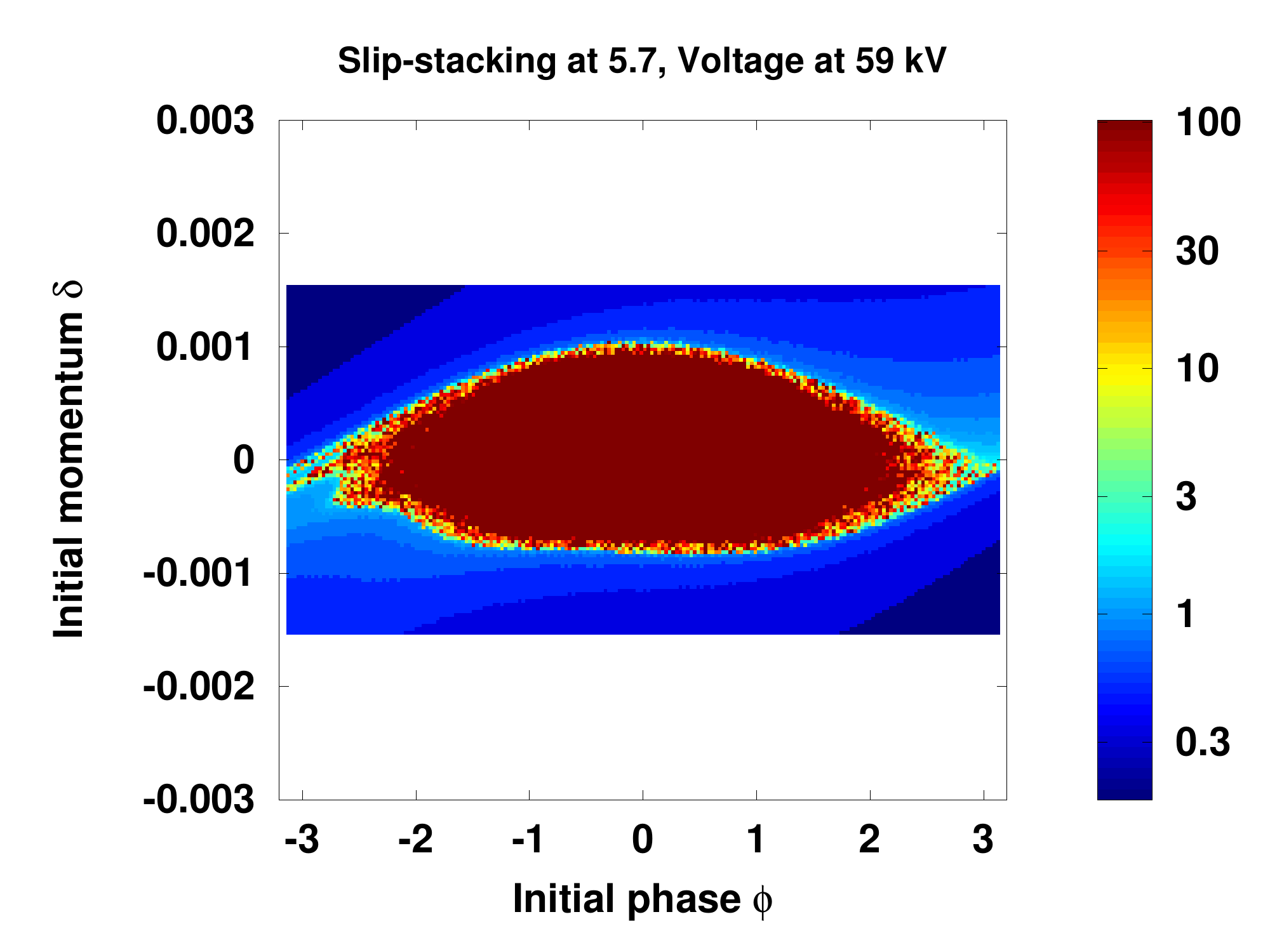}
 \centering
 \includegraphics[scale=0.51]{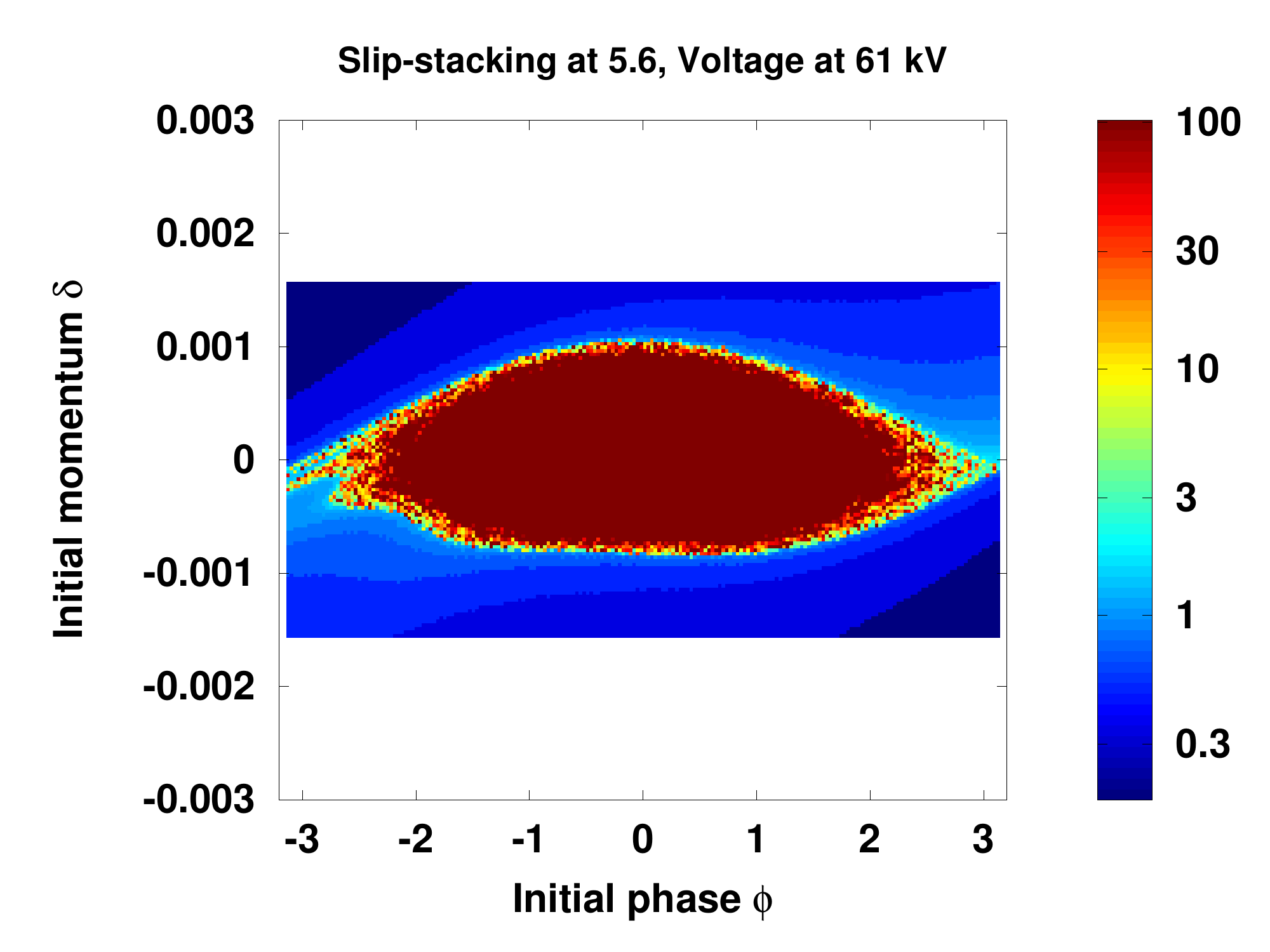}
 \centering
 \includegraphics[scale=0.51]{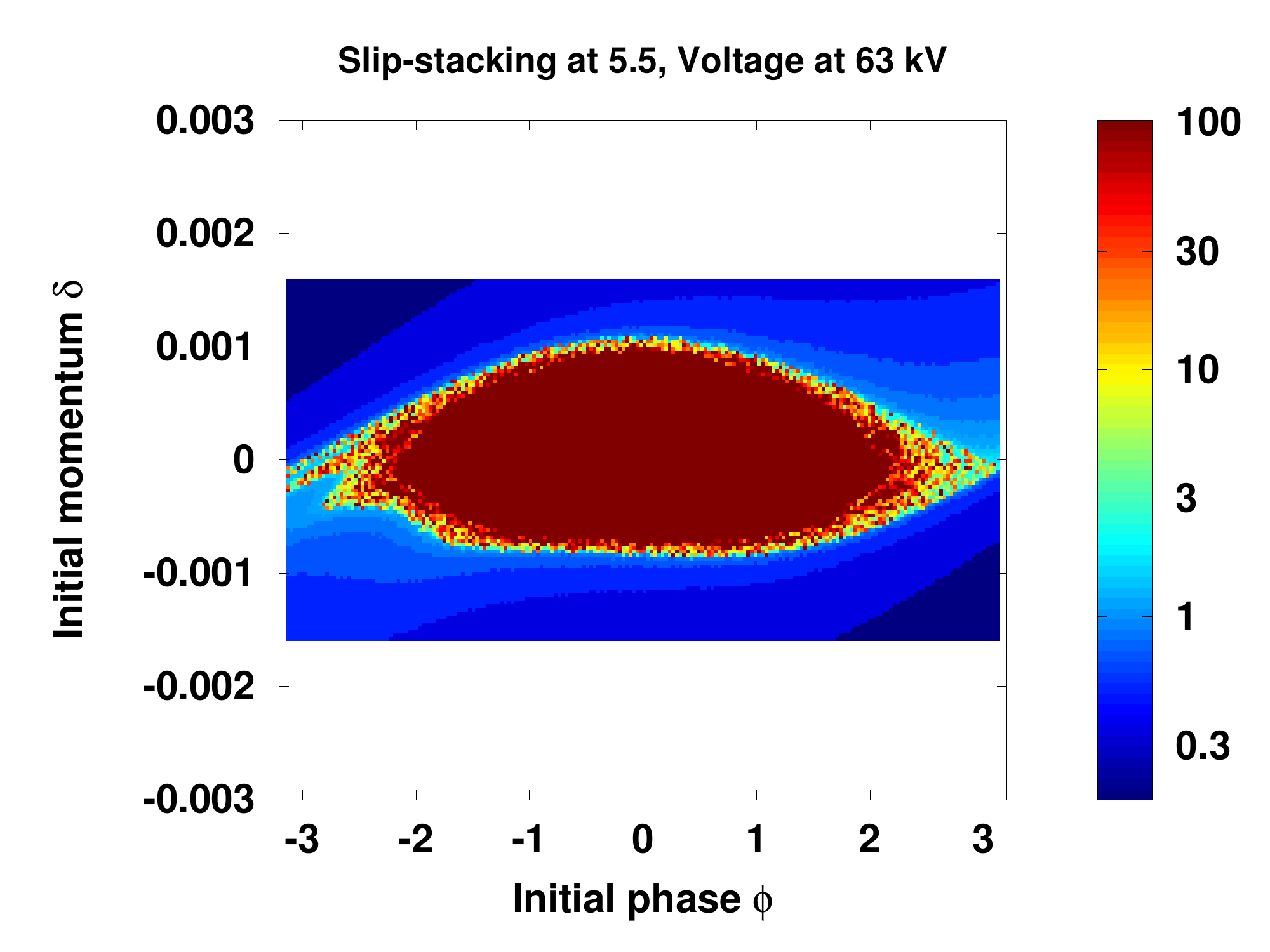}
 \centering
 \includegraphics[scale=0.51]{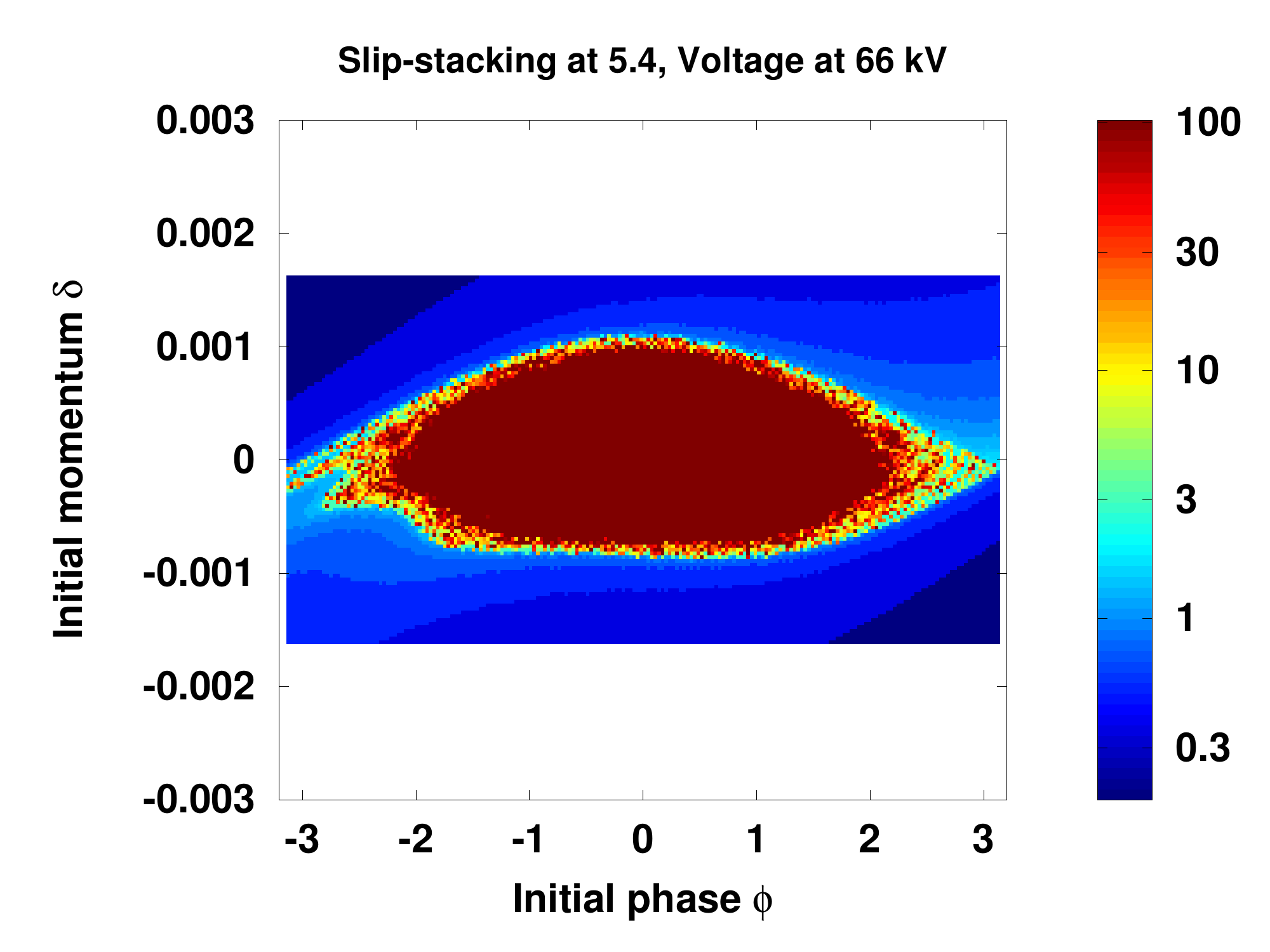}
 \centering
 \includegraphics[scale=0.51]{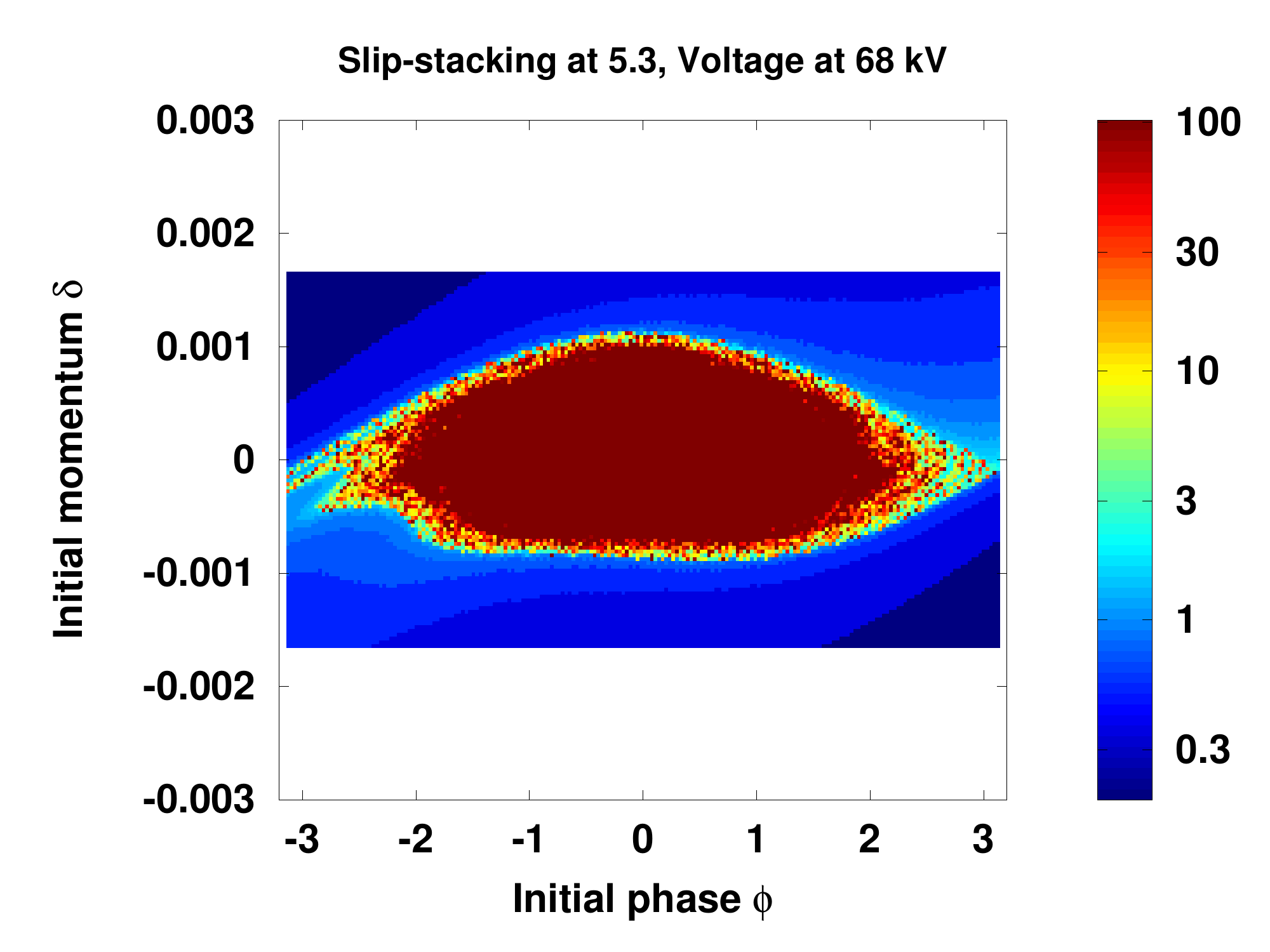}
 \centering
 \includegraphics[scale=0.51]{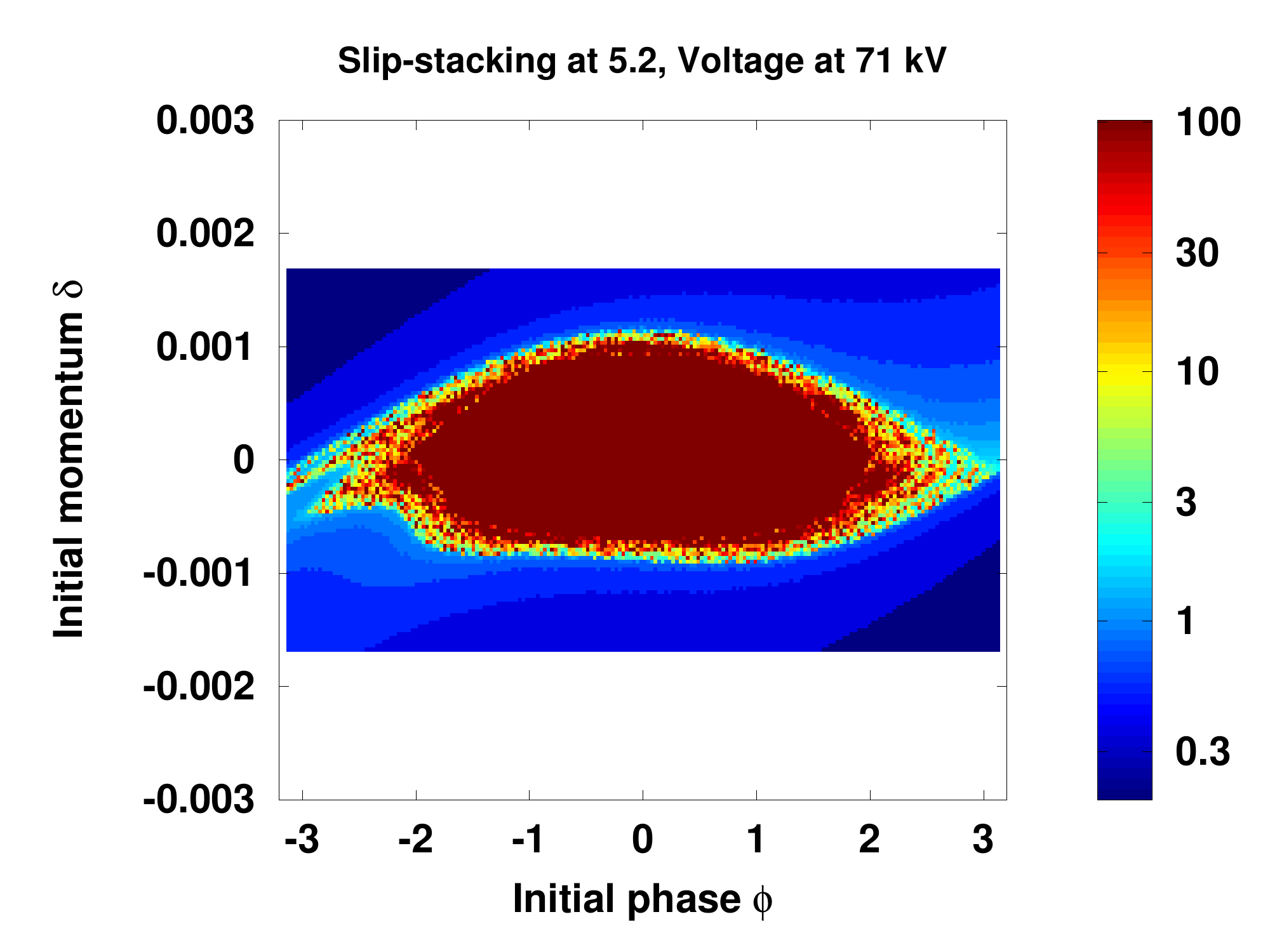}
 \centering
 \includegraphics[scale=0.51]{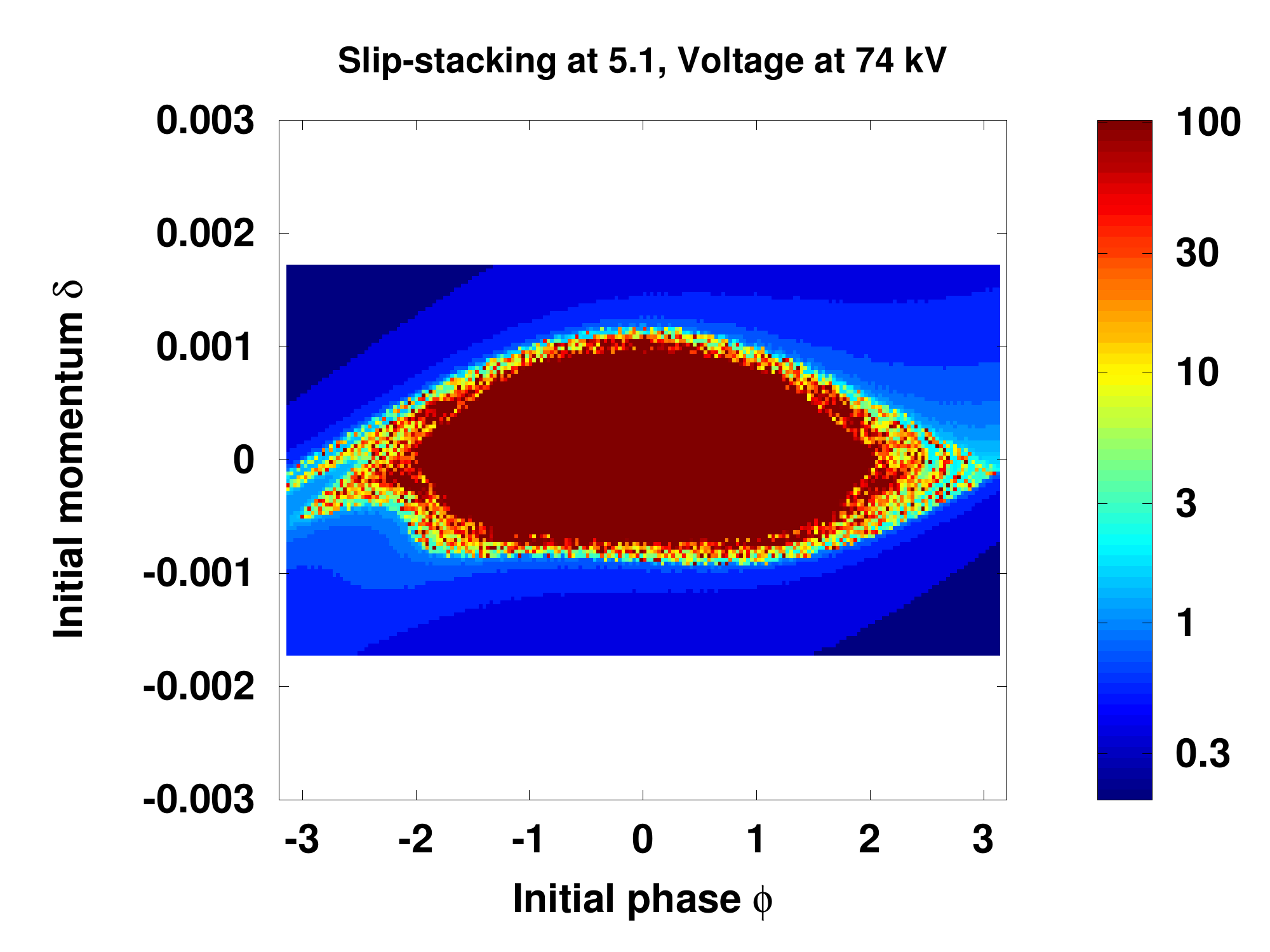}
 \centering
 \includegraphics[scale=0.51]{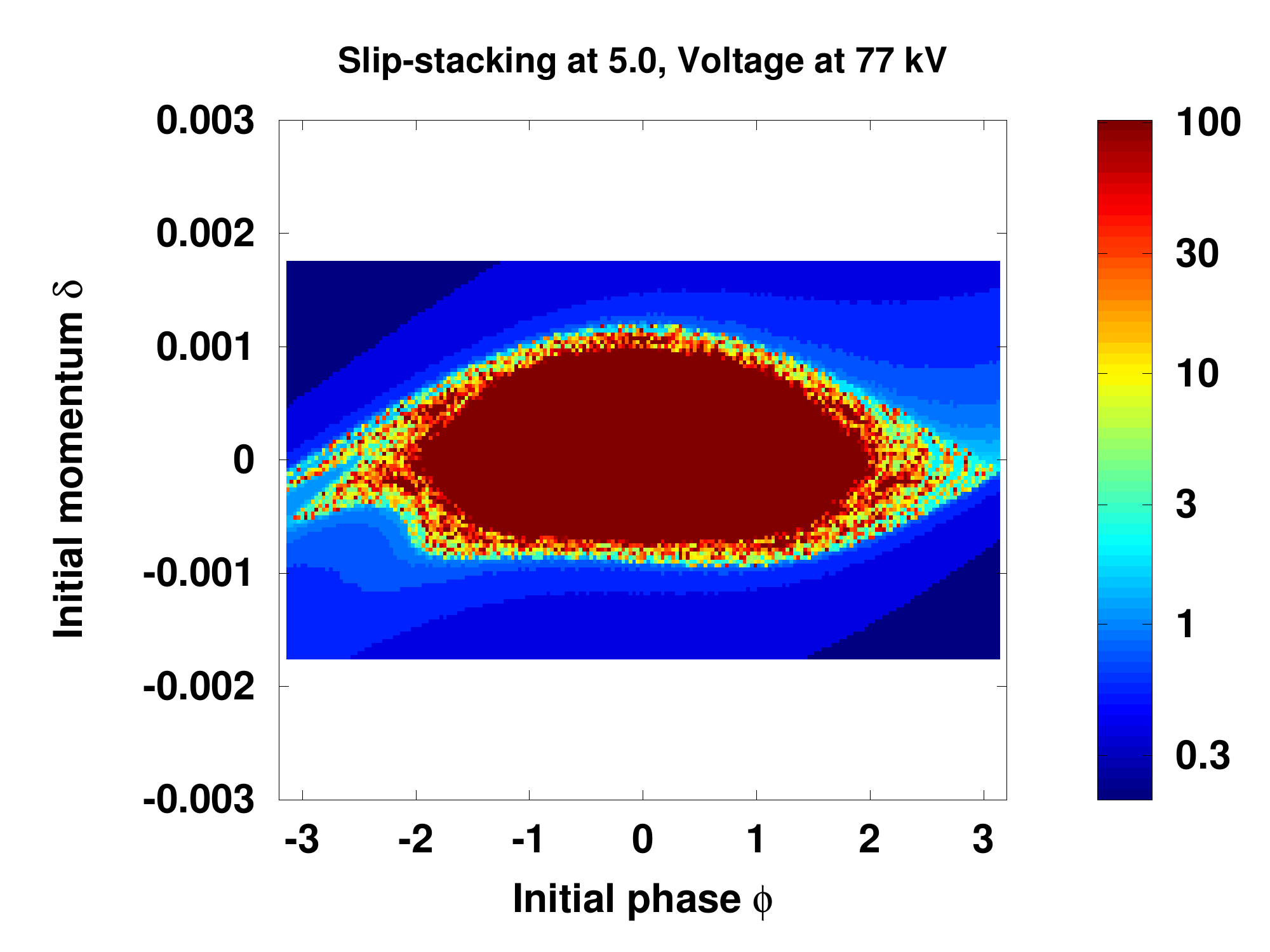}
 \centering
 \includegraphics[scale=0.51]{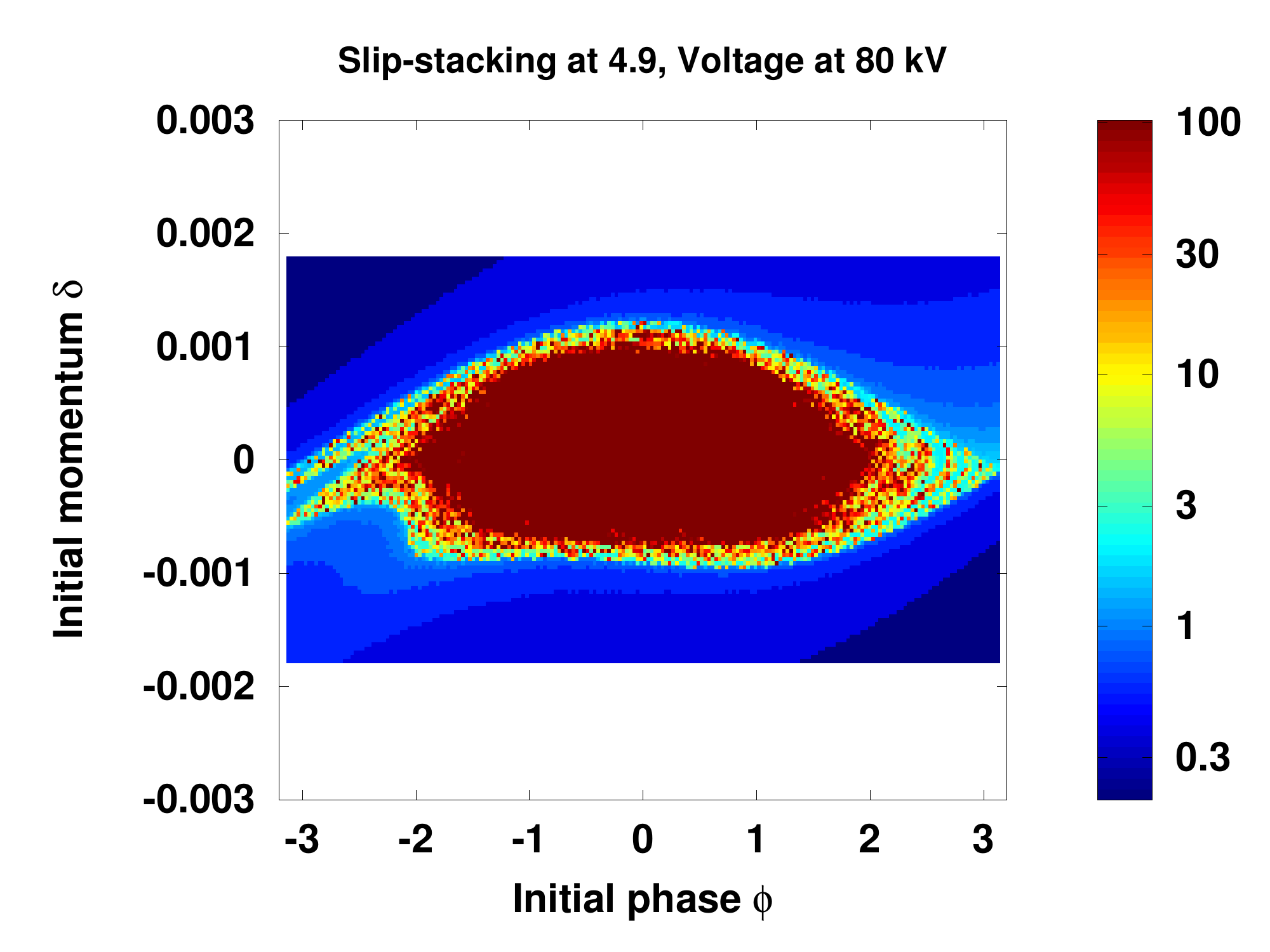}
 \centering
 \includegraphics[scale=0.51]{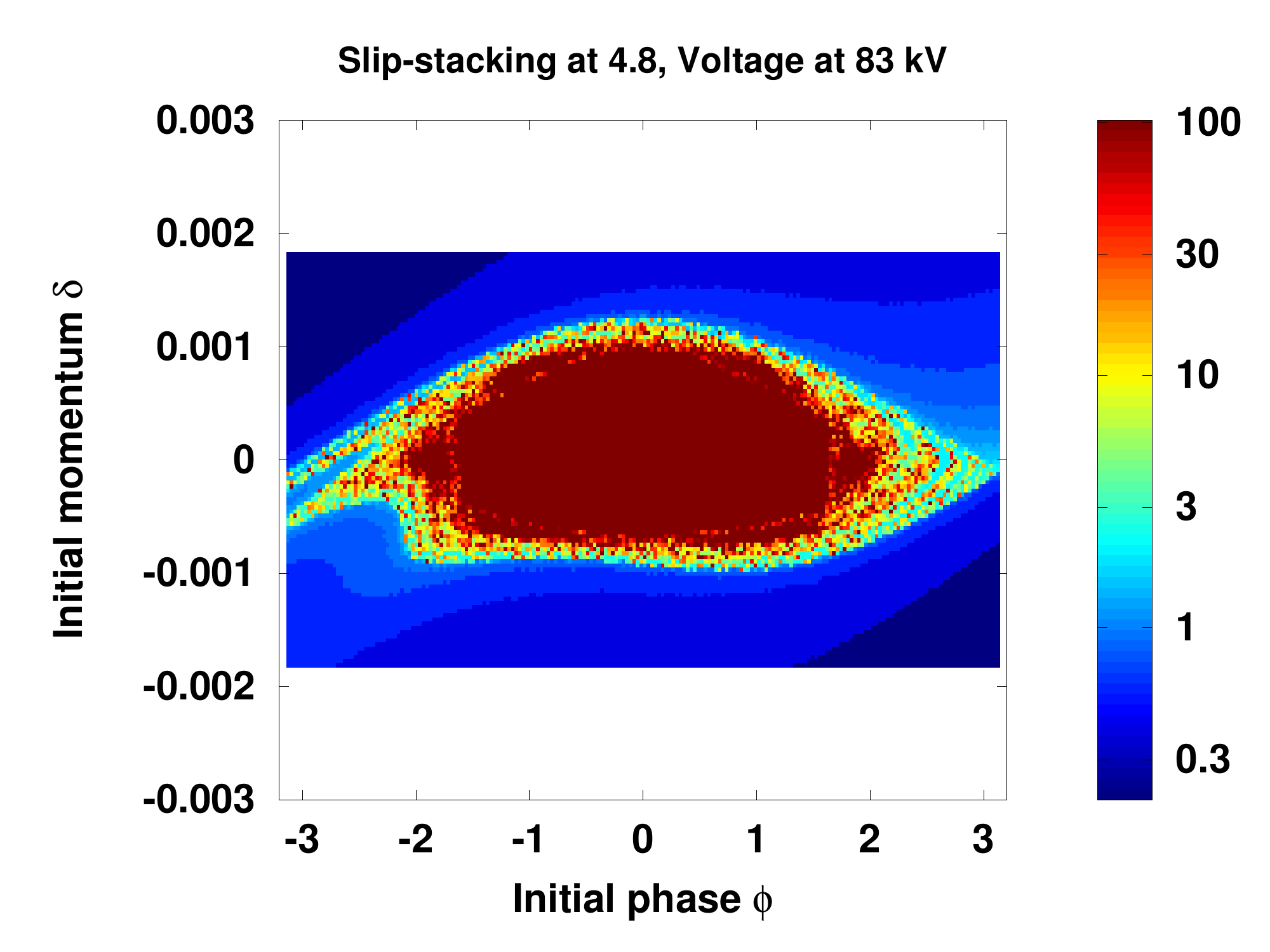}
 \centering
 \includegraphics[scale=0.51]{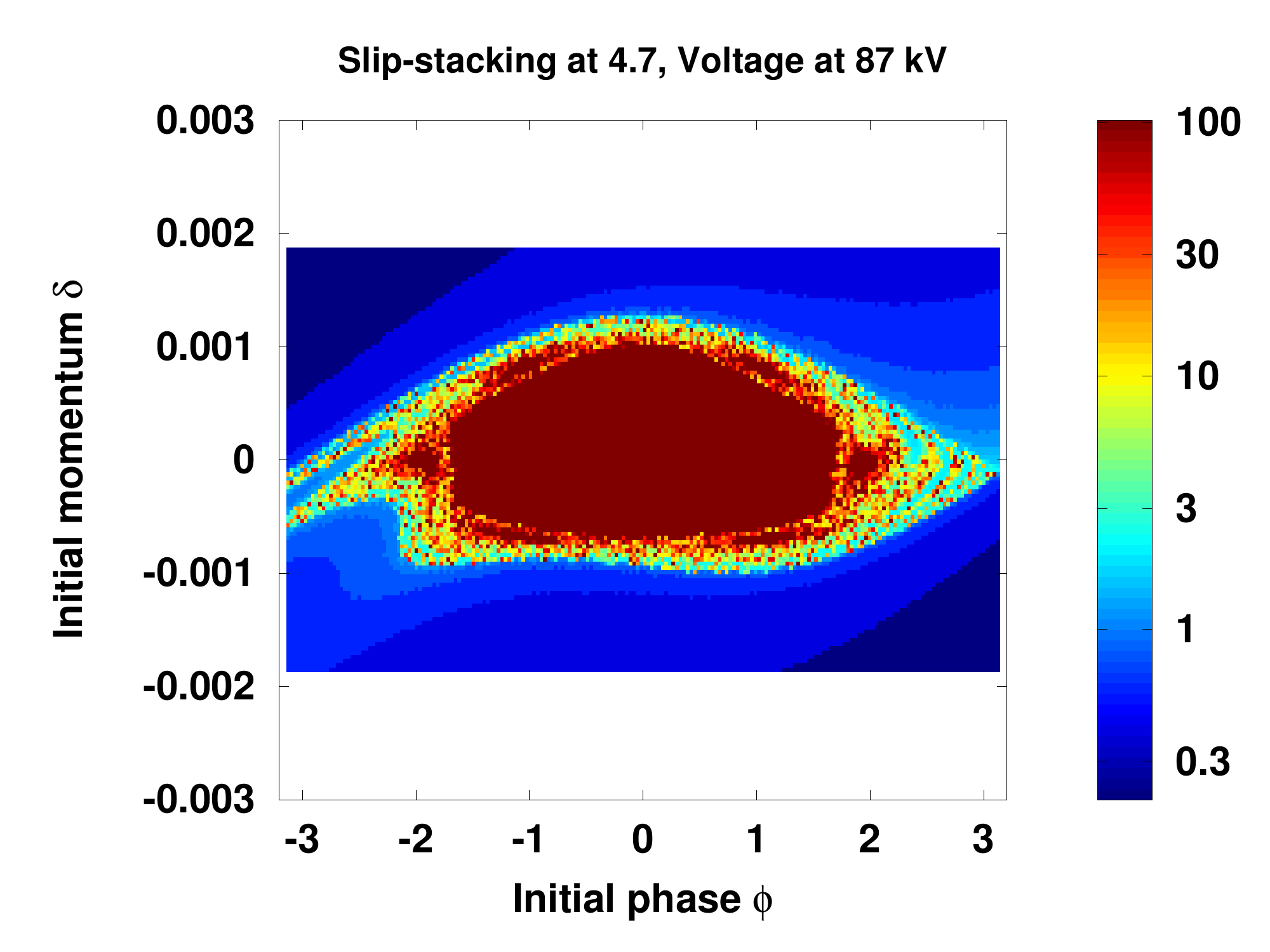}
 \centering
 \includegraphics[scale=0.51]{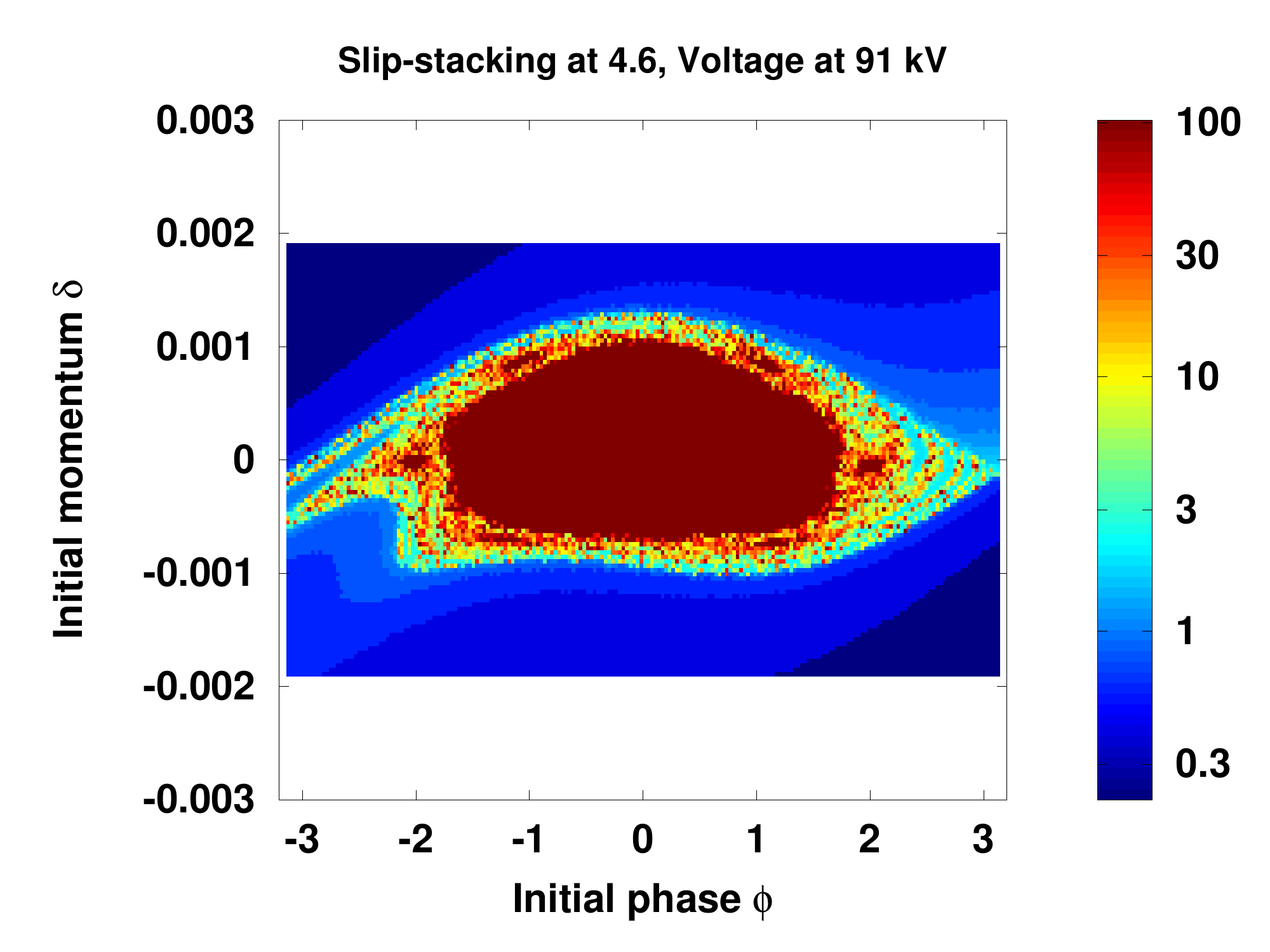}
 \centering
 \includegraphics[scale=0.51]{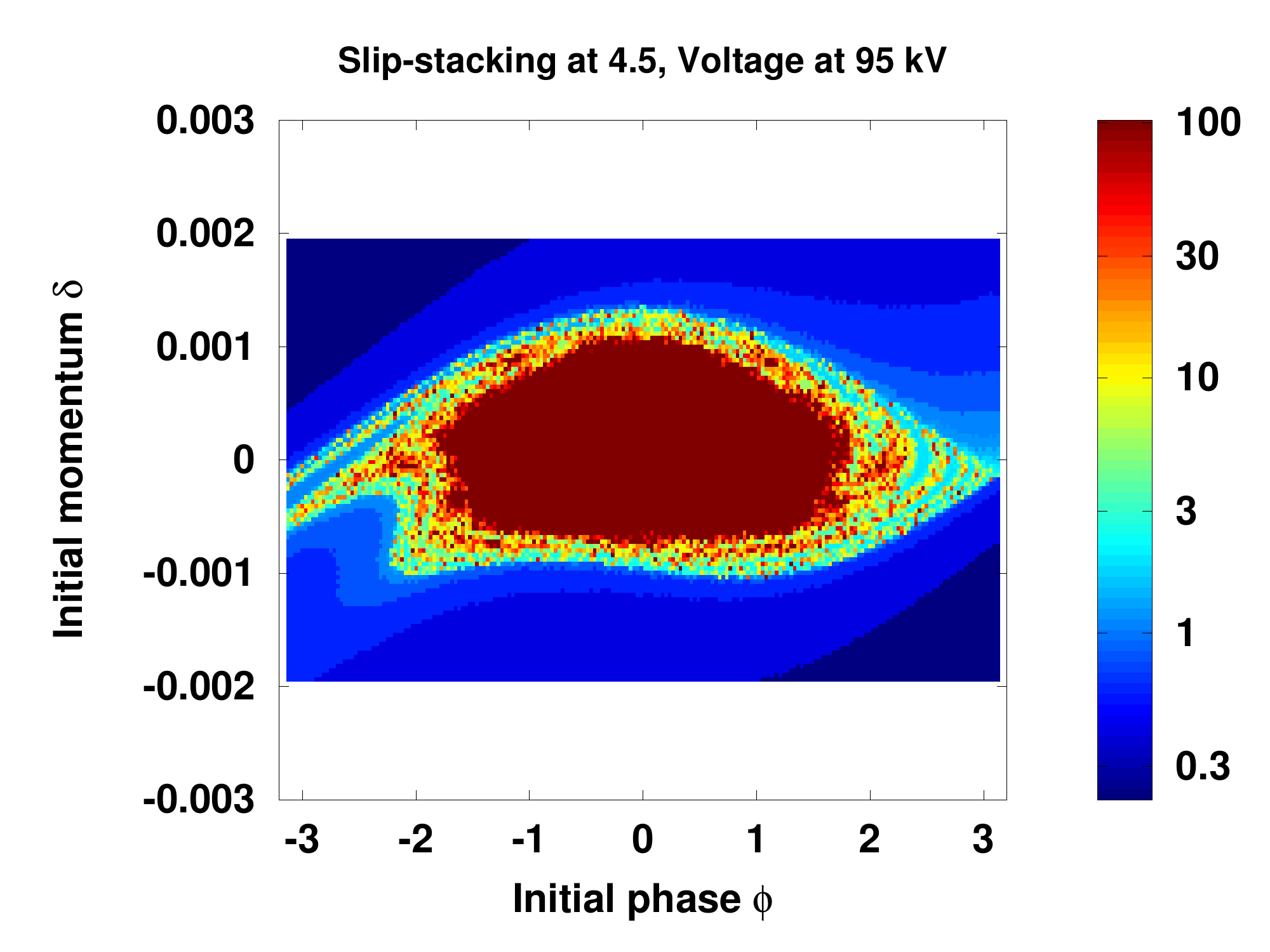}
 \centering
 \includegraphics[scale=0.51]{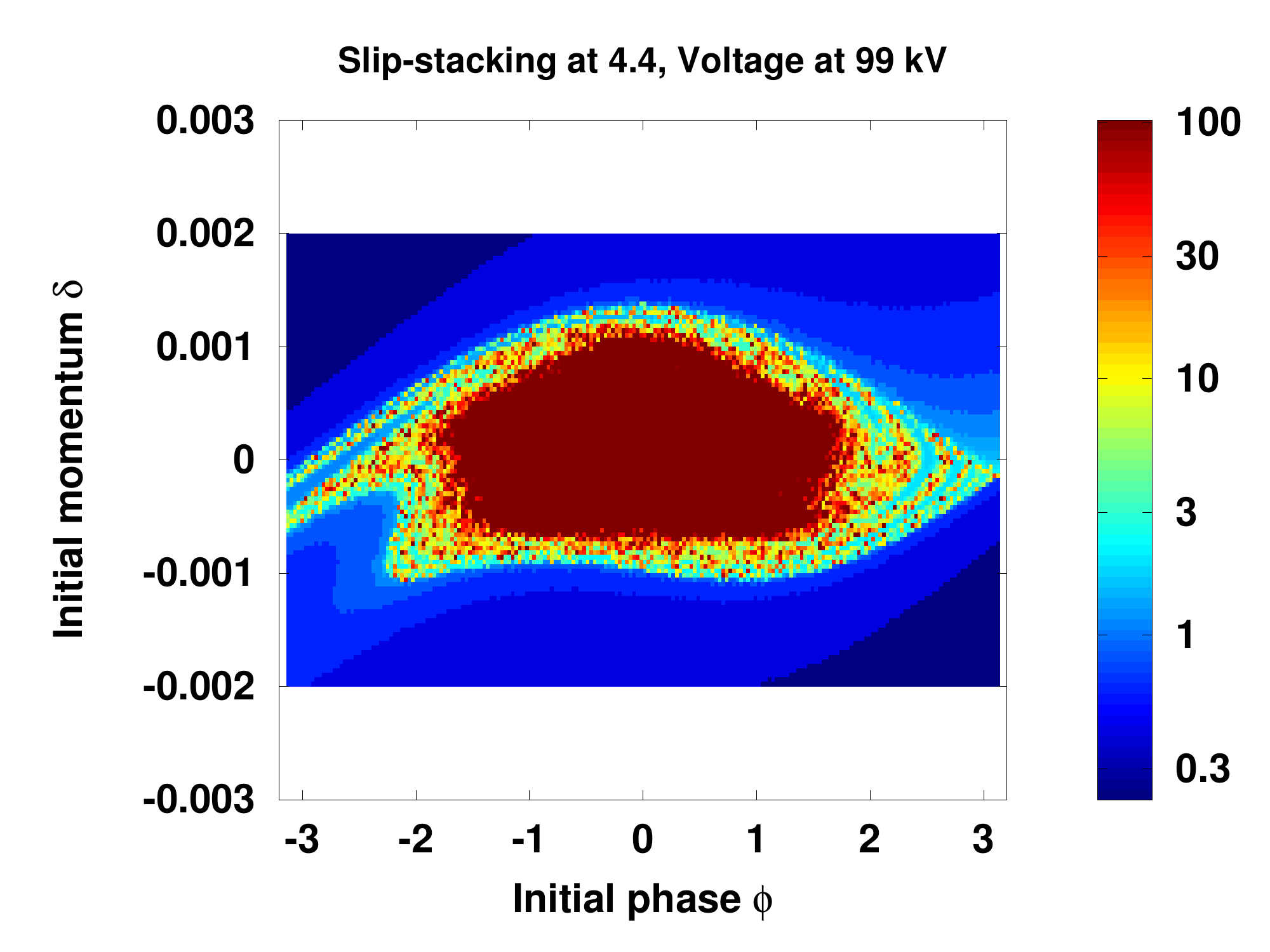}
 \centering
 \includegraphics[scale=0.51]{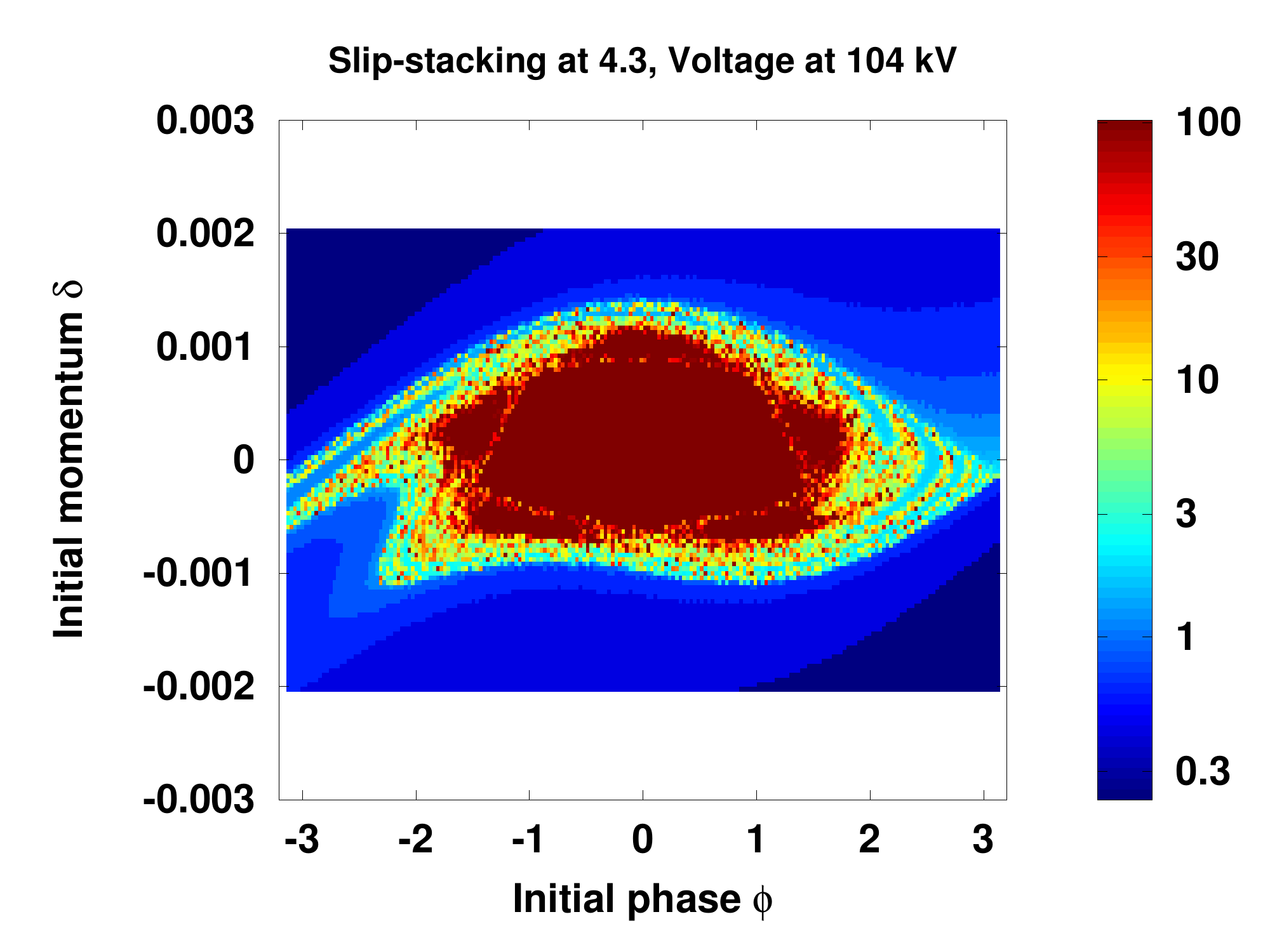}
 \centering
 \includegraphics[scale=0.51]{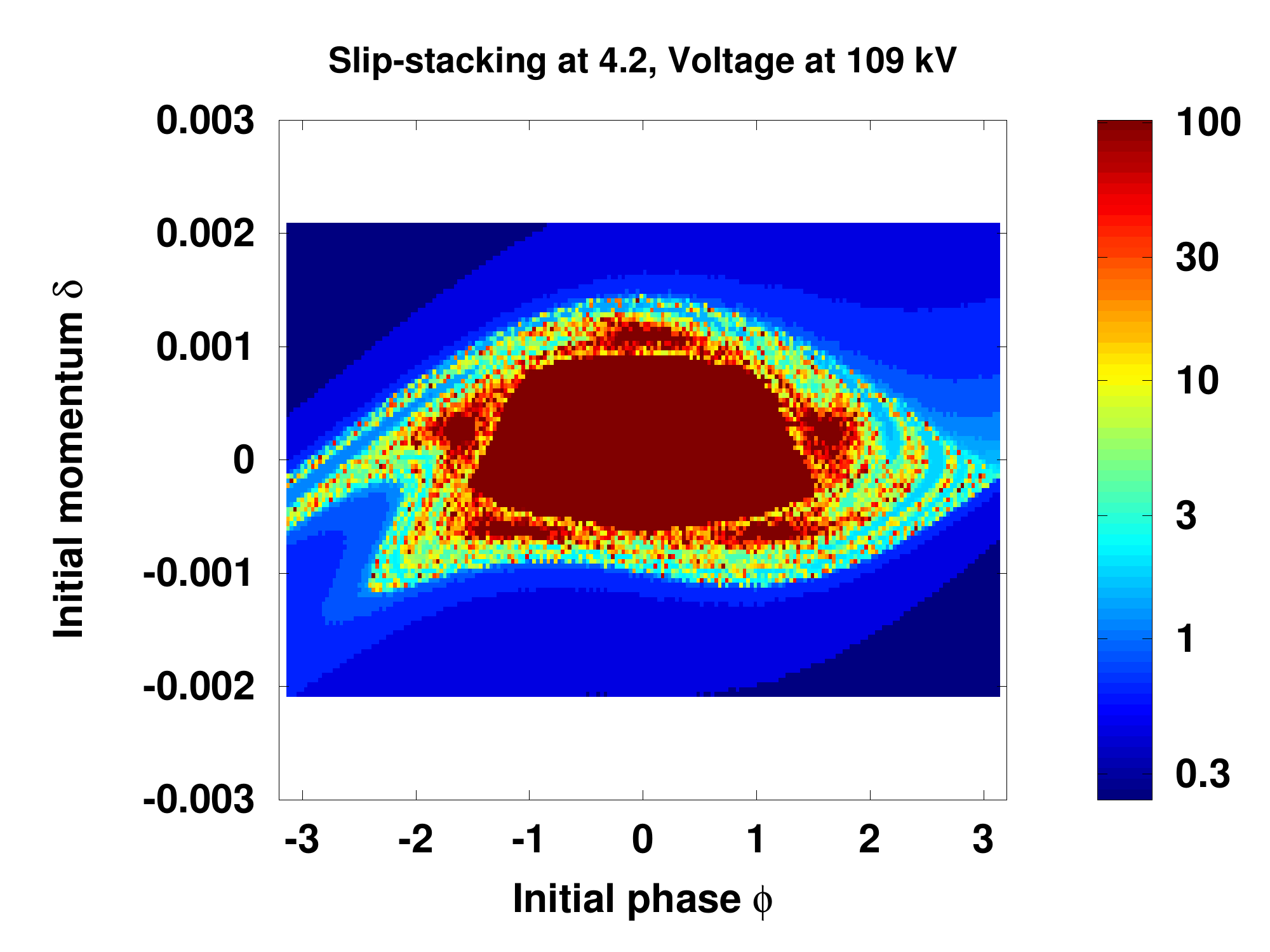}
 \centering
 \includegraphics[scale=0.51]{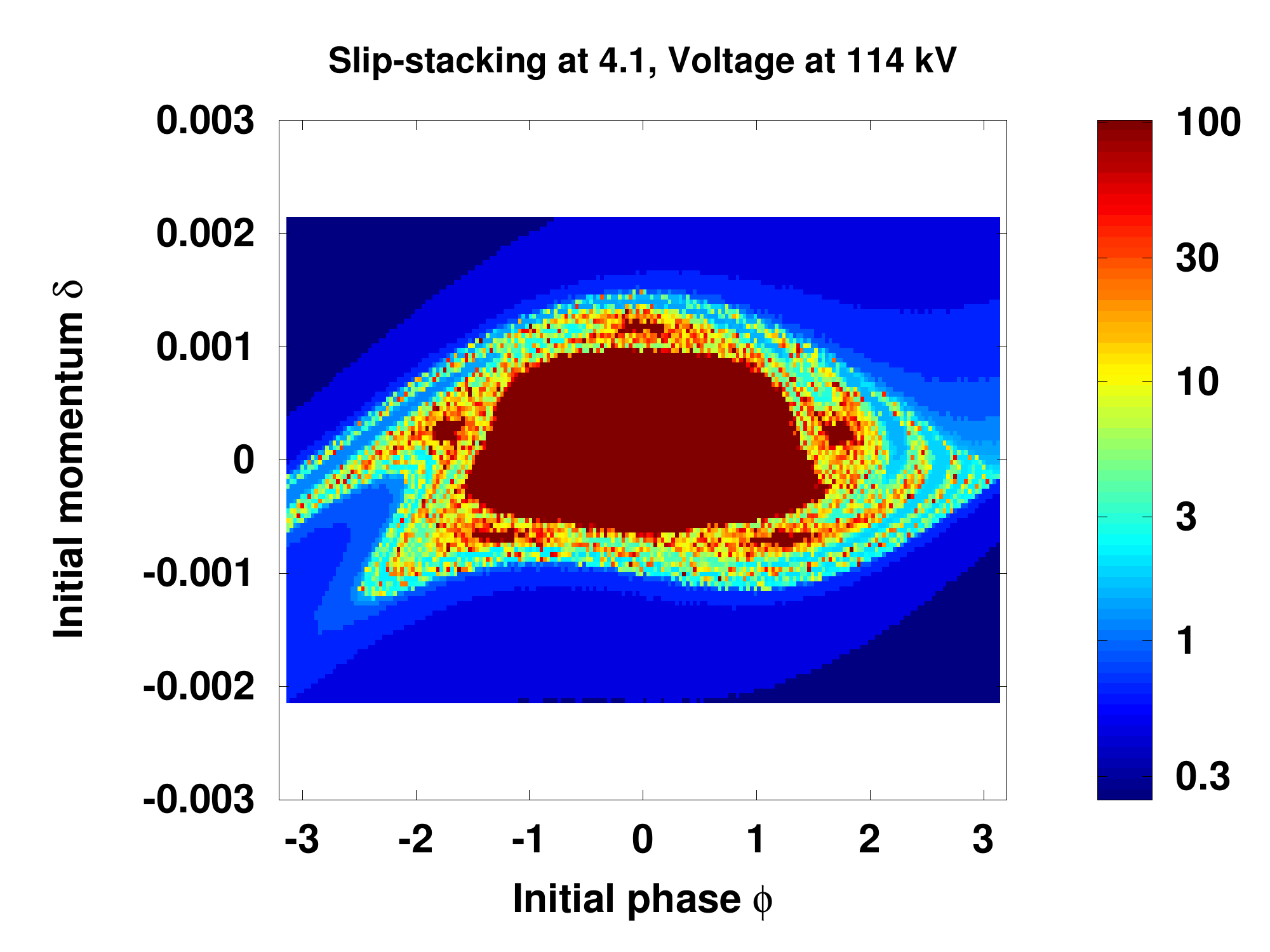}
 \centering
 \includegraphics[scale=0.51]{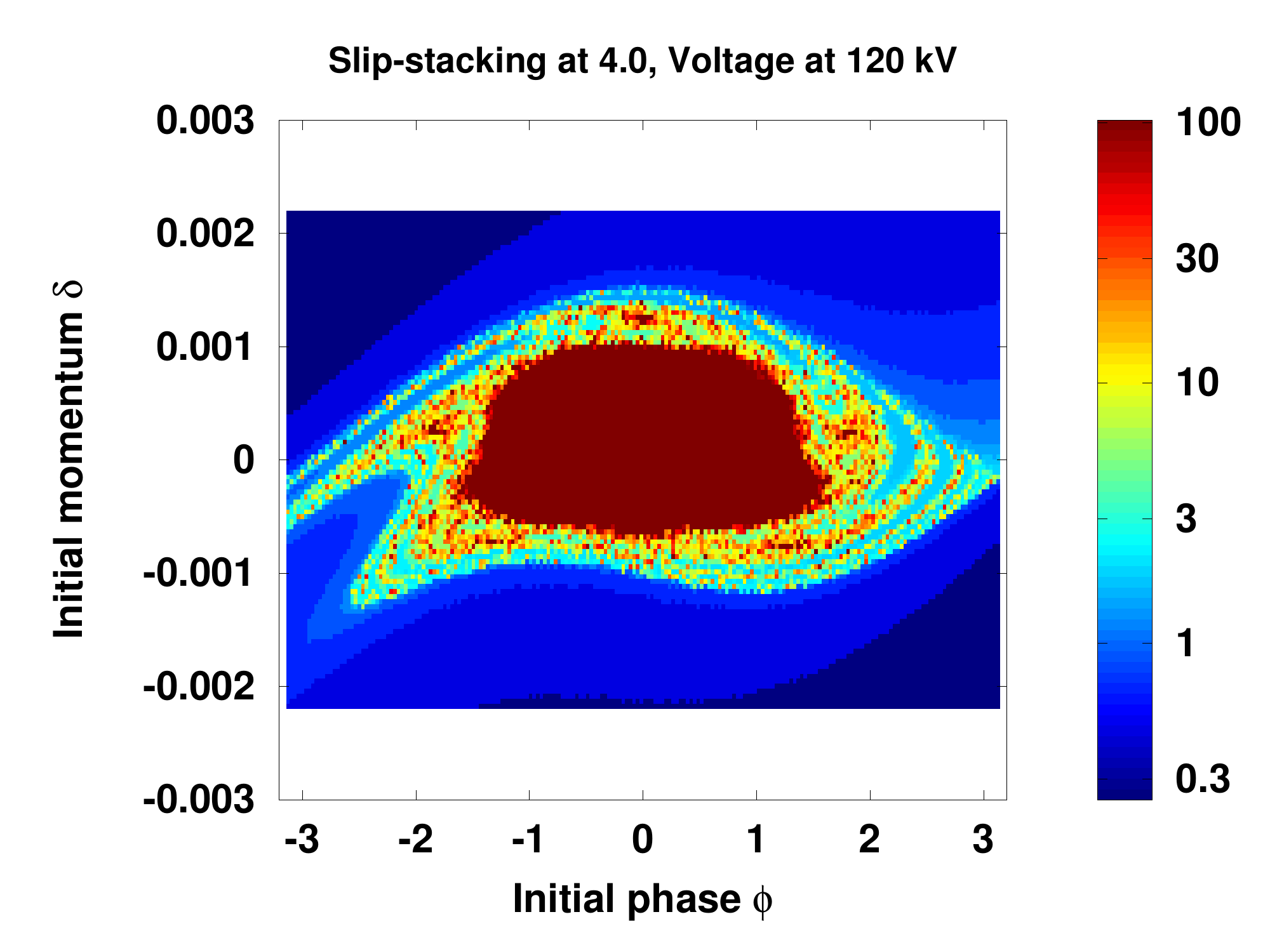}
 \centering
 \includegraphics[scale=0.51]{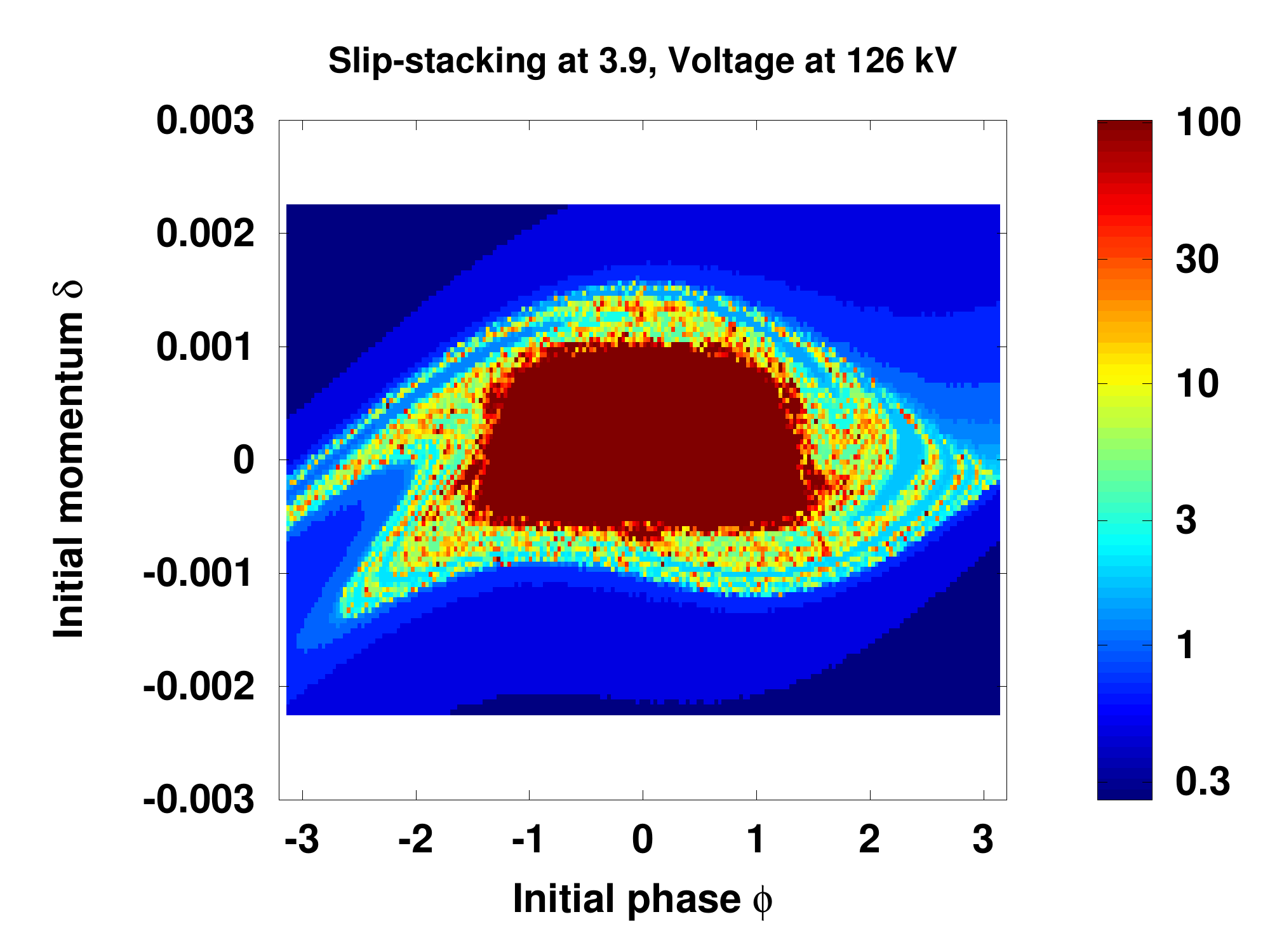}
 \centering
 \includegraphics[scale=0.51]{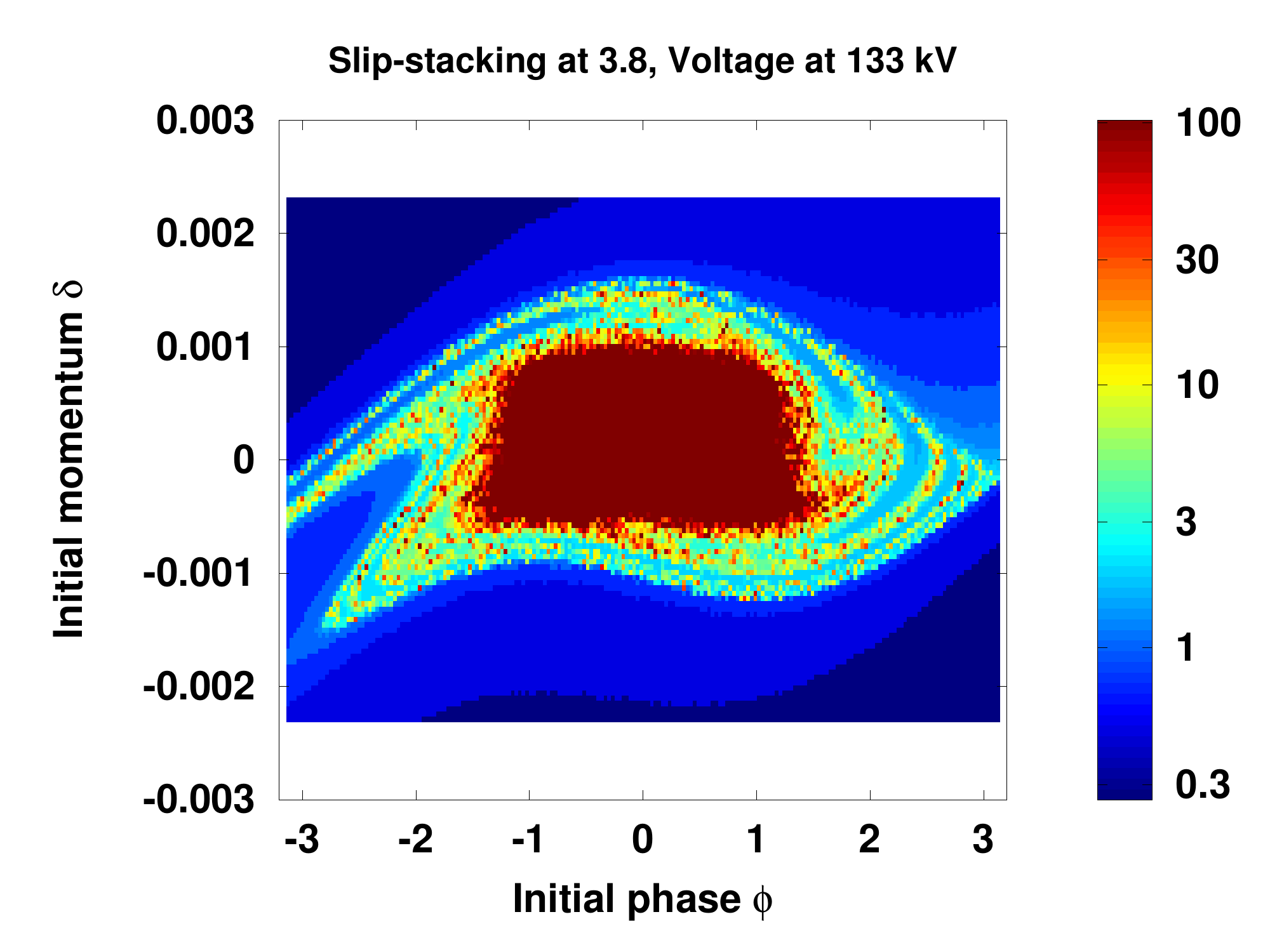}
 \centering
 \includegraphics[scale=0.51]{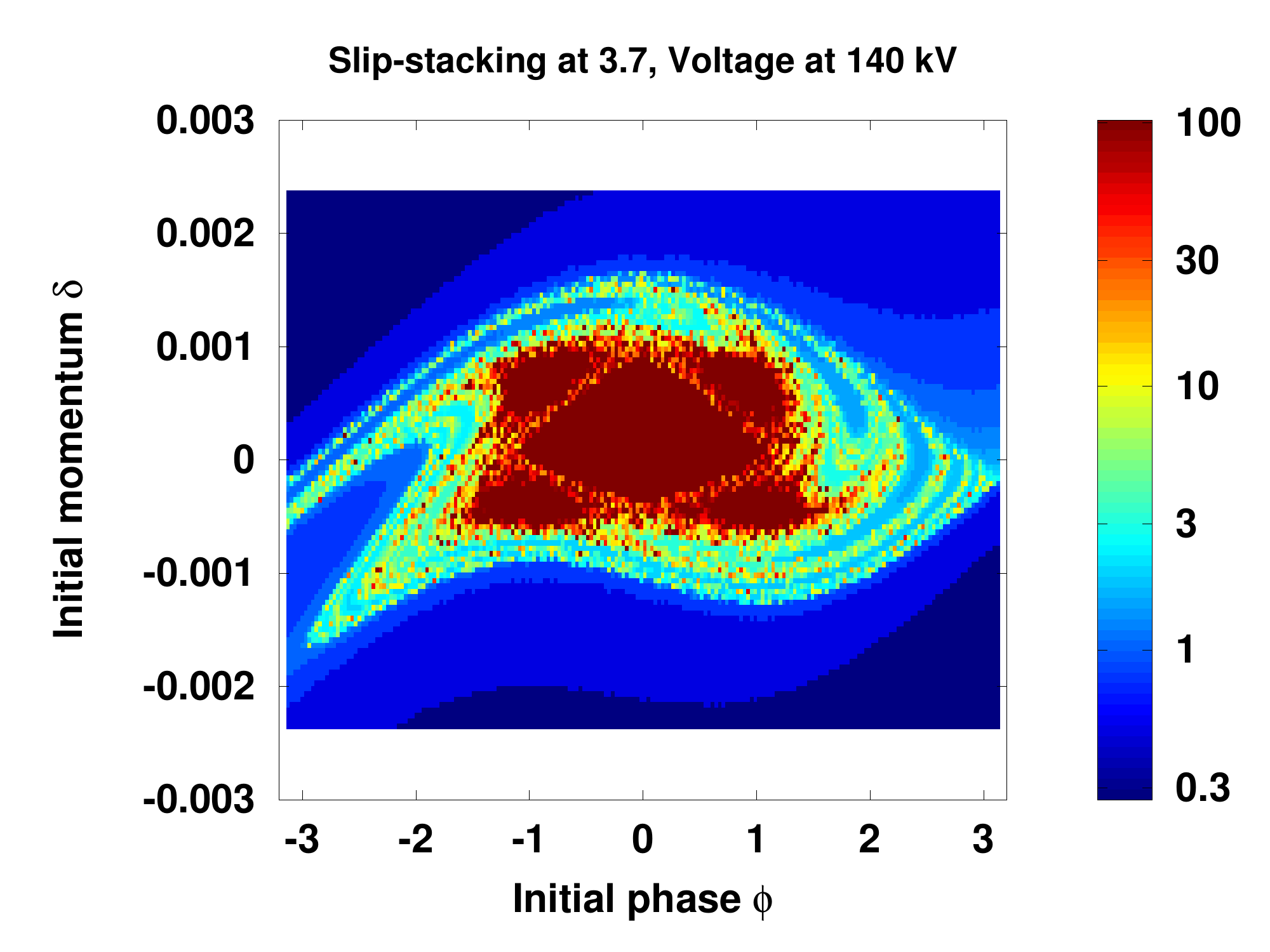}
 \centering
 \includegraphics[scale=0.51]{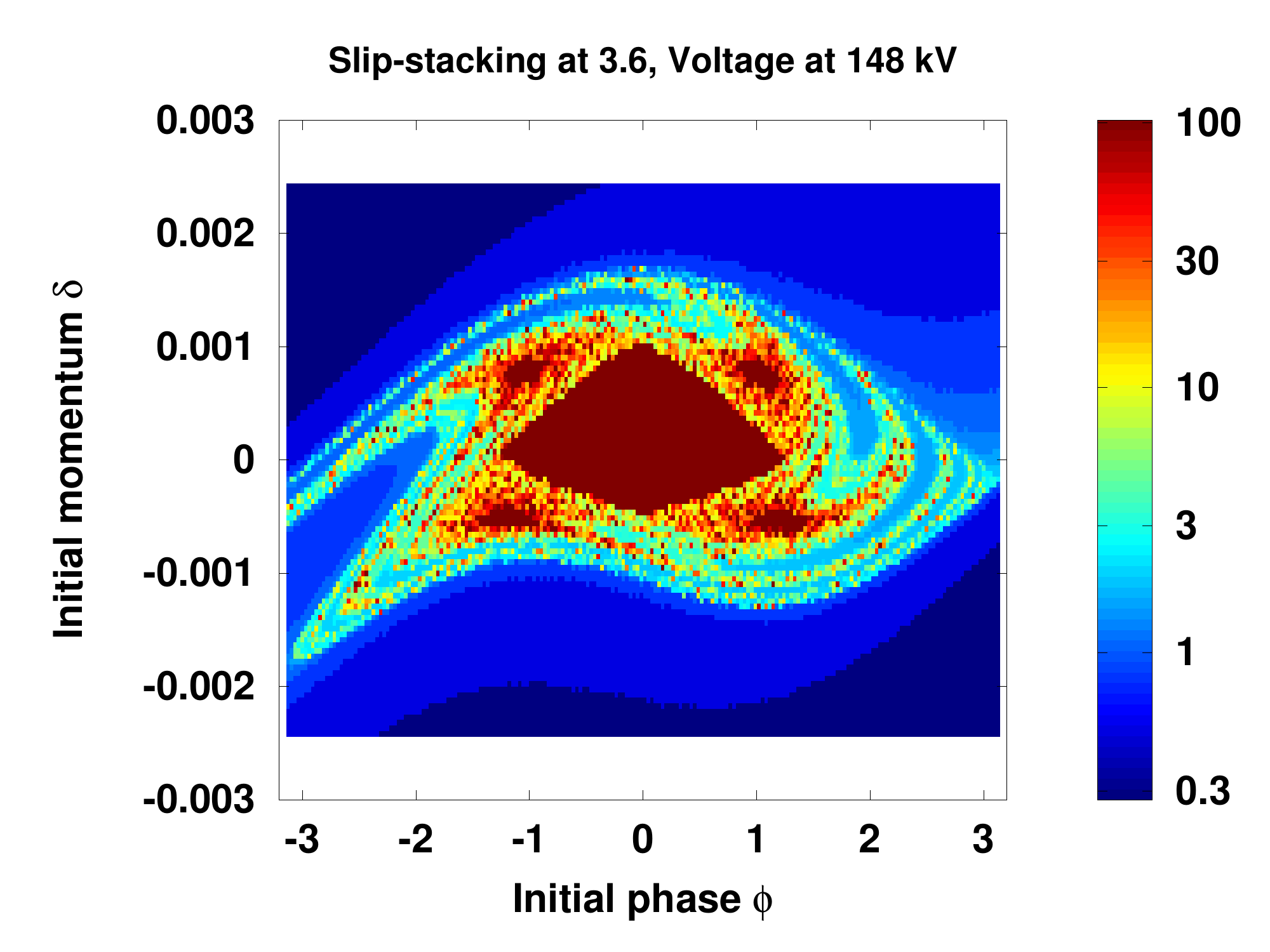}
 \centering
 \includegraphics[scale=0.51]{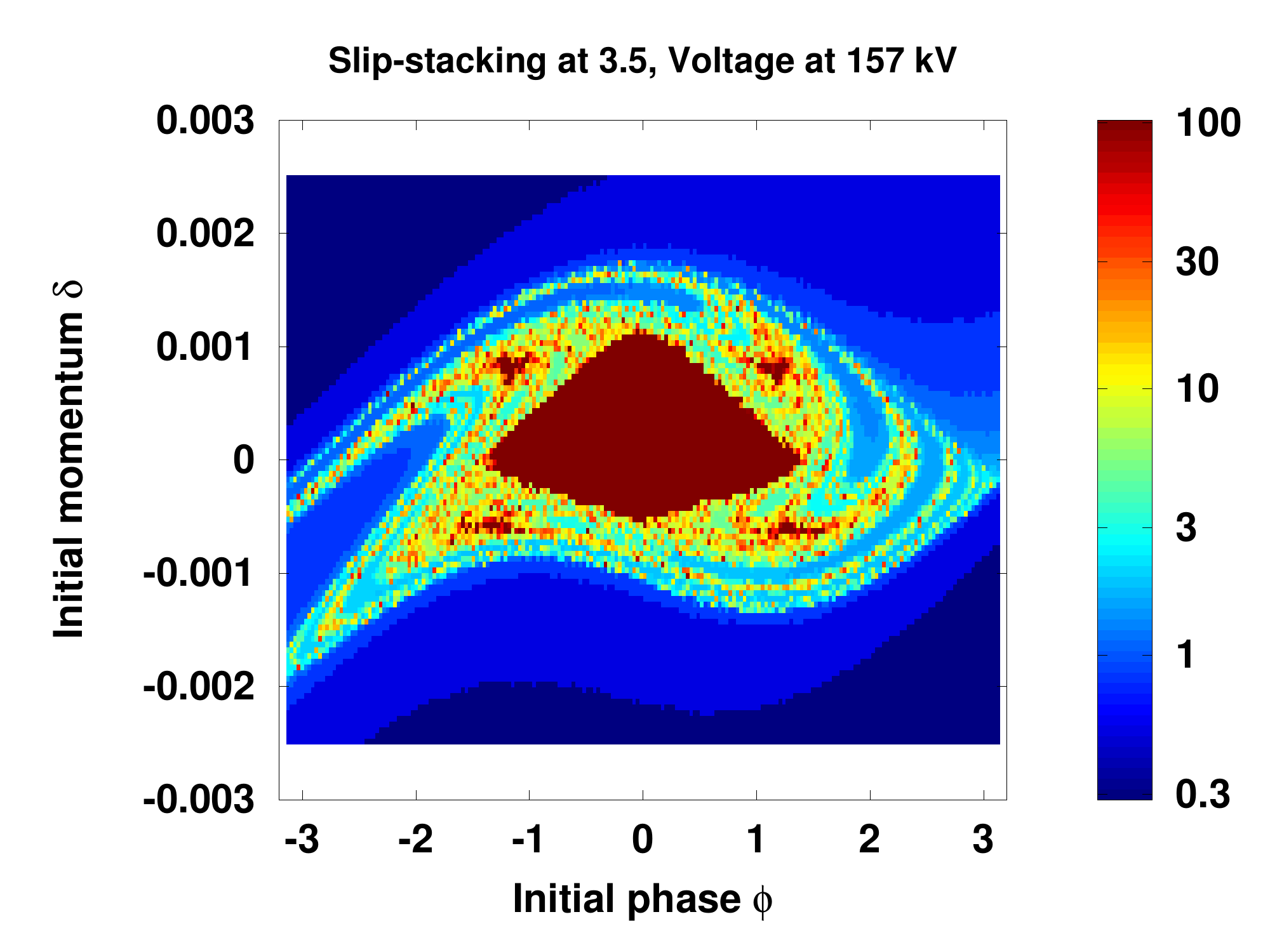}
 \centering
 \includegraphics[scale=0.51]{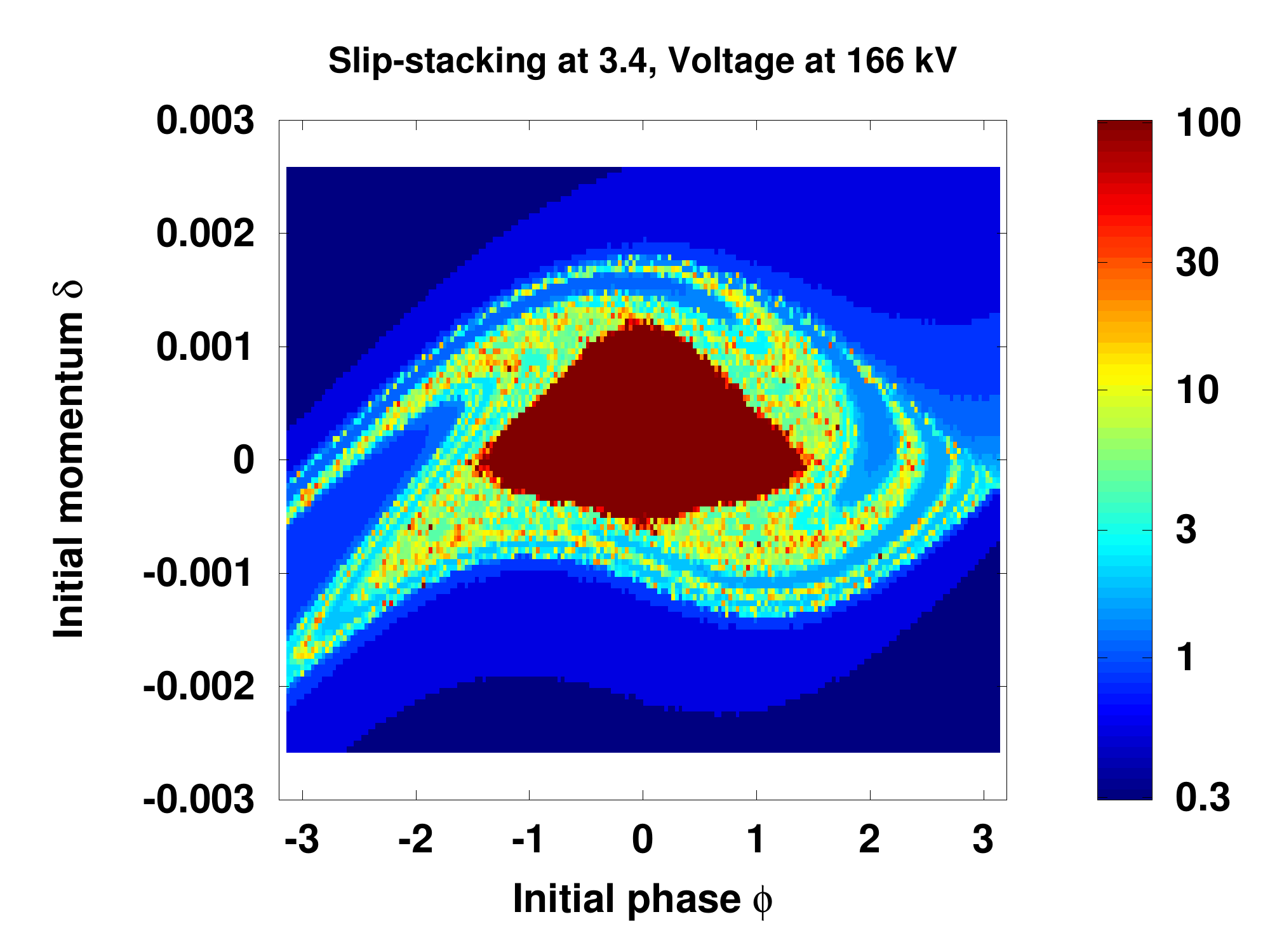}
 \centering
 \includegraphics[scale=0.51]{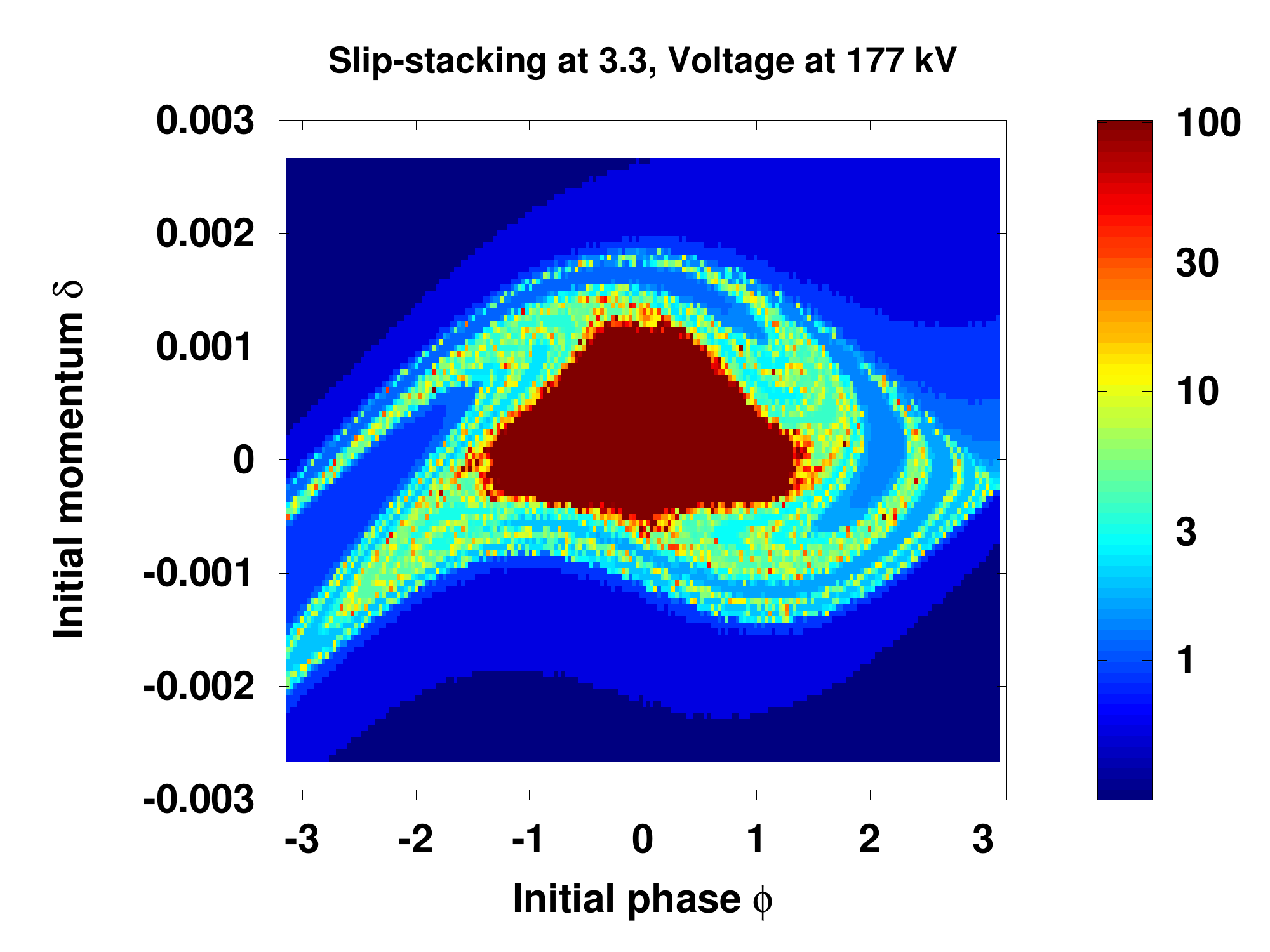}
 \centering
 \includegraphics[scale=0.51]{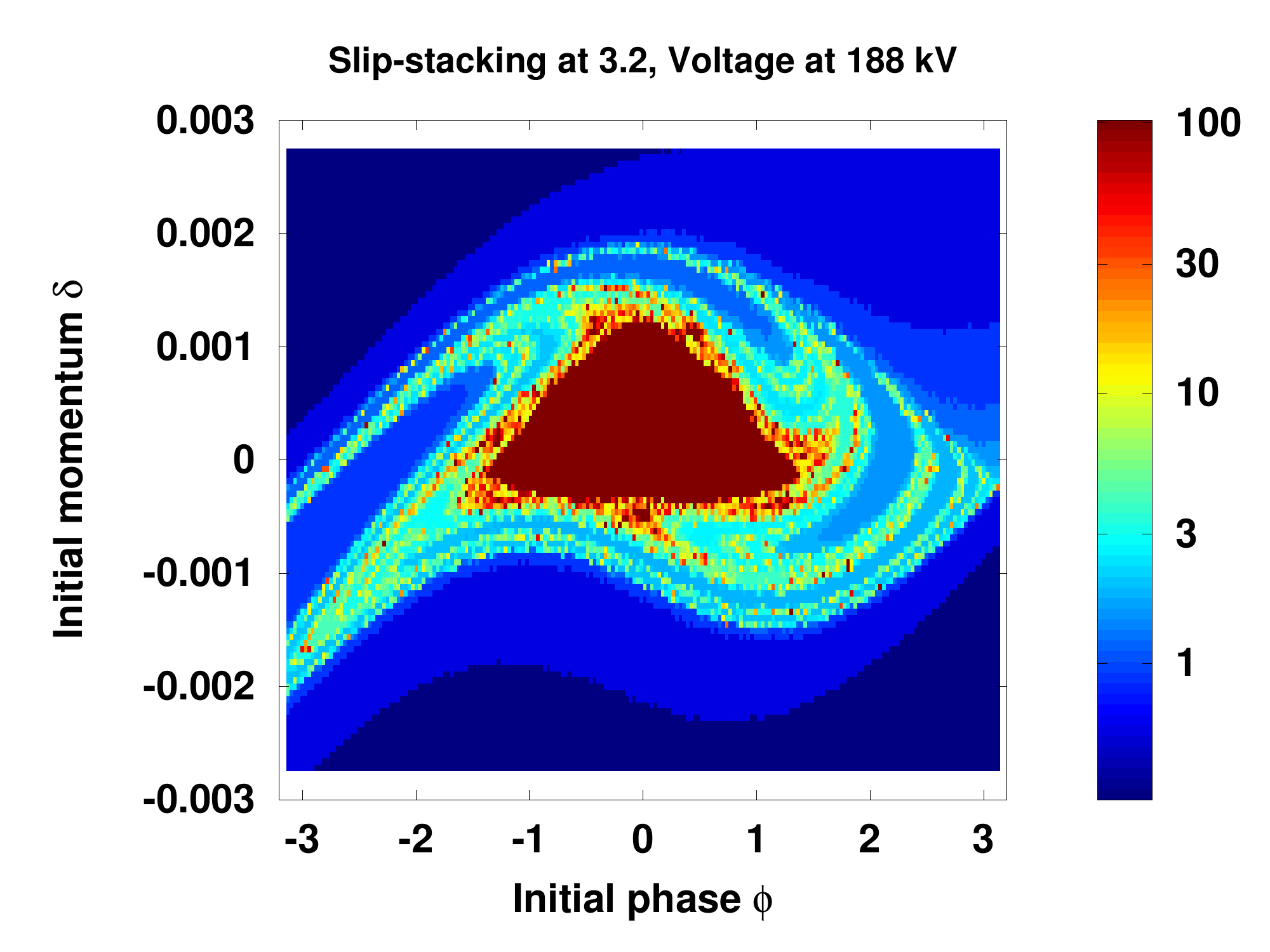}
 \centering
 \includegraphics[scale=0.51]{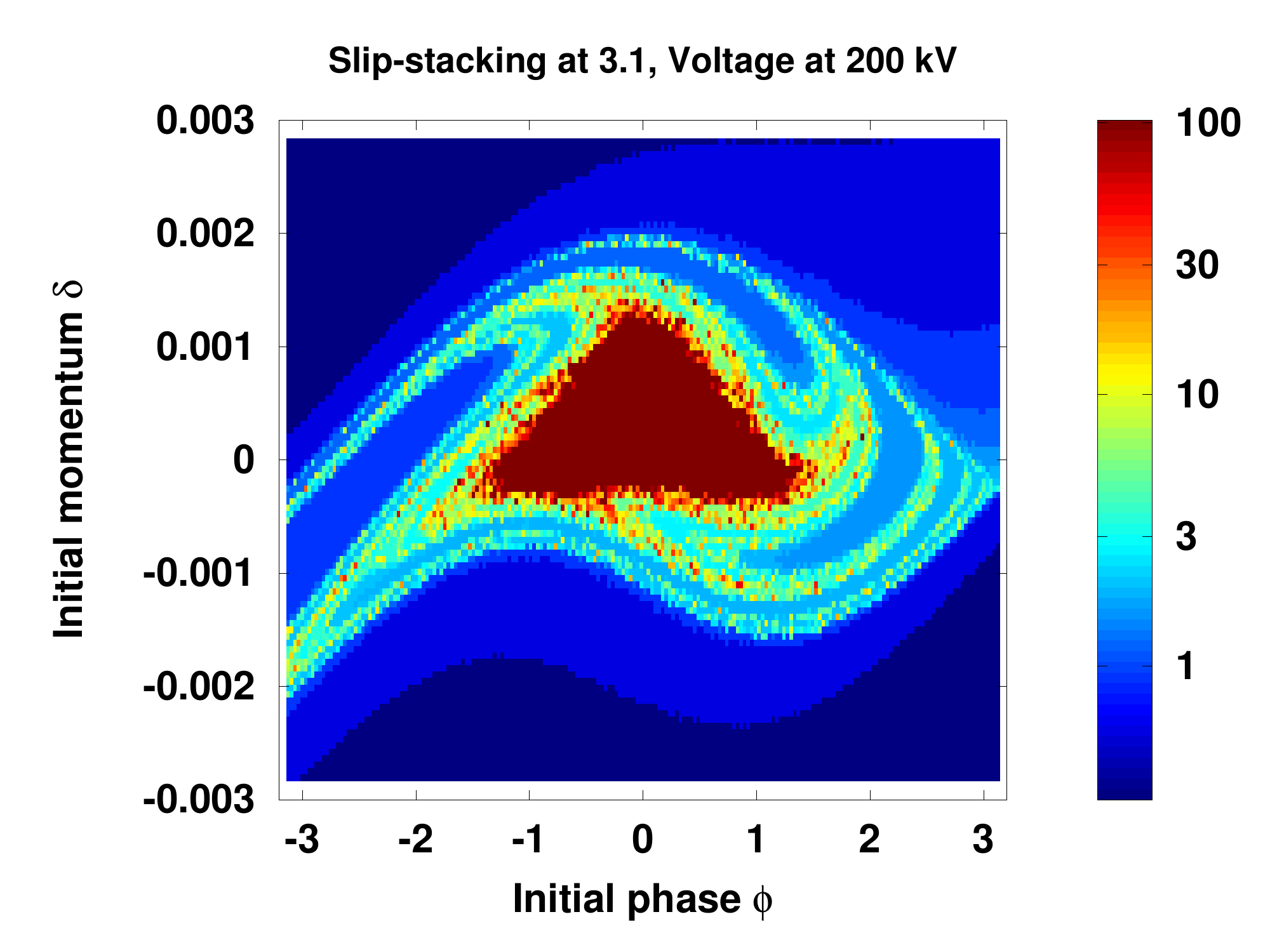}
 \centering
 \includegraphics[scale=0.51]{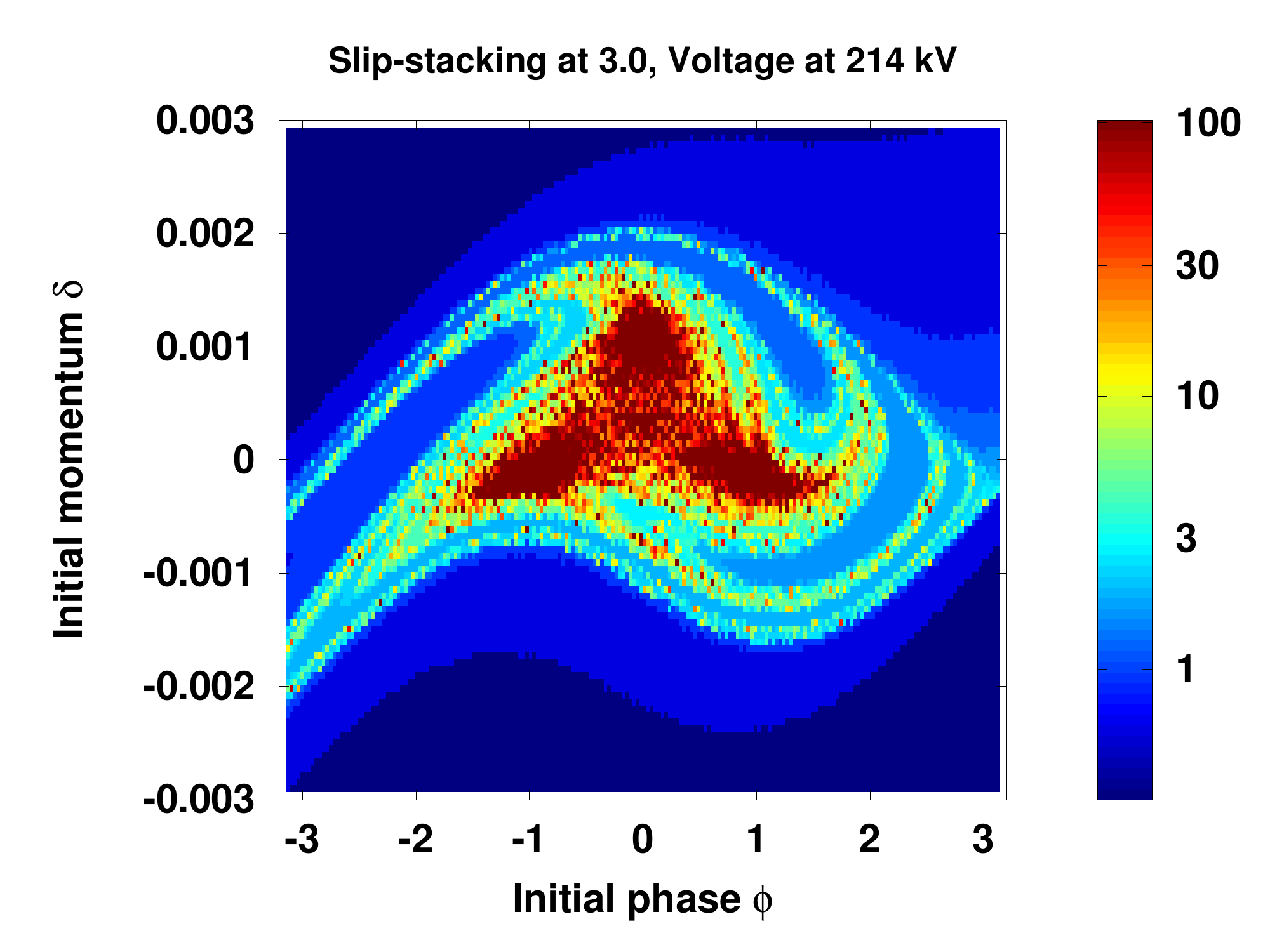}
 \centering
 \includegraphics[scale=0.51]{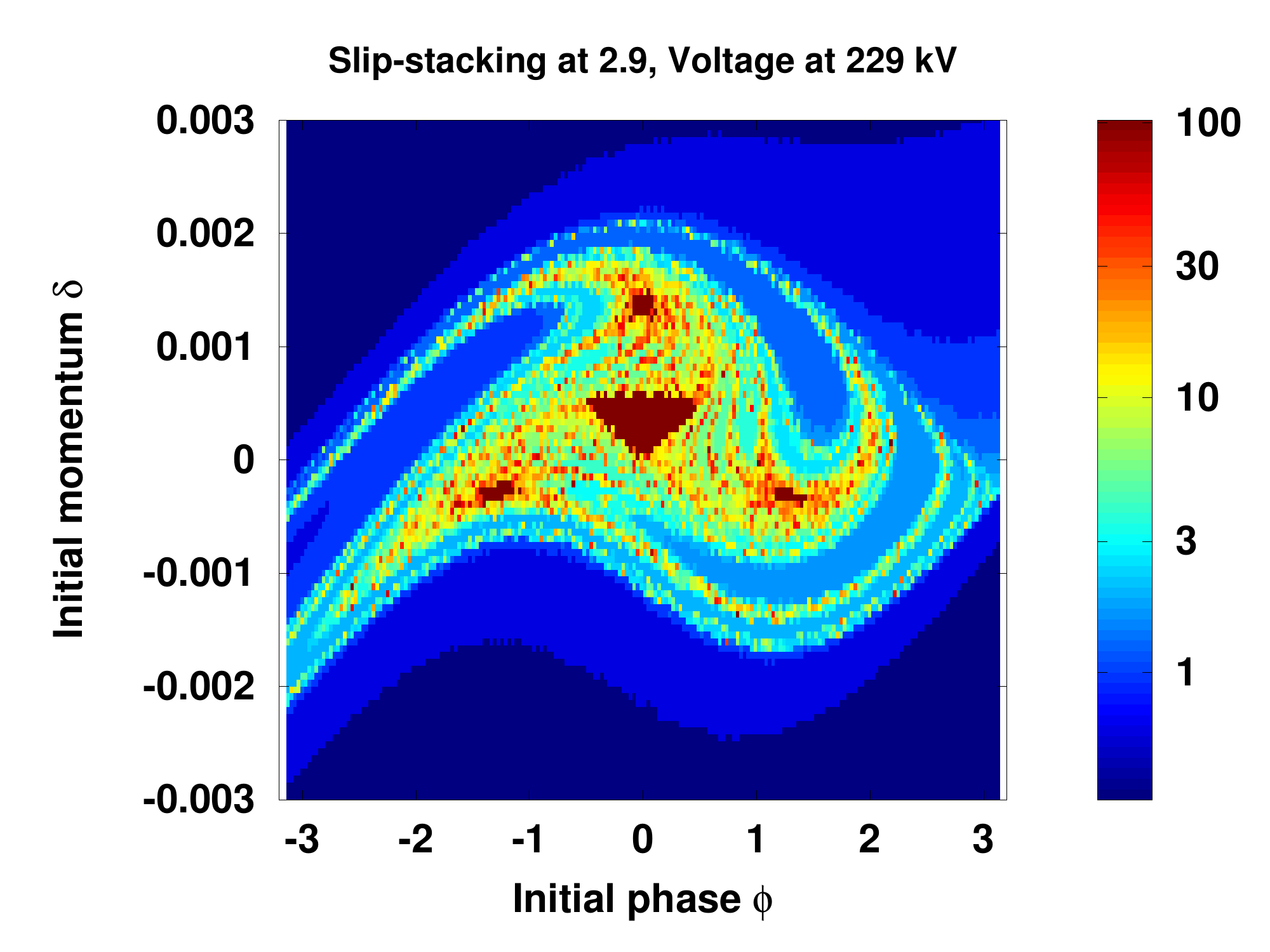}
 \centering
 \includegraphics[scale=0.51]{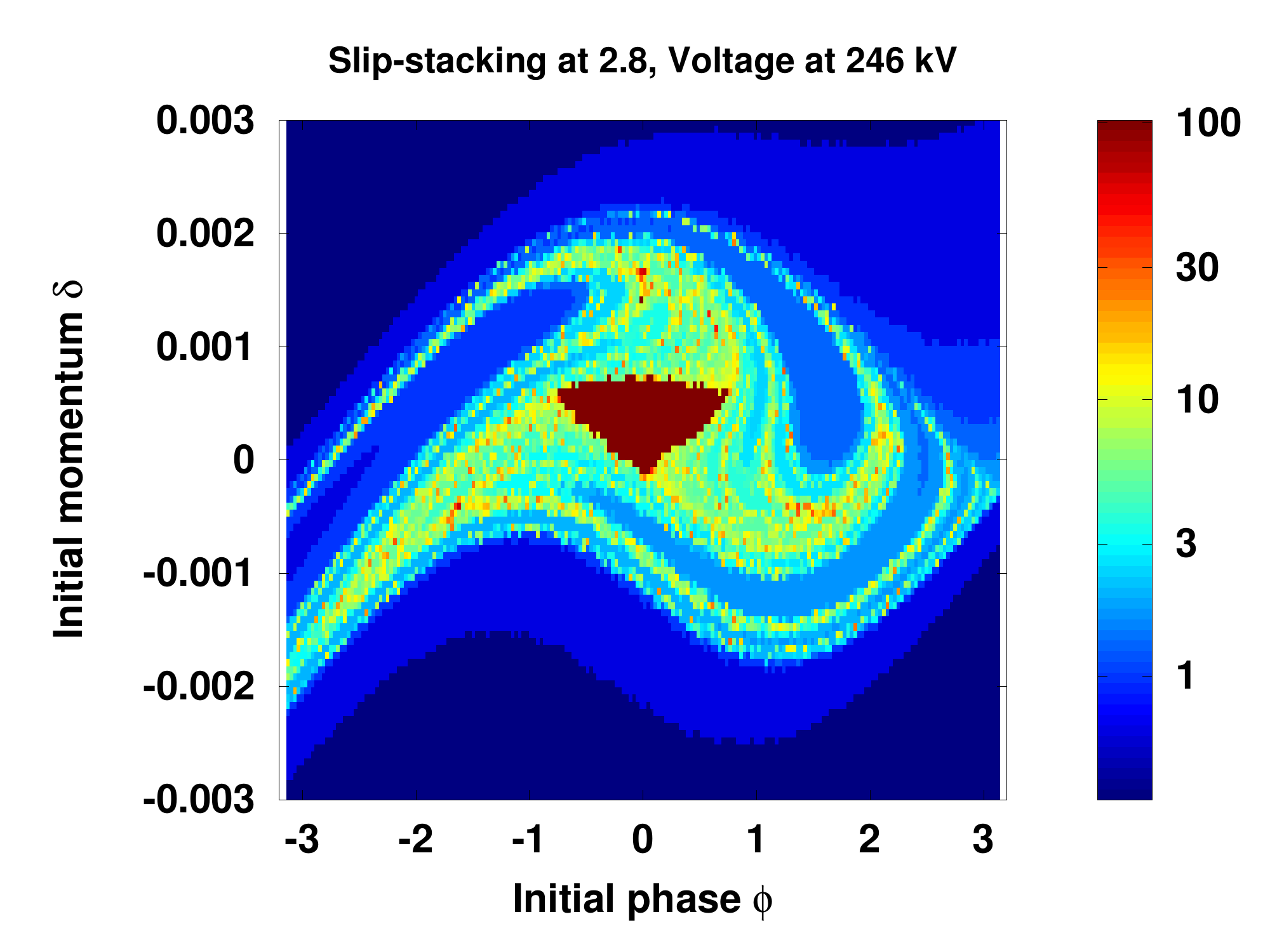}
 \centering
 \includegraphics[scale=0.51]{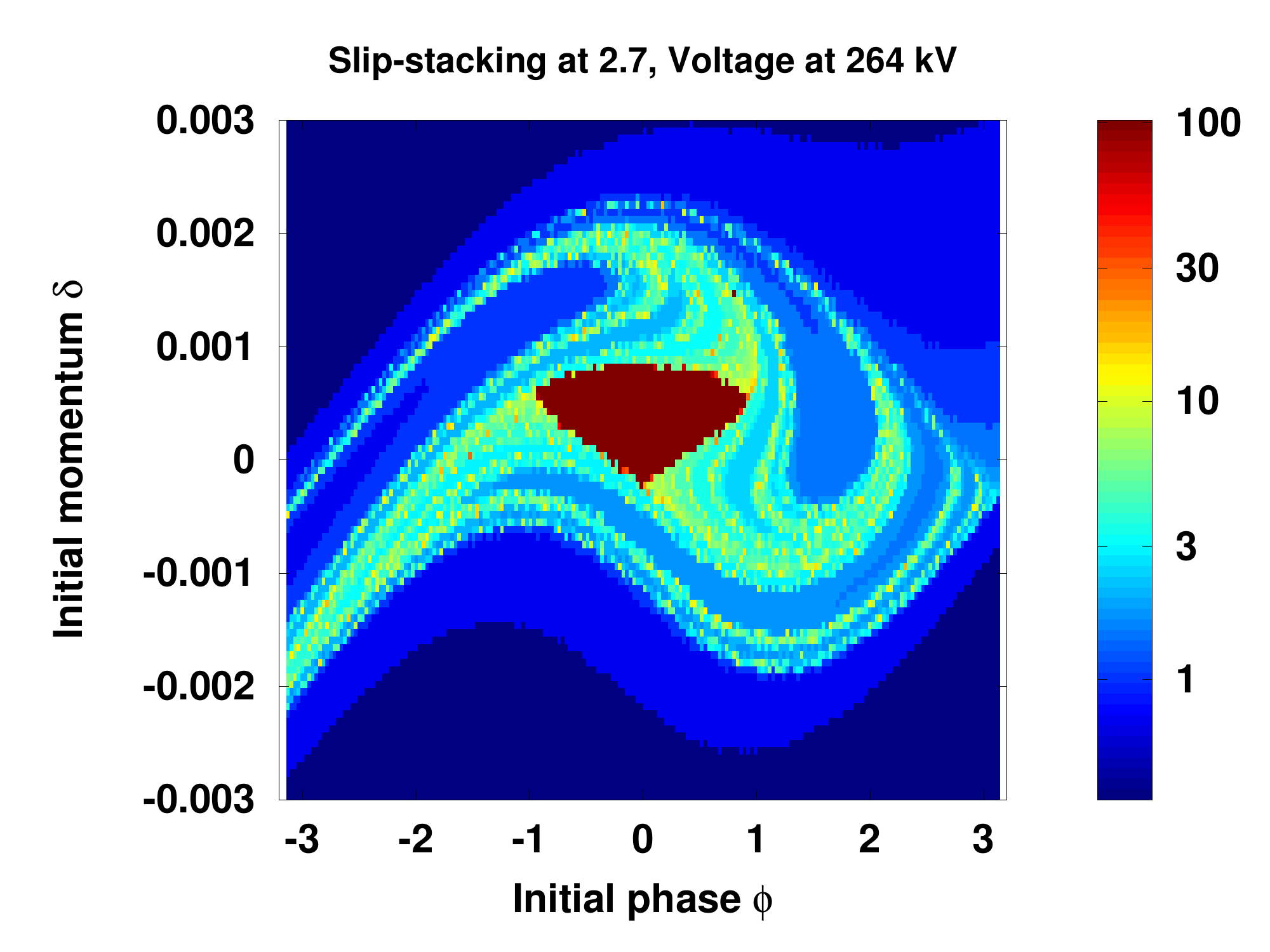}
 \centering
 \includegraphics[scale=0.51]{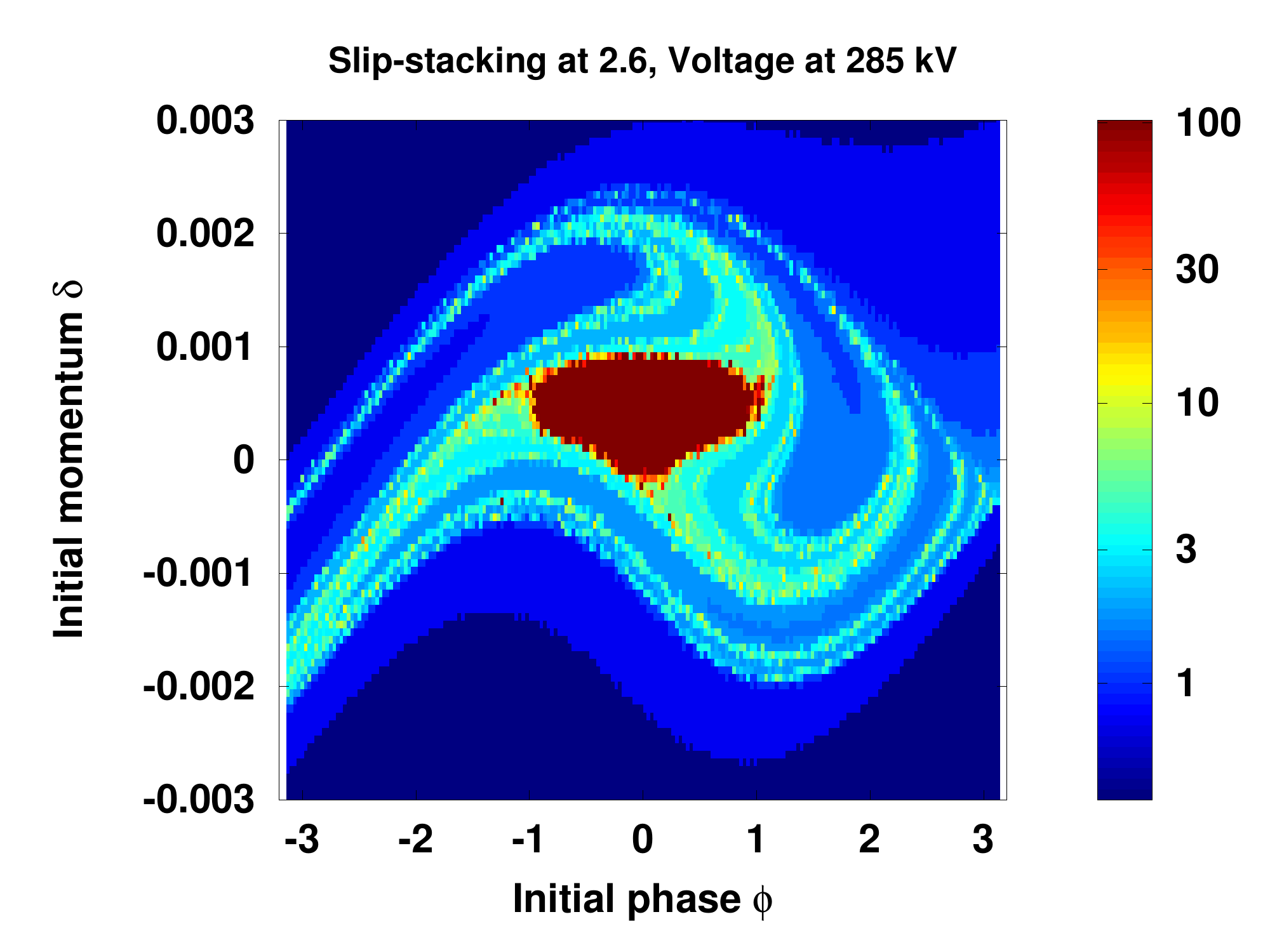}
 \centering
 \includegraphics[scale=0.51]{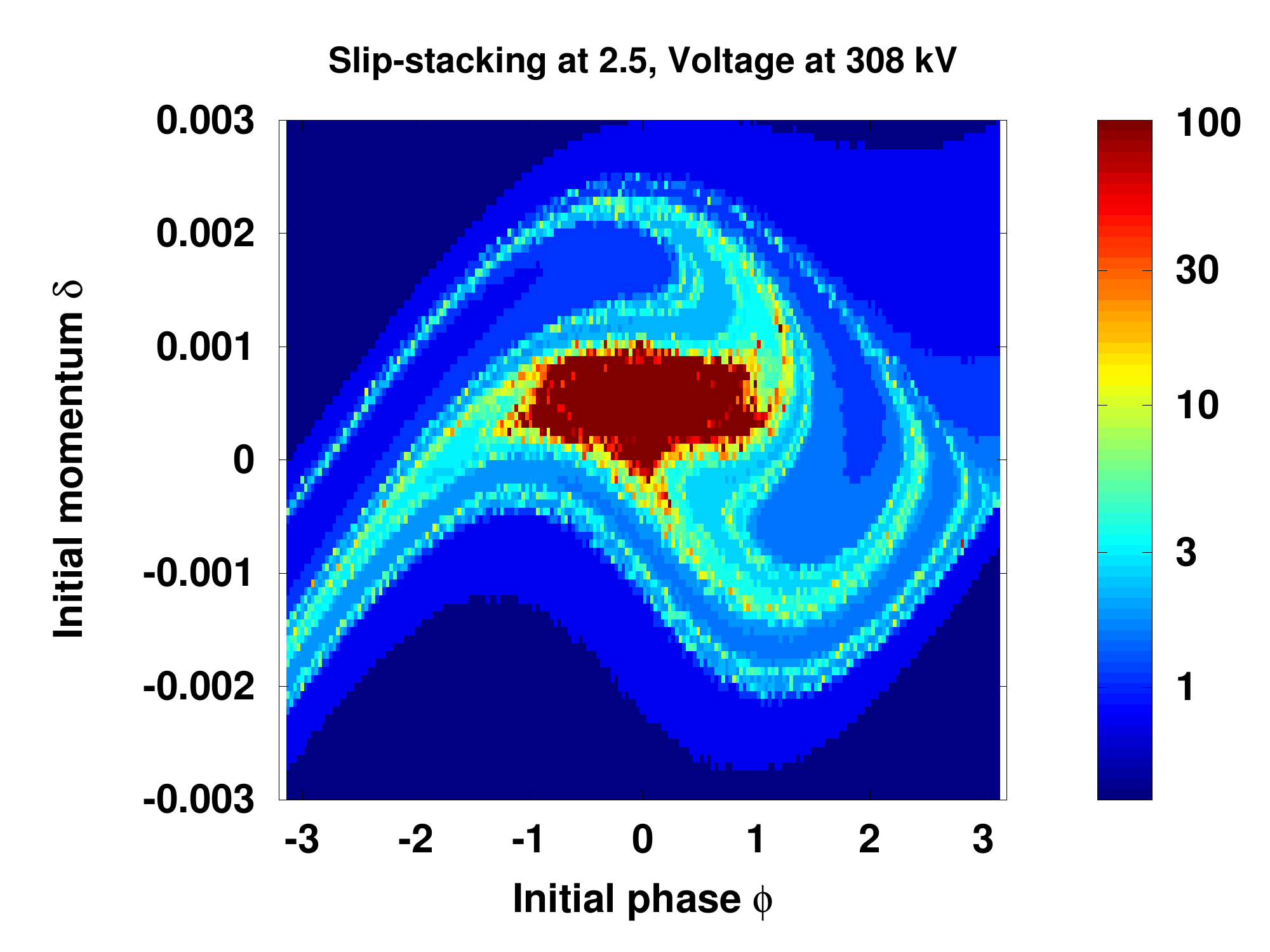}
 \centering
 \includegraphics[scale=0.51]{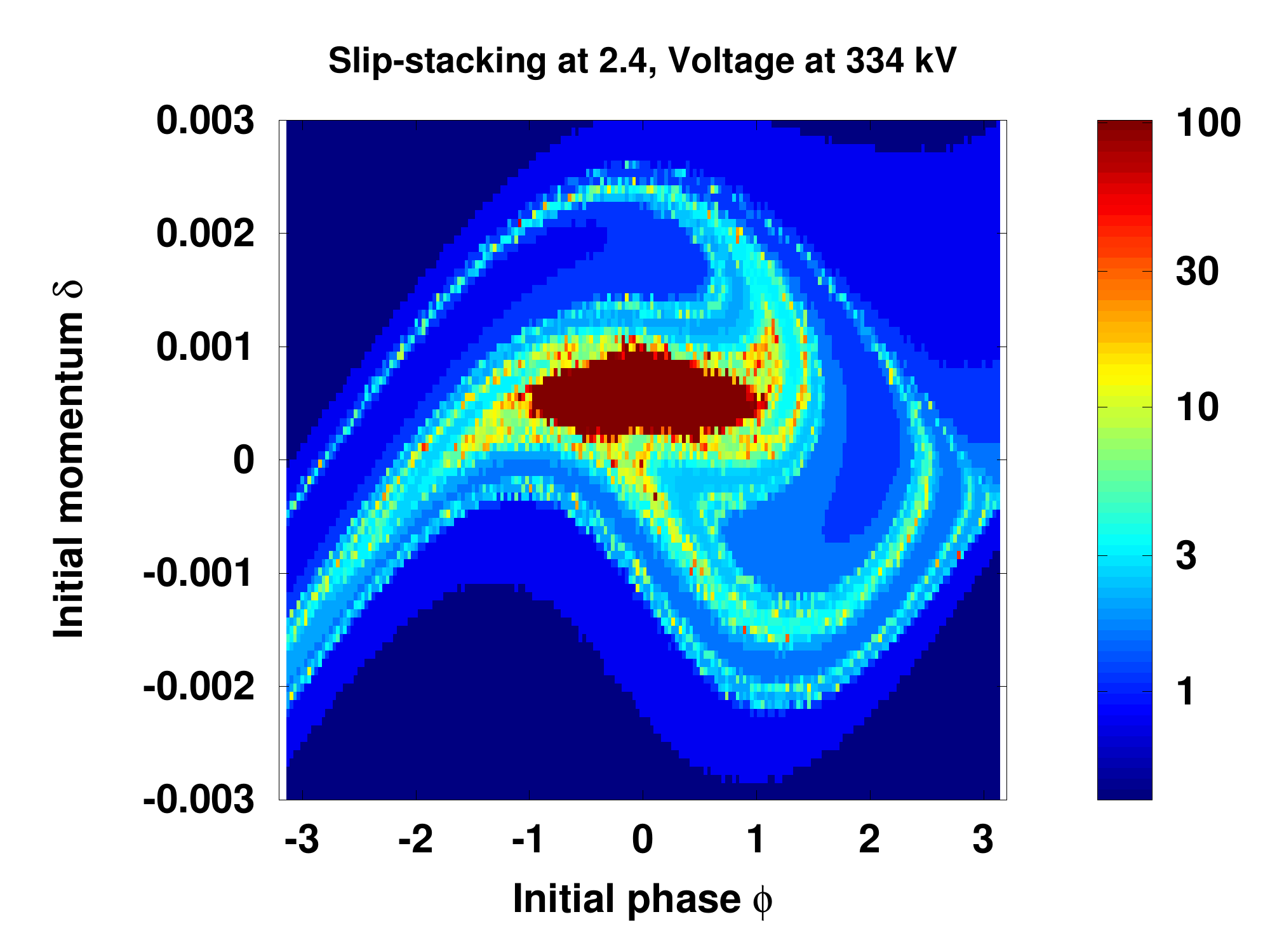}
 \centering
 \includegraphics[scale=0.51]{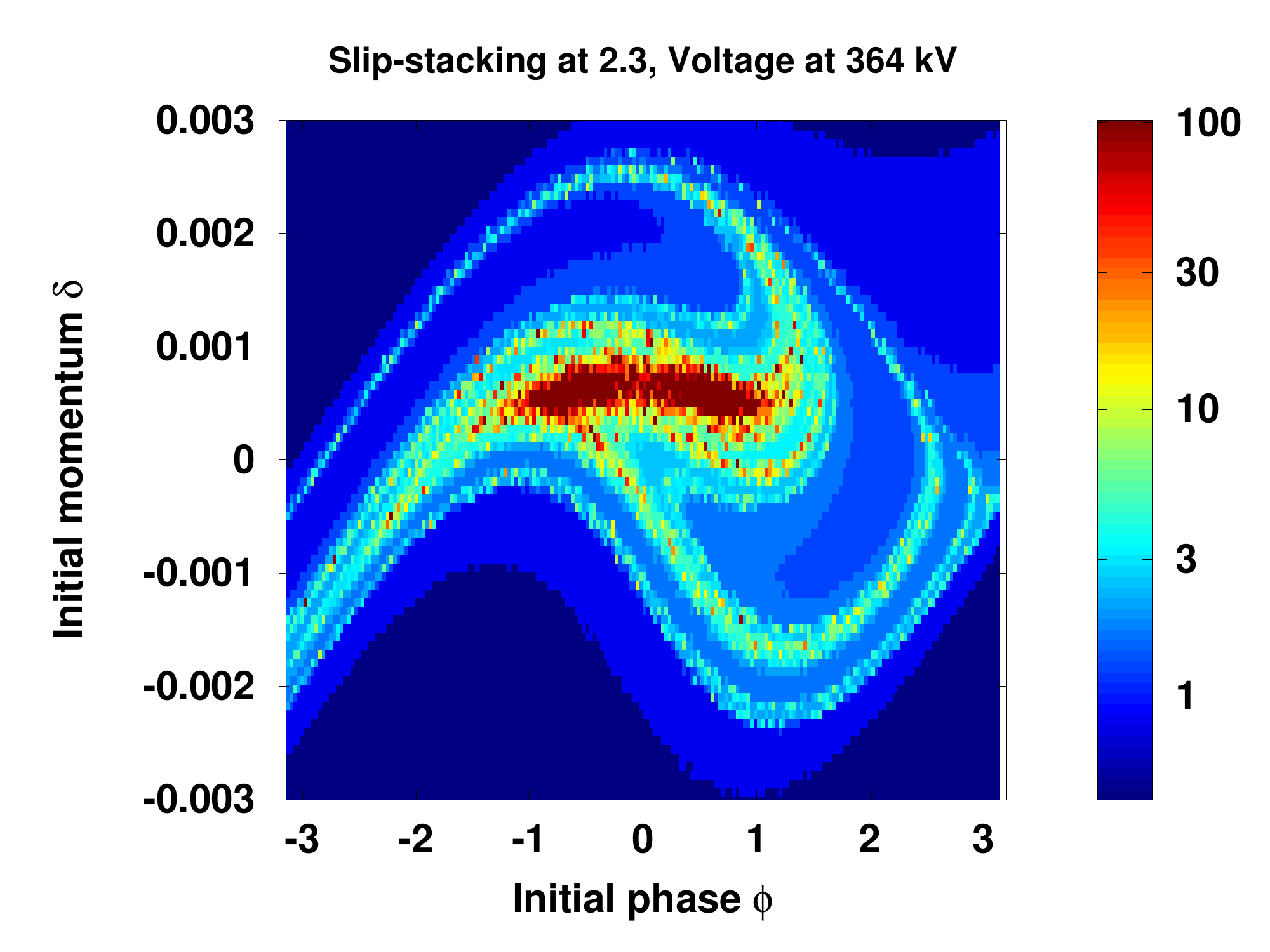}
 \centering
 \includegraphics[scale=0.51]{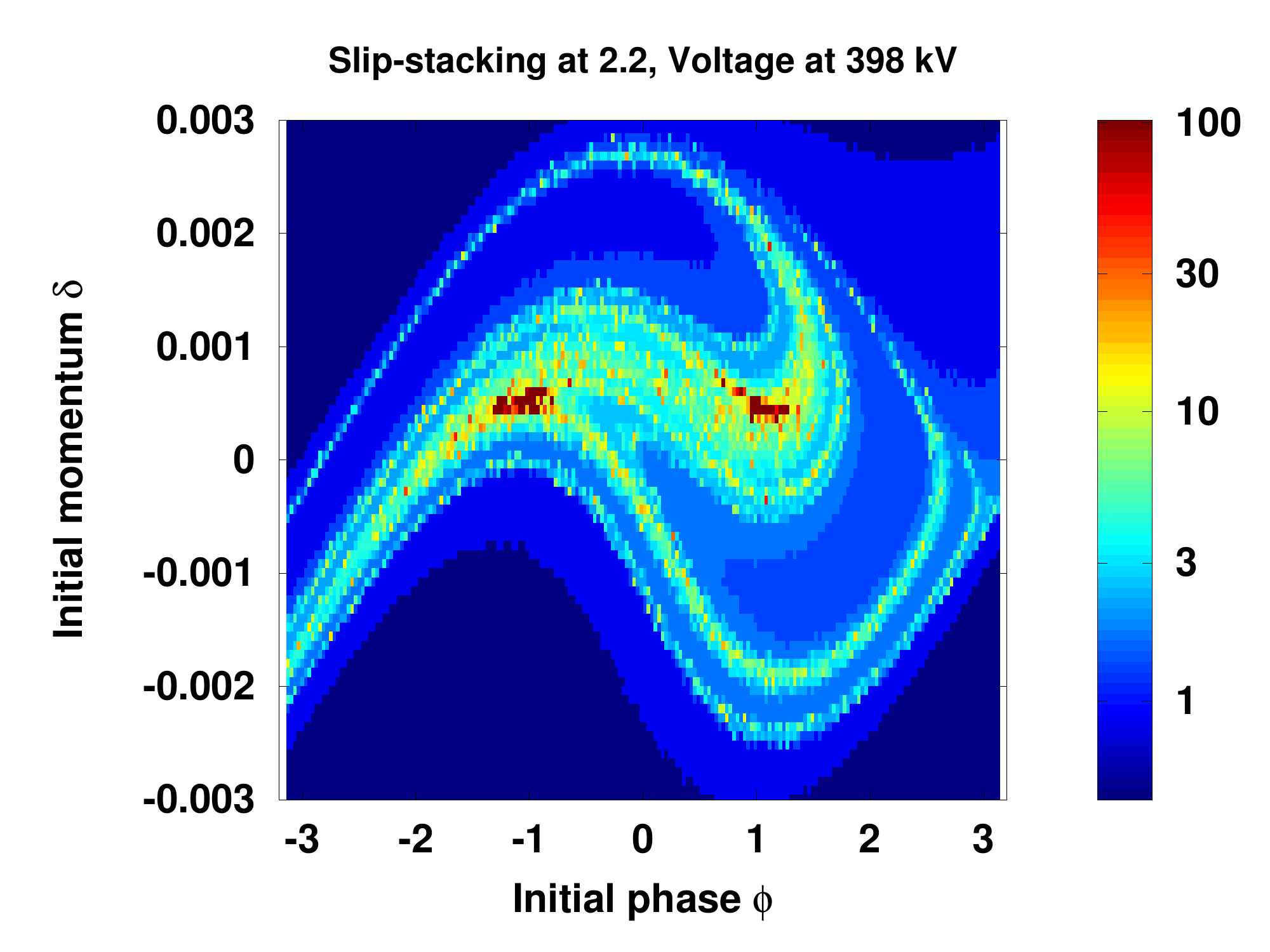}
 \centering
 \includegraphics[scale=0.51]{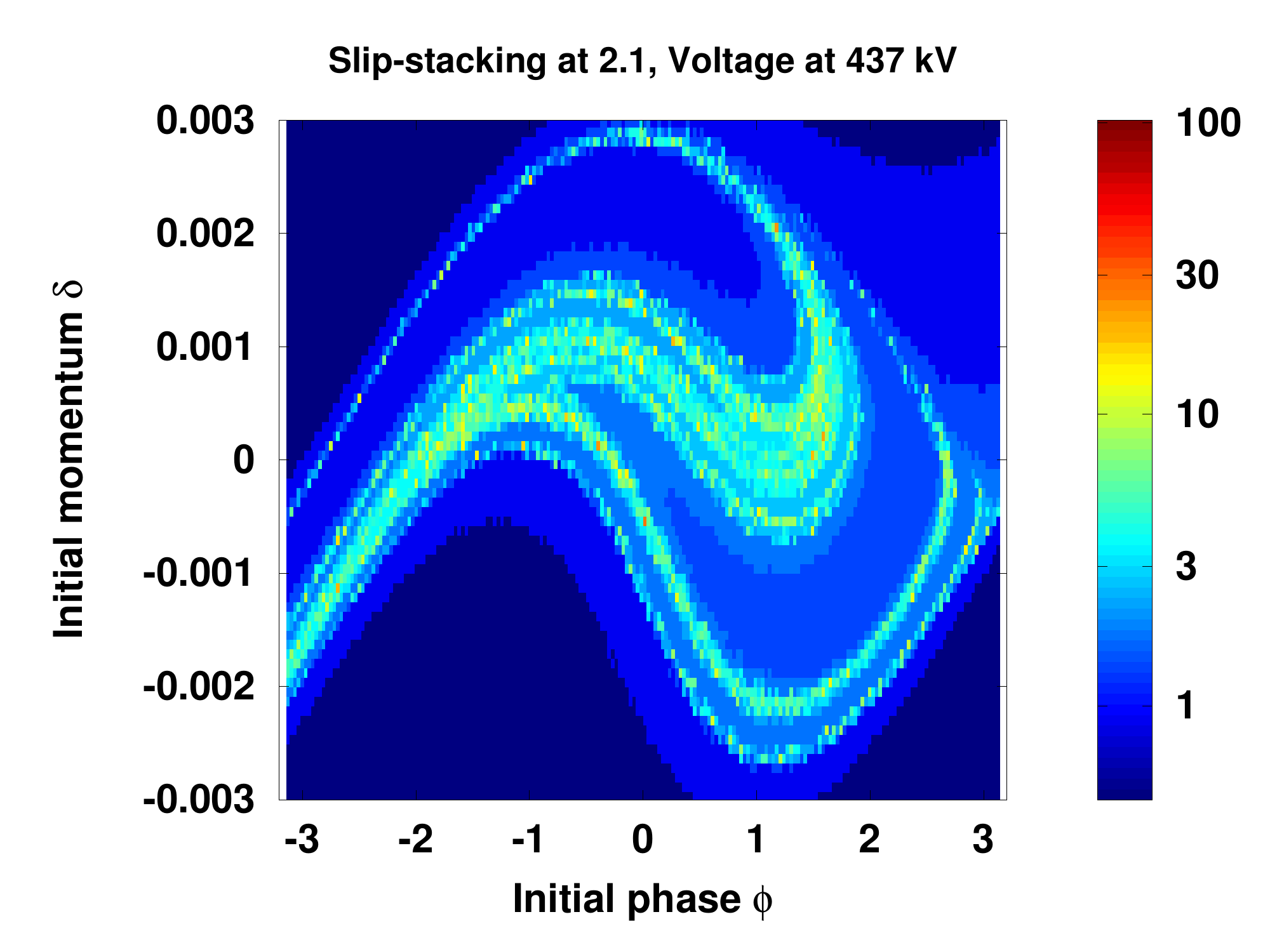}
 \centering
 \includegraphics[scale=0.51]{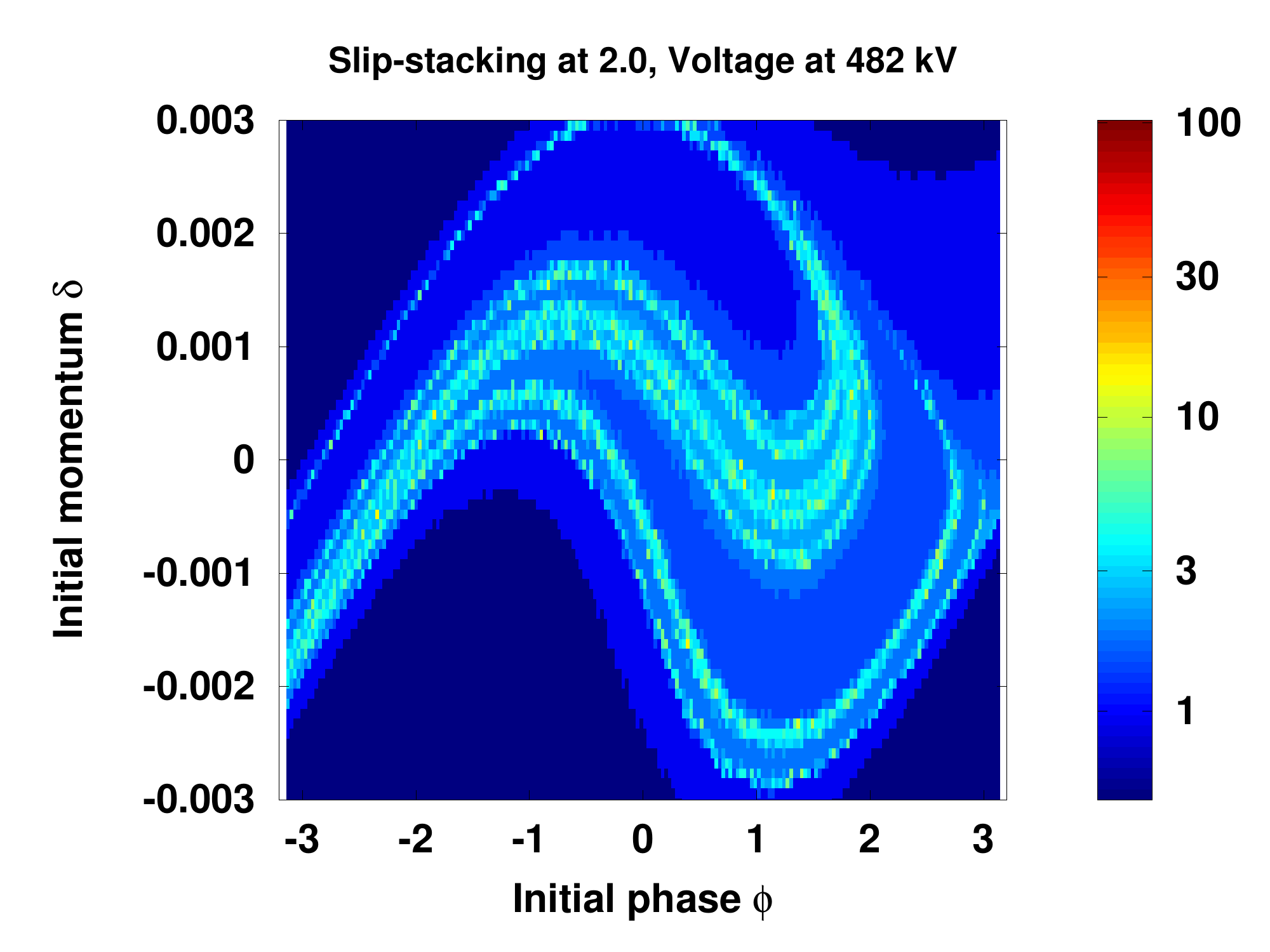}